\documentclass{article}

\usepackage[a4paper, left=3cm, right=3cm]{geometry}
\usepackage{amsmath}
\usepackage{amssymb}
\usepackage{amsthm}
\usepackage{graphicx}
\usepackage{caption}
\usepackage{subcaption}
\usepackage[english]{babel}
\usepackage[colorlinks=true, allcolors=blue]{hyperref}
\usepackage{bbm}
\usepackage[numbers,sort]{natbib}
\usepackage{xcolor}
\usepackage{cleveref}
\usepackage{booktabs}
\usepackage{tabularx}
\usepackage{authblk}
\usepackage{ulem}
\usepackage[usestackEOL]{stackengine}
\usepackage{doi}

\theoremstyle{definition}

\newtheorem{ex}{Example}
\newtheorem{rem}{Remark}

\title{Data-driven Reduction of Transfer Operators \newline for Particle Clustering Dynamics}

\author[1,2]{Nathalie Wehlitz}
\author[3]{Grigorios A. Pavliotis}
\author[1,2]{Christof Schütte}
\author[1]{Stefanie Winkelmann\footnote{Email of corresponding author: \href{mailto:winkelmann@zib.de}{winkelmann@zib.de}}}

\affil[1]{Zuse Institute Berlin, Berlin, Germany}
\affil[2]{Freie Universität Berlin, Berlin, Germany}
\affil[3]{Imperial College London, London, UK}

\date{}

\begin{document}

\maketitle

\begin{abstract}
We develop an operator-based framework to coarse-grain interacting particle systems that exhibit clustering dynamics. Starting from the particle-based transfer operator, we first construct a sequence of reduced representations: the operator is projected onto concentrations and then further reduced by representing the concentration dynamics on a geometric low-dimensional manifold and an adapted finite-state discretization. The resulting coarse-grained transfer operator is finally estimated from dynamical simulation data by inferring the transition probabilities between the Markov states. Applied to systems with multichromatic and Morse interaction potentials, the reduced model reproduces key features of the clustering process, including transitions between cluster configurations and the emergence of metastable states. Spectral analysis and transition-path analysis of the estimated operator reveal implied time scales and dominant transition pathways, providing an interpretable and efficient description of particle-clustering dynamics.

\quad

\end{abstract}

\section{Introduction}

Interacting particle systems in which clustering plays a significant role arise in a wide range of applications, including opinion dynamics \cite{garnier2016, helfmann2021interacting}, swarming and flocking phenomena \cite{topaz2008, carrillo2010swarm}, and biomolecular dynamics \cite{sadeghi2021}.
Such particle dynamics---driven by pairwise interactions and Brownian noise---can exhibit complex clustering behavior, with the specific patterns determined by the form of the interaction potential. For example, locally attractive interaction potentials on a periodic domain give rise to the formation and coalescence of clusters, mass exchange between them, and microscopic reversibility of clustering dynamics~\cite{gerber2025}. 
Related clustering phenomena appear in kinetic (underdamped) Langevin systems, where local attraction leads to metastable multi-cluster states and friction-dependent clustering times, as shown in \cite{leimkuhler2025cluster}. Moreover, on unbounded domains, interacting particle systems display metastable clustering behavior along with a clear separation between the timescales of cluster formation and dissolution~\cite{adams2025}.
Other classes of interactions, such as the globally attractive–repulsive dynamics of multichromatic or Kuramoto-type models~\cite{bertoli2025}, likewise exhibit aggregation and pattern formation, giving rise to a broad family of systems in which effective coarse-grained models are of interest.   

Some aspects of clustering dynamics can also be captured in the mean-field limit via the \textit{McKean--Vlasov PDE}~\cite{dawson1983, gaertner1988}, but crucial effects such as cluster coalescence typically cannot, which motivates the use of stochastic partial differential equations (SPDEs) in form of the \textit{Dean--Kawasaki equation}~\cite{dean1996, kawasaki1998} as an intermediate, continuum-level description.  
In its regularized form~\cite{cornalba2023dean, helfmann2021interacting}, the SPDE provides a scalable model for studying clustering at the level of particle concentrations~\cite{wehlitz2025}. 
Beyond its role as an intermediate continuum description, the Dean–Kawasaki SPDE also enters our study directly: we use it both as a model and as a practical tool for generating concentration data.  Since the SPDE already provides a coarse-grained description of the particle system, a natural question is how to construct an additional, principled coarse-graining of the resulting concentration dynamics.  This motivates the operator-based reduction framework adopted in the present work.

\begin{figure}
    \centering
    \begin{subfigure}{.95\textwidth}
    \includegraphics[width=\textwidth]{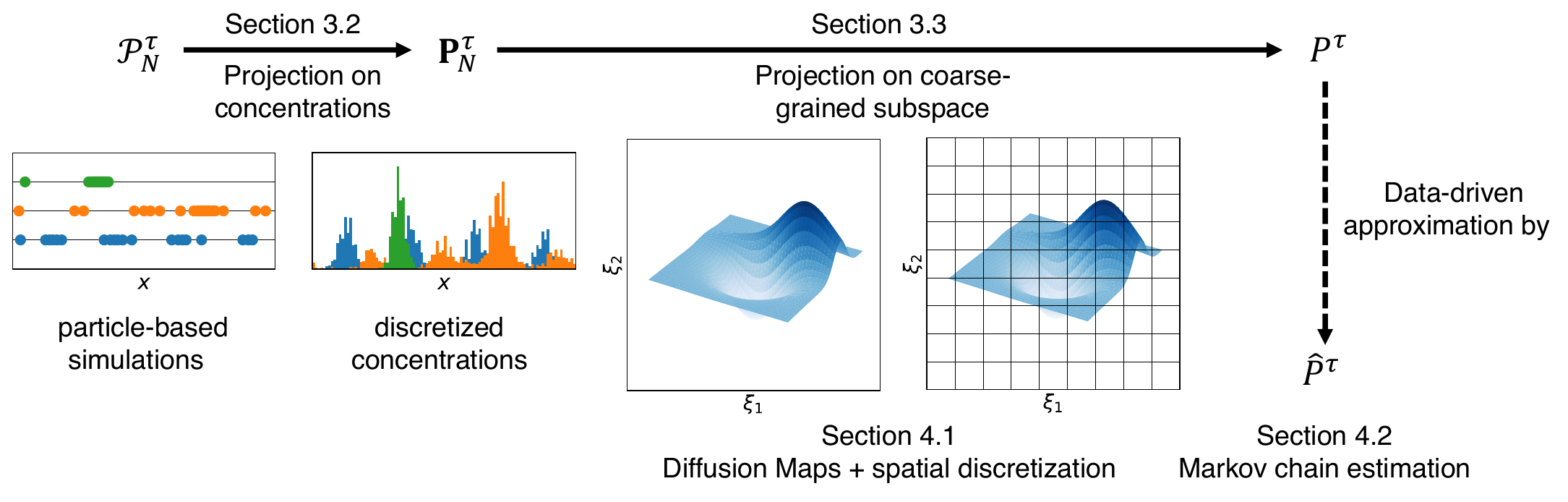}
    \end{subfigure}
    \caption{\textbf{Hierarchy of transfer operators.}
Starting from the Perron--Frobenius operator $\mathcal{P}_N^\tau$ of the particle dynamics~\eqref{eq:Perron-Frobenius} (Section~\ref{sec:TO_exact}), projection onto discretized concentrations (Section~\ref{sec:TO_hist}) yields the finite-dimensional operator $\mathbf{P}_N^\tau$~\eqref{matrix_T}. 
A second projection onto a coarse partition of the concentration space (Section~\ref{sec:TO_coarse}) produces the reduced operator $P^\tau$~\eqref{eq:P_Galerkin}. 
In Section~\ref{sec:model_reduction}, this reduction is implemented in a data-driven manner using concentration data from numerical solutions of the Dean--Kawasaki SPDE: Diffusion Maps provide a low-dimensional geometric embedding (Section~\ref{sec:galerkin_diffmap}), followed by discretization and estimation of the transition matrix $P^\tau$ (Section~\ref{sec:galerkin_msm}).}  
    \label{fig:overview_scheme}
\end{figure}

In this work, we study model reduction for clustering dynamics by following the general transfer-operator paradigm for metastable stochastic processes as formalized in many articles in the literature starting with~\cite{DeuflhardDellnitzJungeSchuette1999,SchuetteFischerHuisingaDeuflhard1999}, see~\cite{schuette2023} for a recent review: the transfer operator associated with the particle dynamics is first projected and reduced to a suitable coarse representation, and the resulting reduced operator is then estimated from dynamical data, see~\Cref{fig:overview_scheme} for an overview. 
In our setting, the first part of this procedure consists of projecting the particle-level transfer operator onto the space of concentrations and equipping this space with an abstract spatial discretization. These steps constitute a purely analytical reduction of the operator and yield a mathematically well-defined coarse operator whose approximation error relative to the full operator does not depend on data.

For a concrete practical implementation based on configuration data, we use \textit{Diffusion Maps}~\cite{coifman2006diffusion, coifman2005geometric} as an exemplary tool to construct a geometry-based reduction of the concentration space. This provides us with a low-dimensional structure on which we define a Markov-state partition. By using dynamical simulation data, we estimate the transition probabilities between the Markov states, resulting in a data-driven approximation of the reduced transfer operator.   
We apply the data-based two-step procedure---dimensionality reduction followed by dynamics estimation---to two representative examples of interaction: the globally attractive--repulsive multichromatic potential~\cite{bertoli2025} and the locally attractive Morse potential~\cite{carrillo2019, wehlitz2025}. However, we point out that our approach is equally applicable to a broad class of particle systems exhibiting clustering behavior and thus provides a general framework for model reduction in this setting.

Our operator-oriented viewpoint relates our approach to the methodologies in~\cite{helfmann2021, koltai2020}, which likewise combine geometric reduction of state space with data-driven estimation of reduced finite-state dynamics, although in different application domains. Methods based on collective variables in molecular dynamics~\cite{Sadeghi2023} similarly rely on predefined or data-informed low-dimensional representations before estimating effective Markovian dynamics in the reduced coordinates. The application of Diffusion Maps to reduce histogram data is related in spirit to~\cite{evangelou2024}, although in that work the reduced coordinates are used to fit a data-driven ODE rather than to construct a coarse-grained transfer operator. Dimensionality reduction applied to time series data is also employed in the study of temporal networks~\cite{blaskovic2025} to identify clusters of time snapshots characterized by similar network structures.

Data-driven manifold-learning methods have also been applied in other contexts involving particle or tracer data. For example, diffusion-map approaches have been used to extract coherent flow structures in fluid dynamics, such as in the quantification of scalar mixing from particle tracks~\cite{kluenker2025} and in the study of Lagrangian coherent sets in turbulent Rayleigh--Bénard convection~\cite{schneide2019}. These works share with ours the use of nonlinear dimensionality reduction to uncover low-dimensional structure, but focus on advective transport and mixing rather than on coarse-grained dynamics of clustering. 

Our approach contrasts with the data-driven approximation in~\cite{ioannou2025data}, where \textit{extended dynamic mode decomposition (EDMD)} is used to build a finite-dimensional approximation of the transfer operator associated with the mean-field (decoupled McKean–Vlasov) stochastic differential equation, 
and where the resulting operator is analyzed spectrally to identify coherent or metastable behavior. In our setting, the final Markov chain plays an analogous role: given the Markov process on the reduced space, we analyze its spectral properties to identify metastable behavior, implied timescales, and transition pathways, following the approaches of~\cite{roeblitz2013, metzner2009}.

From an applied-mathematics perspective, the main strength of the proposed framework is that the reduced dynamics are mathematically defined (on the level of transfer operators) before any data-driven approximation is introduced. The method preserves the collective structure of clustering dynamics---namely the number, size, and spatial arrangement of clusters---without relying on predefined reaction coordinates. 
The use of Diffusion Maps provides an intrinsic low-dimensional representation adapted to the concentration data, and the final Markov approximation turns the reduced dynamics into an interpretable probabilistic object on which standard tools such as spectral analysis, implied timescales, PCCA+, and transition-path methods can be applied directly. Altogether, the framework combines analytical structure, data-adapted geometry, and computational tractability in a way that is broadly transferable to other systems with emergent collective organization.

The paper is structured as follows.  The particle-based model and its SPDE approximation are formulated in~\Cref{sec:Model setup}, followed by the analytical reduction of the transfer operator in~\Cref{sec:TO}. \Cref{sec:model_reduction} explains the data-driven approximation of the reduced transfer operator using the two-step procedure of Diffusion Maps and Markov-chain construction. The analysis of the reduced model for metastability and implied timescales is presented in \Cref{sec:analyzeMSM}.

\section{Model formulation}
\label{sec:Model setup}

In~\Cref{sec:model}, we introduce the particle-based model for the stochastic interaction–diffusion dynamics together with its approximation by the Dean--Kawasaki SPDE. The two representative interaction potentials used throughout this work are presented in~\Cref{sec:examples}.

\subsection{Particle-based dynamics and SPDE approximation}\label{sec:model}

We study a system of $N\in\mathbb{N}$ particles moving on the one-dimensional torus $\mathbb{T}$ of length $L>0$, 
$\mathbb{T} := \mathbb{R}/(L\mathbb{Z})$ which we identify with $[-\tfrac{L}{2},\tfrac{L}{2}).$

The configuration of the system at time $t\ge 0$ is given by $\boldsymbol{X}(t)=(X_1(t),\dots,X_N(t))\in \mathbb{X}$ for $\mathbb{X}:=\mathbb{T}^N$, where the coordinate $X_i(t)\in\mathbb{T}$ describes the position of particle $i\in\{1,\dots,N\}$. Their motion is governed by the coupled stochastic differential equations
\begin{equation}
\label{eq:PBD_alt}
dX_i(t) = - \frac{1}{N}\sum_{j=1}^N U'\big(X_i(t)-X_j(t)\big)\, dt \;+\; \sigma\, dW_i(t),
\qquad i=1,\dots,N,
\end{equation}
where the solution $\boldsymbol{X}(t)$ is understood modulo $L$. Here, $U:\mathbb{R}\to \mathbb{R}$ denotes an interaction potential and $U'(x)=\tfrac{d}{dx}U(x)$. The processes $W_1,\dots,W_N$ are independent standard Brownian motions, and $\sigma>0$ is a fixed diffusion parameter.
We will refer to the stochastic system~\eqref{eq:PBD_alt} as the \textit{particle-based dynamics}. The formulation readily extends to higher-dimensional spatial domains, but we focus on the one-dimensional case for clarity of presentation.

The setting is motivated by membrane-mediated receptor kinetics, as discussed in~\cite{sadeghi2021}. However, analogous clustering phenomena also arise in other systems, such as social interaction kinetics, where clustering could correspond to consensus formation in opinion dynamics. The dynamics considered here are mass-conserving and non-reactive, in contrast to biochemical reaction–diffusion systems in which particle numbers vary due to chemical reactions~\cite{winkelmann2020stochastic}. We impose periodic boundary conditions, implying that the modeled domain represents a small region of a much larger system and that curvature and edge effects can be neglected.

The interacting particle system \eqref{eq:PBD_alt} belongs to the class of interacting overdamped Langevin dynamics, which has been studied in various contexts; see, for example, \cite{garnier2016, helfmann2021interacting} for opinion dynamics, \cite{topaz2008, carrillo2010swarm} for swarming phenomena, and \cite{sadeghi2021, wehlitz2025} for biomolecular dynamics. 
Building on this well-established modeling framework, we develop an operator-based coarse-graining approach tailored to clustering phenomena in such systems.

When studying cluster formation and evolution, the exact positions of individual particles are of secondary importance; instead, all relevant information is captured by the population state, defined by the number (or concentration) of particles as a function of position. This motivates formulating the dynamics directly at this coarse-grained level. To retain stochastic effects, which play a crucial role in clustering, we consider a stochastic partial differential equation (SPDE) rather than the corresponding mean-field limit given by a deterministic partial differential equation (PDE).

\paragraph{Approximation by the Dean--Kawasaki equation.} 
Let $c(x,t)$ denote the particle concentration as a function of spatial location $x\in \mathbb{T}$ and time $t\geq 0$. For our numerical experiments, we approximate the particle-based dynamics~\eqref{eq:PBD_alt}  
by the corresponding \textit{stochastic partial differential equation (SPDE)}
\begin{align} \label{eq:Dean-Kawasaki}
    \partial_t c(x,t) = \partial_x (c(x,t) (U' * c(\cdot, t))(x)) + \frac{\sigma^2}{2}  \partial_{xx} c(x,t) + \frac{\sigma}{\sqrt{N}}\partial_x \big(\sqrt{c(x,t)}\,Z(x,t)\big),
\end{align}
where 
\begin{equation}
    (U'* c(\cdot, t))(x):= \int_{\mathbb{T}} U'(x-y)c(y,t)\, dy
\end{equation}
is the convolution between the interaction force $U'$ and the concentration $c$. Here, $Z(x,t)$ denotes space-time white noise, i.e., a spatiotemporal (generalized) Gaussian random field with
\begin{equation}
 \mathbb{E}\left(Z(x,t)\right)=0, \quad  \quad  \mathbb{E}\left(Z(x,t)Z(x',t')\right)= \delta(x-x')\delta(t-t'),   \quad \quad \forall t,t'\geq 0, \forall x,x'\in  \mathbb{T},
\end{equation}
where $\delta(x)$ denotes the Dirac delta distribution. 
Equation~\eqref{eq:Dean-Kawasaki} is commonly referred to as the \textit{Dean--Kawasaki equation}~\cite{dean1996,kawasaki1998}. 
In the SPDE literature, its solutions are typically called \textit{fluctuating densities}. 
In this manuscript, however, we use the term \textit{concentration} (or concentration profiles) to avoid confusion with probability densities on configuration space.

We recall that the Dean--Kawasaki equation is mathematically ill-defined as an SPDE, since the multiplicative noise term $\partial_x(\sqrt{c} \,Z)$ is not well posed on the level of function. 
In fact, the only formal (martingale) solutions are given by empirical measures of the underlying particle system~\cite{konarovskyi2019}. Consequently, one typically works with regularized or coarse-grained versions of~\eqref{eq:Dean-Kawasaki}. 
The fact that regularized solutions of this SPDE are practical tools for replicating particle-based clustering dynamics has been demonstrated in~\cite{wehlitz2025}, where it was shown that such models reproduce both the initial cluster formation and the long-term merging dynamics that deterministic mean-field approaches fail to capture.

\subsection{Exemplary interaction potentials}\label{sec:examples}

In this work, we consider the following two exemplary types of interaction rules, both introducing local clustering behavior of the particles. 

\begin{figure}
    \centering
    \begin{subfigure}{0.49\textwidth}
    \includegraphics[width=\textwidth]{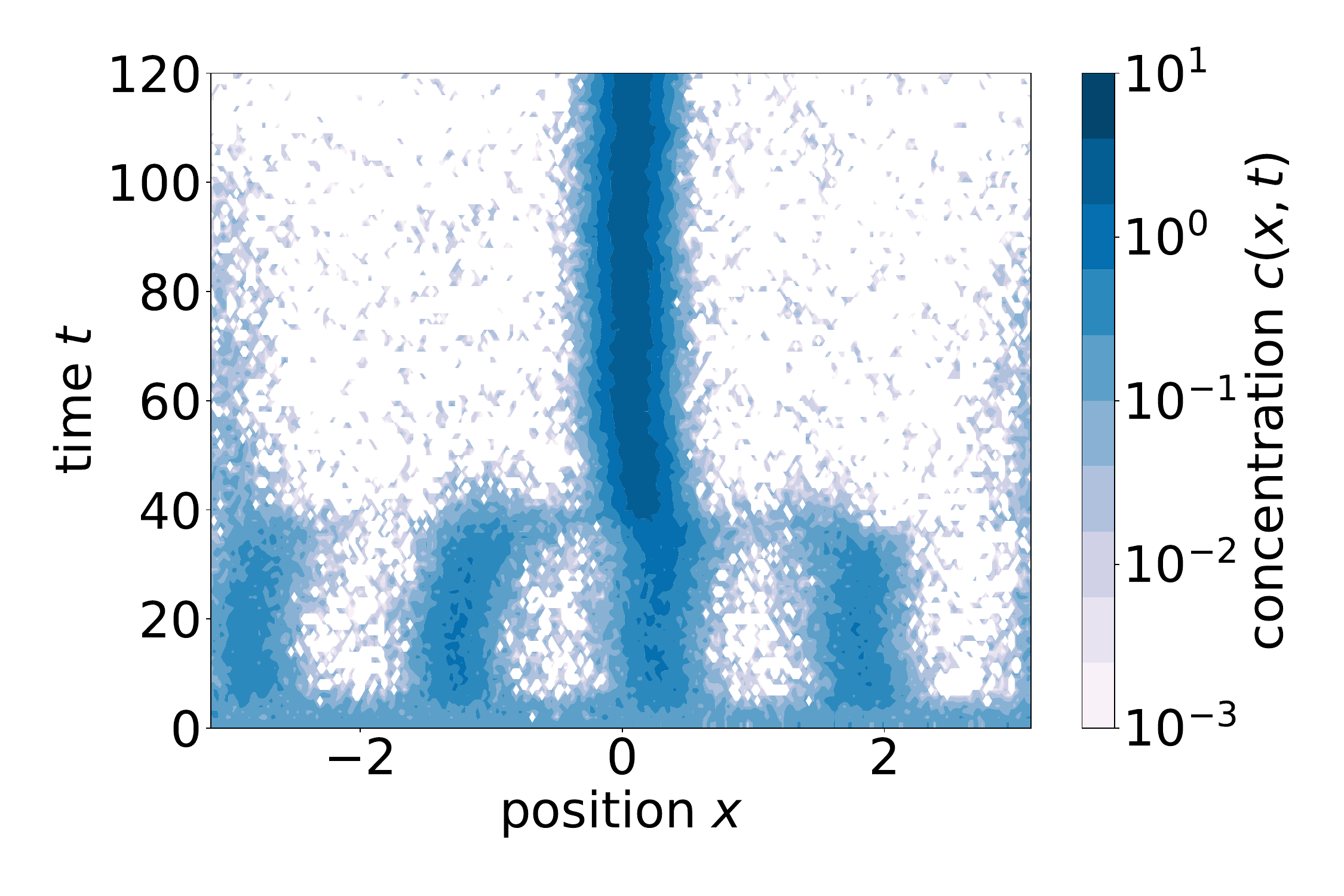} 
    \caption{Multichromatic potential (\Cref{ex1})}
    \label{fig:diffusion_maps_SPDE_multichrom_a}
    \end{subfigure}
        \begin{subfigure}{0.49\textwidth}
    \includegraphics[width=\textwidth]{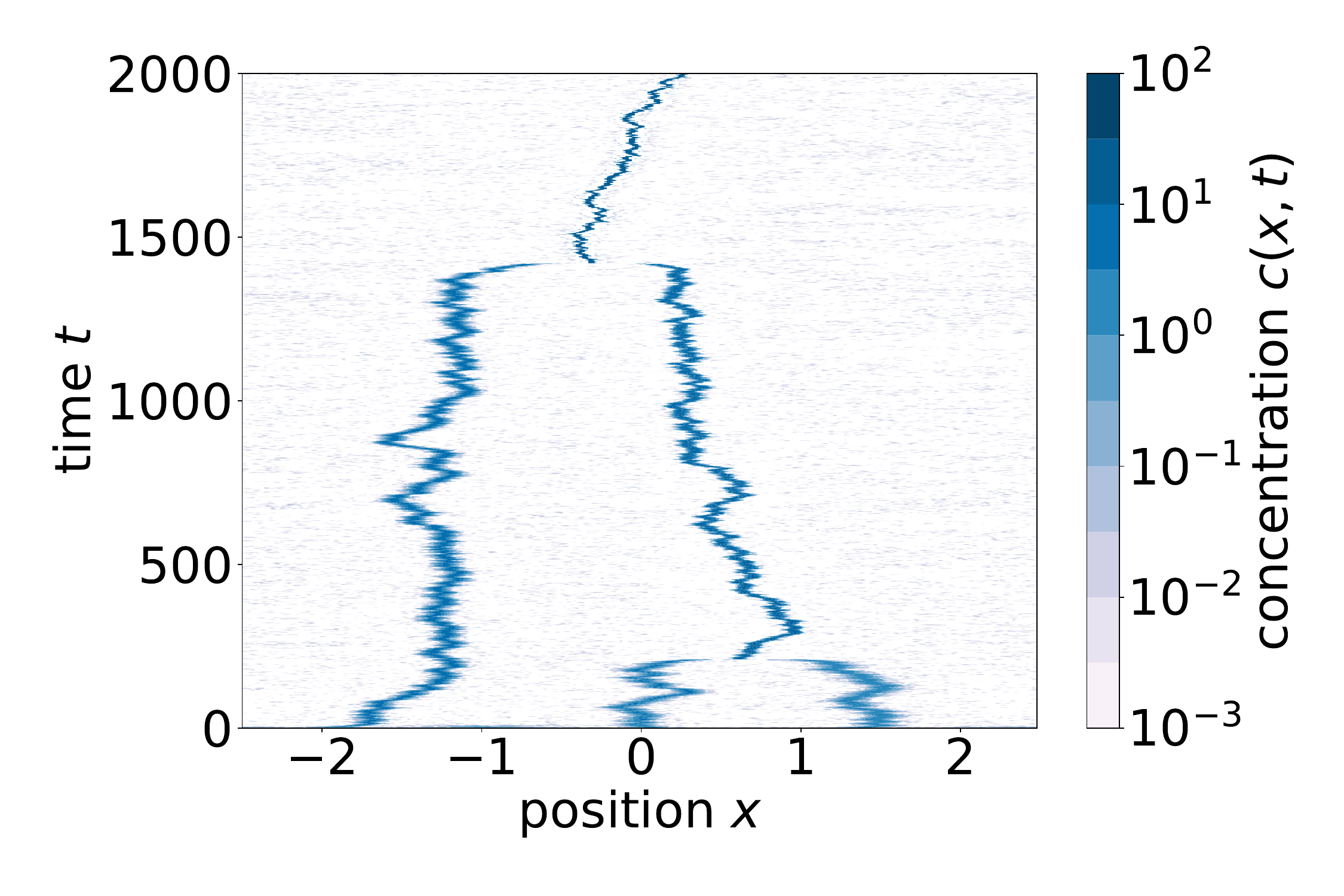}
    \caption{Morse potential (\Cref{ex2})}
    \label{fig:diffusion_maps_SPDE_Morse_a}
    \end{subfigure}
    \caption{\textbf{Exemplary trajectories for the interaction potentials in Examples~\ref{ex1} and \ref{ex2}.} 
    Single SPDE simulations of clustering dynamics using (a) the multichromatic potential given in~\eqref{eq:F} up to $T=120$ and (b) the Morse potential given in~\eqref{eq:F_Morse} up to $T=2000$, both starting from a uniform distribution at $t=0$. The parameter values used in the simulations are given on page~\pageref{parameter_values}.}  \label{fig:diffusion_maps_SPDE_trajectories}
\end{figure}

\begin{ex}\label{ex1}\textbf{Multichromatic interaction potential.} 
Motivated by the analysis in~\cite{bertoli2025}, we employ a multichromatic interaction potential of the form  
\begin{equation}\label{eq:F}
    U(x)=1-\cos(x)-a\cos(4x), \qquad x\in \mathbb{T}, \; a >0,
\end{equation}
which combines the first and fourth Fourier modes on a torus whose length $L$ is a multiple of $2\pi$. In contrast, there are the monochromatic potentials $U(x)=-\cos(kx), k\in\mathbb{N}$, which define the \textit{generalized Kuramoto model}. In~\cite{bertoli2025}, it was shown that multichromatic interaction potentials can give rise to rich phase behavior and multipeak stationary states, with the number of peaks linked to the nonzero Fourier modes of the interaction. Inspired by this mechanism, we adopt a similar potential to introduce competing length scales in the particle interactions: the $\cos(x)$ term promotes aggregation at a characteristic distance, while the higher harmonic $\cos(4x)$ introduces a finer structure that can stabilize multiple clusters or subclusters. 
Throughout this manuscript, we use the parameter values $L=2\pi$ and $a=0.25$ for this example. 
With these values, the interaction force induced by the potential $U$ combines short-range attraction with secondary repulsive effects at intermediate distances,
leading to richer clustering behavior than a purely monochromatic potential. An exemplary trajectory of the dynamics is plotted in~\Cref{fig:diffusion_maps_SPDE_multichrom_a}.
\end{ex}

\begin{ex}\label{ex2}\textbf{Morse potential.} As a second example, and in contrast to the periodic multichromatic potential of \Cref{ex1}, we consider the generalized \textit{Morse potential}~\cite{carrillo2019, dorsogna2006}, 
\begin{equation}\label{eq:F_Morse}
    U(x)=-C_a \, e^{-|x|/l_a} + C_r \, e^{-|x|/l_r}, \qquad x\in \mathbb{T},
\end{equation}
where $l_a, l_r > 0$ denote the length scales of the attraction and repulsion, respectively, and ${C_a, C_r > 0}$ are their corresponding strengths.
In general, this potential is used for parameter values that induce short-range repulsion combined with long-range attraction---a canonical mechanism underlying self-organization and swarming behavior \cite{leverentz2009, topaz2008}. 
For our numerical investigations, however, we choose parameters that lead to effective local attraction, where the repulsive component primarily regulates the attraction. 
Throughout this manuscript, we fix 
$L=5$, $C_a=4$, $l_a=\frac{1}{40}L$, $C_r=1$, and $l_r=\frac{1}{100}L$. 
This parameter regime yields clustering dynamics as analyzed in~\cite{wehlitz2025}. An exemplary trajectory of the dynamics is plotted in~\Cref{fig:diffusion_maps_SPDE_Morse_a}.
\end{ex}

\paragraph{Simulation parameter values.}\label{parameter_values} All numerical simulations of particle concentrations are performed for $N=10^3$ particles moving on a torus of length $L>0$ with periodic boundary conditions. Regularized solutions to the SPDE~\eqref{eq:Dean-Kawasaki} are obtained via a finite difference scheme \cite{cornalba2023dean,wehlitz2025} with grid size $h=L\cdot 2^{-8}$ and time step $dt_{\text{sim}}=0.001$. In Examples~\ref{ex1} and \ref{ex2}, the diffusion coefficient is set to $\sigma=0.4$, which in both cases is below the respective critical noise strengths, thus enabling clustering behavior~\cite{carrillo2020, bertoli2025}. 

\quad

In both settings we observe characteristic clustering dynamics, see~\Cref{fig:diffusion_maps_SPDE_multichrom_a} and \Cref{fig:diffusion_maps_SPDE_Morse_a}. Starting from an initially uniform distribution, the particles rapidly aggregate into several clusters, which then persist over substantial time intervals. In the multichromatic case of~\Cref{ex1}, the cluster positions and separations remain comparatively stable, reflecting the structure of the interaction force~\eqref{eq:F}. By contrast, under the Morse potential~\eqref{eq:F_Morse} of~\Cref{ex2}, the cluster centers continue to move in space. Over time, clusters may merge either through the dissolution of one cluster, whose particles are absorbed by others, or through the direct collision and coalescence of two clusters. The characteristic times between successive cluster merges increase roughly exponentially as the system evolves, reflecting the progressive slowdown of the dynamics as the number of clusters decreases and the remaining clusters become larger and more stable.
Ultimately, the system evolves toward a single surviving cluster, while reverse events of cluster splitting are highly unlikely and have never been observed throughout the simulation time. 

These observations highlight the emergence of slow, low-dimensional structures in the dynamics, governed by a few collective variables such as the number and relative positions of clusters. To systematically characterize these structures and their evolution, we next introduce the transfer operator framework, which enables a probabilistic and reduced description of the dynamics and forms the basis for the subsequent coarse-grained analysis.

\section{Analytical reduction of the transfer operator}
\label{sec:TO}

To analyze the long-term and collective behavior of the stochastic particle system, we adopt the transfer-operator (Perron--Frobenius) perspective.
This framework describes the time evolution of probability distributions rather than individual trajectories and thus provides a natural bridge between microscopic dynamics and coarse-grained, population-level descriptions. The constructions in this section are classical in transfer-operator theory. Their purpose here is not to introduce new operator-theoretic results, but to formulate the analytical framework underlying our data-driven reduction. In particular, this formulation clarifies which reduced operator is approximated in Section~\ref{sec:model_reduction} and how the SPDE-based implementation relates to the underlying particle dynamics.

We proceed in three conceptual steps (see also~\Cref{fig:overview_scheme}). 
First, we introduce the exact Perron--Frobenius operator associated with the particle-based dynamics (\Cref{sec:TO_exact}), which is formally consistent with the empirical-measure interpretation of the Dean--Kawasaki SPDE (Section~2.1). 
Second, we project this operator onto a finite-dimensional space of discretized concentrations via a spatial Galerkin discretization (\Cref{sec:TO_hist}), mirroring the numerical discretization of the SPDE. 
Finally, we apply a second Galerkin projection onto a coarse partition of the concentration space (\Cref{sec:TO_coarse}), yielding the reduced operator that is approximated in Section~\ref{sec:model_reduction}.

\subsection{Transfer operator of the particle-based system}
\label{sec:TO_exact}

For a time-homogeneous Markov process $\boldsymbol{X}(t)\in\mathbb{X}$ with transition density
$p(\boldsymbol{y},\tau \, |\, \boldsymbol{x})$, the Perron--Frobenius (transfer) operator
$\mathcal{P}_N^\tau : L^1(\mathbb{X}) \to L^1(\mathbb{X})$ \cite{lasota1994}
propagates probability distributions according to
\begin{equation} \label{eq:Perron-Frobenius}
(\mathcal{P}_N^\tau \rho)(\boldsymbol{y})
= \int_{\mathbb{X}} p(\boldsymbol{y},\tau \,|\,\boldsymbol{x})\,\rho(\boldsymbol{x})\,d\boldsymbol{x}.
\end{equation}
The operators $\mathcal{P}_N^\tau$ form a Markov semigroup with infinitesimal generator
$\mathcal{L}^{\!*}$, i.e.,
\[
\rho_{t+\tau} = \mathcal{P}_N^\tau \rho_t,
\qquad
\mathcal{P}_N^\tau = e^{\tau \mathcal{L}^{\!*}} .
\]

In our $N$-particle setting on $\mathbb{X}=\mathbb{T}^N$ with pairwise interaction potential~$U$ and noise amplitude~$\sigma$ (see \Cref{eq:PBD_alt}), the generator takes the explicit form~\cite{gardiner2009stochastic} 
\begin{equation}
(\mathcal{L}^{\!*}\rho)(\boldsymbol{x})
= \sum_{i=1}^N \partial_{x_i}\!\left[\frac{1}{N}\sum_{j=1}^N U'(x_i-x_j)\,\rho(\boldsymbol{x})\right]
\;+\;\frac{\sigma^2}{2}\sum_{i=1}^N \partial_{x_i x_i}\rho(\boldsymbol{x}), 
\end{equation}
where $\boldsymbol{x}=(x_1,\dots,x_N)$ with $x_i$ denoting the position of particle $i$.
Thus, the transfer operator $\mathcal{P}^\tau_N$ propagates the joint probability distribution of the $N$-particle system forward in time according to the Fokker--Planck equation.

On a formal level, the Dean--Kawasaki SPDE introduced in \Cref{sec:model} provides a stochastic evolution equation for the empirical density of the interacting particle system. 
Its (formal) solutions coincide with the empirical measure associated with the particle process and therefore describe the same evolution of probability measures. Thus, the particle-level transfer operator introduced here can also be regarded as the operator governing the SPDE dynamics at the measure level.

\subsection{Projection onto discretized concentrations} \label{sec:TO_hist}

The continuous transfer operator~$\mathcal{P}_N^\tau$ introduced above acts on probability distributions over the $N$-particle configuration space~$\mathbb{T}^N$. To obtain a computable representation of this operator, we project it onto a finite-dimensional function space associated with a spatial discretization of the physical domain~$\mathbb{T}$. Concretely, we replace the full particle configuration by a coarse-grained concentration. 

We introduce a uniform partition $(B_k)_{k=1}^K$ of the torus~$\mathbb{T}$ into boxes of 
width~$\Delta = L/K$. The discretized concentration is the piecewise constant function 
\begin{equation}
    c_\Delta(x,t) = \frac{1}{N \,\Delta}\sum_{i=1}^N \mathbf{1}_{B(x)}(X_i(t))
\end{equation}
where $B(x)$ denotes the unique box containing $x\in \mathbb{T}$. By construction, $c_\Delta(x,t)$ is nonnegative and integrates to $1$. Since it depends
on the particle process $\boldsymbol{X}(t)$, the concentration $c_\Delta$ is itself
a stochastic process.
 
For fixed $N$ and spatial resolution $\Delta$, only finitely many distinct concentrations $c_\Delta$ can occur, because they are determined by the integer particle counts in the $K$ boxes. Let these distinct states be denoted by $c^{(1)}, \dots, c^{(n_c)}$, and let
\[
\mathbb{F} := \{c^{(1)},\dots,c^{(n_c)}\}
\]
be the resulting finite concentration state space. There exists a mapping
$f : \mathbb{T}^N \to \mathbb{F}$ that associates each particle configuration
$\boldsymbol{x}$ with its coarse-grained concentration.

We define the finite-dimensional space of functions over $\mathbb{F}$ as
\[
    \mathcal{F}_{n_c} := \mathrm{span}\{\chi_i : i=1,\dots,n_c\},
\]
where each $\chi_i$ is the characteristic function of the state $c^{(i)}$:
\[
    \chi_i(\boldsymbol{x}) =
    \begin{cases}
        1, & f(\boldsymbol{x}) = c^{(i)}, \\
        0, & \text{otherwise}.
    \end{cases}
\]

The Galerkin projection $Q : L^1(\mathbb{T}^N) \to \mathcal{F}_{n_c}$ is the projection onto $\mathcal{F}_{n_c}$ obtained by averaging over the elements of the concentration partition. It is given by
\[
    Q\rho 
    = \sum_{i=1}^{n_c}
      \frac{\langle \chi_i, \rho \rangle}{\langle 1, \chi_i \rangle} \, \chi_i,
\]
where $\langle f,g\rangle := \int_{\mathbb{T}^N} f(\boldsymbol{x})\, g(\boldsymbol{x}) \, d\boldsymbol{x}$ denotes the dual pairing between $L^\infty$ and $L^1$.
Since our basis functions are indicators, they belong to both spaces.

The projected transfer operator admits a matrix representation $\mathbf{P}^\tau_{\!N} \in \mathbb{R}^{n_c\times n_c}$
with entries
\begin{equation} \label{matrix_T}
   [\mathbf{P}^\tau_{\!N}]_{ij}
   = \frac{\langle \chi_i, \mathcal{P}_N^\tau \chi_j \rangle}
           {\langle 1, \chi_j \rangle}.
\end{equation}
This matrix is column-stochastic and thus represents the Perron–Frobenius
operator on the finite state space $\mathbb{F}$:
\[
    \mathbf{P}^\tau_{\!N} : \ell^1(\mathbb{F}) \to \ell^1(\mathbb{F}).
\]

\begin{rem}
    For very large $N$, the number of distinct coarse-grained concentration states
is high because each spatial bin can assume many closely spaced
values.  A further reduction could be achieved by discretizing the
value range of the concentrations themselves, i.e., by grouping nearby
concentration levels into larger bins.
\end{rem}

\paragraph{Relation to the SPDE discretization.}
The numerical solutions of the Dean--Kawasaki SPDE~\eqref{eq:Dean-Kawasaki} are represented on the same spatial grid $\{B_k\}_{k=1}^{K}$ and thus evolve piecewise constant concentration fields. This discretization mirrors the Galerkin projection defined above in that it describes the dynamics in terms of concentration values associated with the spatial boxes.

While the projected particle dynamics evolves concentration vectors with discrete values in multiples of $1/(N\Delta)$, the grid-discretized SPDE yields real-valued concentration vectors and thus defines a Markov process on a continuous concentration space. The corresponding transfer operator acting on observables of the discretized concentration field can be regarded as a continuous counterpart of the projected particle-level operator $\mathbf P_N^\tau$. In this sense, the SPDE discretization provides a practical surrogate for the first reduction step at the numerical level.

\subsection{Projection onto a coarse-gained subspace} \label{sec:TO_coarse}

Building on the concentration-based discretization from 
Section~\ref{sec:TO_hist}, we now apply a second Galerkin projection to 
obtain a coarse-grained transfer operator acting on a reduced subspace.
This step aggregates the discretized concentration states into a smaller 
number of coarse sets and thereby yields a more compact representation of 
the dominant long-term dynamics.

\paragraph{Abstract coarse partition.}
Let $\mathbb{F}=\{c^{(1)},\dots,c^{(n_c)}\}$ denote the finite state space of 
discretized concentrations obtained in Section~\ref{sec:TO_hist}.  
To define a coarse-grained representation, we introduce an arbitrary 
measurable assignment
\begin{equation}
    \label{eq:assign}
       \kappa : \mathbb{F} \to \{1,\dots,n_S\}=:\mathbb{S},
\end{equation}
which associates each concentration state $c\in\mathbb{F}$ with one of 
$n_S$ coarse states.  
This assignment induces a finite partition 
\[
   \mathbb{F} = \bigcup_{k=1}^{n_S} \mathbb{F}_k,
   \qquad 
   \mathbb{F}_k := \{\, c\in \mathbb{F} : \kappa(c)=k \,\}.
\]
No structure is assumed for the partition: it may arise, for example, from geometric, statistical, or problem-specific considerations. A concrete, data-driven construction of the map~$\kappa$ will be introduced later in Section~\ref{sec:model_reduction}.

\paragraph{Coarse basis functions and subspace.}
For each coarse state $k\in \mathbb{S}$, define the characteristic function
$\phi_k : \mathbb{F} \to \{0,1\}$ by
\[
   \phi_k(c) = 1_{\mathbb{F}_k}(c)
   =
   \begin{cases}
      1, & c\in \mathbb{F}_k,\\
      0, & \text{otherwise}.
   \end{cases}
\]
The functions $\phi_k$ form a basis of the coarse subspace
\[
   \mathcal{F}_{n_S} := \mathrm{span}\{\phi_k:k=1,\ldots,n_S\}
   \subset \ell^\infty(\mathbb{F}),
   \qquad 
   \dim(\mathcal{F}_{n_S}) = n_S.
\]

\paragraph{Galerkin projection.}
Equipped with the standard pairing
\[
   \langle f,g\rangle_{\mathbb{F}}
   := \sum_{c\in\mathbb{F}} f(c)\,g(c),
\]
the Galerkin projection onto $\mathcal{F}_{n_S}$ is given by
\[
   Q_S f 
   = \sum_{k=1}^{n_S} 
     \frac{\langle \phi_k, f\rangle_{\mathbb{F}}}
          {\langle 1, \phi_k\rangle_{\mathbb{F}}}\, \phi_k .
\]
Applying $Q_S$ to the discrete transfer operator 
$\mathbf{P}^\tau_{\!N}$ from Section~\ref{sec:TO_hist} yields a 
coarse-grained transfer operator with matrix representation
\begin{equation}
    \label{eq:P_Galerkin}
       P^\tau = (P^\tau_{kl})_{k,l=1}^{n_S},
   \qquad
   P^\tau_{kl}
   = 
   \frac{
      \big\langle \phi_k,\; \mathbf{P}^\tau_{\!N} \phi_l \big\rangle_{\mathbb{F}}
   }{
      \big\langle 1,\; \phi_l \big\rangle_{\mathbb{F}}
   }.
\end{equation}
The resulting matrix $P^\tau$ is column-stochastic and propagates 
probability distributions over the coarse partition $\mathbb{S}$
\[
    P^\tau : \ell^1(\mathbb{S}) \to \ell^1(\mathbb{S}).
\]

\paragraph{Role in the full framework.}
The construction above is purely analytical: it defines an abstract projection of the fine-scale transfer operator onto a reduced partition of the concentration space.  
In Section~\ref{sec:model_reduction}, we realize this second projection in a data-driven manner by (i) selecting a coarse partition using static configuration data and (ii) estimating the transition probabilities of $P^\tau$ from dynamical simulation data.

\section{Data-driven approximation of the coarse-grained transfer operator} 
\label{sec:model_reduction}

Section~\ref{sec:TO_coarse} introduced an abstract second Galerkin projection, in which the discrete concentration space $\mathbb{F}$ is partitioned into finitely many sets $\{\mathbb{F}_k\}_{k=1}^{n_S}$. While this formulation specifies how a coarse-grained transfer operator $P^\tau$ acts once the sets $\mathbb{F}_k$ are given, it does not prescribe how such a partition should be chosen in practice.

In this section, we construct the partition based on simulation data and obtain a numerical approximation of the associated coarse-grained operator.  
The data-driven procedure has two components. 
First, in~\Cref{sec:galerkin_diffmap}, we use static configuration data to reveal the intrinsic geometry of the concentration space via the \textit{Diffusion Maps} method~\cite{coifman2005geometric,coifman2006diffusion}. This geometry then guides the choice of a suitable partition.  
Second, in~\Cref{sec:galerkin_msm}, we use dynamical simulation data to estimate the transition probabilities between the resulting coarse sets by counting transitions at lag time~$\tau$, yielding a concrete matrix approximation of the coarse-grained transfer operator defined above.

\subsection{Geometric discretization from data}
\label{sec:galerkin_diffmap}

We begin by extracting a low-dimensional geometric representation of sampled concentration profiles from SPDE simulations. This embedding provides a coordinate system in which a coarse partition can be defined.  
The construction of the embedding is described in~\Cref{subsec:emb_proc} and applied to the two examples in~\Cref{sec:CV}. The resulting partition of the embedded space (using either a uniform grid or Voronoi cells) is described in Section~\ref{sec:partition}. 

\subsubsection{Geometric embedding via Diffusion Maps} 
\label{subsec:emb_proc}

We use the \textit{Diffusion Maps} algorithm~\cite{coifman2005geometric,coifman2006diffusion} to obtain a small number of intrinsic coordinates that parametrize the sampled concentration profiles in a low-dimensional but geometrically meaningful way.

\paragraph{Data.} 
The data consist of discretized concentration profiles $c(x,t)$ obtained from SPDE simulations on a spatial grid. Each snapshot is represented as a vector $c \in \mathbb{F}$, where $\mathbb{F}$ now denotes the space of continuous-valued concentration profiles on the domain $\mathbb{T}$.  
These SPDE-based concentrations play the same conceptual role as the coarse-grained concentrations introduced in Section~\ref{sec:TO_hist}, but arise directly from the continuum formulation and therefore represent its empirical realizations.

\paragraph{Low-dimensional parametrization.} 
If the sampled concentration profiles lie close to a low-dimensional manifold in the high-dimensional space $\mathbb{F}$, they can be represented effectively by a small number of \textit{embedding coordinates}\footnote{Also named \textit{collective variables}, \textit{reaction coordinates}, or \textit{order parameters} in other contexts.}. This corresponds to a map
\[
   \xi : \mathbb{F} \to \mathcal{S}\subset \mathbb{R}^d, \qquad d \ll \dim(\mathbb{F}),
\]
that assigns to each sampled concentration $c_i$ an intrinsic low-dimensional descriptor $\xi(c_i)$. 
Given a collection of data points
\[
   \{c_1,\dots,c_M\} \subset \mathbb{F},
\]
Diffusion Maps computes these coordinates such that concentration profiles that are similar under the chosen metric remain close in the embedding space.

\paragraph{Diffusion Maps construction.}
The Diffusion Maps algorithm builds a weighted graph in which transition probabilities are high between nearby concentrations and negligible between distant ones.
The leading eigenvectors of the resulting Markov transition matrix provide intrinsic coordinates adapted to the sampled concentration ensemble, and these eigenvectors form the
components of the embedding $\xi(c_i)$.  The construction of these coordinates follows the standard Diffusion Maps procedure, which consists of the following steps:
\begin{enumerate}
    \item Choose a kernel
    \begin{equation} 
    \label{kernel}
        k_\varepsilon(c_i,c_j) = \exp\left(-\frac{\delta(c_i,c_j)^2}{\varepsilon}\right),
    \end{equation}
    where $\delta$ denotes a suitable distance between the particle concentration profiles $c_i,c_j$ (which will be specified in~\Cref{sec:CV}), and $\varepsilon>0$ is a scaling parameter that controls the locality of the similarity measure, which can be chosen following standard heuristics~\cite{coifman2008graph}. While the Gaussian kernel is the canonical choice, other nonnegative kernels could in principle be employed.
    \item Define $q_\varepsilon(c_i)=\sum_{m=1}^M k_\varepsilon (c_i,c_m)$ and pre-normalize the kernel via
    \begin{equation}
        \tilde{k}_\varepsilon(c_i,c_j) = \frac{k_\varepsilon(c_i,c_j)}{q_\varepsilon(c_i)\,q_\varepsilon(c_j)}.
    \end{equation}
    This normalization yields an \textit{anisotropic kernel}~\cite{coifman2006diffusion}, which compensates for nonuniform sampling of the data. 
    In particular, it removes the influence of the empirical data density so that the resulting diffusion process reflects the intrinsic geometry of the underlying manifold rather than artifacts of uneven sampling.\footnote{The pre-normalization cancels bias from nonuniform sampling.
Following~\cite{coifman2006diffusion}, this corresponds to the anisotropic normalization with tuning parameter~$\alpha=1$.}
    \item Re-normalize using row sums $s_\varepsilon(c_i)=\sum_{m=1}^M \tilde{k}_\varepsilon(c_i,c_m)$ to obtain the entries of the transition matrix 
    \begin{equation} 
    \label{eq:diff_map_matrix}
        Q_\varepsilon(c_i,c_j)= \frac{\tilde{k}_\varepsilon(c_i,c_j)}{s_\varepsilon(c_i)}.
    \end{equation}
        The matrix $Q_\varepsilon$ represents the transition probabilities of a random walk---or virtual \textit{diffusion}---on the data set, thereby exploring the intrinsic geometry of the manifold.
    \item Compute the eigenvalues $1=\lambda_0 > \lambda_1 \geq \lambda_2 \geq \dots$ and corresponding eigenvectors $\psi_0, \psi_1, \psi_2, \dots$ of the transition matrix $Q_\varepsilon$.\footnote{Since $Q_\varepsilon$ corresponds to a reversible Markov chain with respect to its stationary distribution, it is self-adjoint in the associated weighted inner product. 
Consequently, all eigenvalues are real and lie in $[0,1]$~\cite{coifman2006diffusion}.
} These eigenpairs encode the dominant modes of the random walk and thus reveal the large-scale geometric and dynamical structure of the data.
    \item Define the Diffusion Map embedding of the data points as
    \begin{equation}
        \xi(c_i) = (\xi_{1,i},\ldots,\xi_{d,i}) = \big(\lambda_1 (\psi_1)_i, \, \lambda_2 (\psi_2)_i, \, \dots, \, \lambda_d (\psi_d)_i\big) \in \mathbb{R}^d,
    \end{equation}
    where $d$ is the target embedding dimension and $(\psi_j)_i$ is the $i$th component of the $j$the eigenvector of $Q_\varepsilon$. 
    The coordinates $\xi_1, \xi_2, \dots, \xi_d$ are the \textit{Diffusion Map coordinates} (embedding coordinates), which parametrize the intrinsic geometry of the data.
\end{enumerate}
After removing the trivial constant eigenvector (associated to the eigenvalue $\lambda_0=1$), the remaining Diffusion Maps eigenvectors provide a sequence of coordinates ordered by decreasing eigenvalue. Since a clear spectral gap is not expected in general, the embedding
dimension $d$ is selected pragmatically, for instance by inspecting the eigenvalue decay or by evaluating the quality of the resulting low-dimensional representation. These $d$ coordinates then serve as intrinsic variables characterizing the concentration profiles.

\quad 

The computation of pairwise distances and eigenvectors of $Q_\varepsilon$ becomes prohibitively expensive for large data sets. A common remedy is to work with a suitably chosen sub-sample: the Diffusion Map matrix and its eigenpairs are computed only for this subset, and further data points can then be embedded employing the \textit{out-of-sample extension}~\cite{coifman2006geometric}, see Appendix~\ref{app:out-of-sample}. 
In our setting, the sub-sample consists of representative \textit{snapshots} of the system, i.e., particle concentrations recorded at discrete time points separated by a fixed interval~$dt_{\text{diff}}$, which serves as the effective time step for the Diffusion Maps analysis. 

\subsubsection{Geometry revealed by the embedding: metrics and numerical results} 
\label{sec:CV}

A crucial ingredient of the Diffusion Maps construction is the definition of a distance between concentrations, which determines the notion of similarity in the data.
The choice of the metric must reflect the relevant physical features of the system and may depend on the interaction potential: for example, in the multichromatic case, stable cluster positions are emphasized, whereas in the Morse case, the mobility and merging of cluster centers become more significant.

To apply Diffusion Maps to particle clustering dynamics, we therefore require a distance~$\delta$ suitable for use in~\eqref{kernel}.
Since the dynamics evolve on the torus~$\mathbb{T}=\big[-\frac{L}{2},\frac{L}{2}\big)$, the metric must respect periodic boundary conditions. Moreover, we are primarily interested in the number, sizes, and shapes of clusters rather than in their absolute positions on the torus, which motivates the use of \textit{translation-invariant} distances. 

\paragraph{Translation-invariant $L^2$-distance.} For the multichromatic potential of~\Cref{ex1}, we employ the translation-invariant $L^2$-distance 
\begin{equation}
    \label{eq:L2-dist}
    \delta_{L^2}(c_i,c_j):=\min_{0\leq \ell < L}\big\lVert c_i(g(\cdot+\ell))-c_j\big\rVert_{2}
\end{equation}
where $g:\mathbb{R} \to \mathbb{T}=\big[-\frac{L}{2}, \frac{L}{2}\big)$ is the projection onto the torus, $g(x):= (x+\frac{L}{2})\text{mod}\, L - \frac{L}{2}$. 
This choice is natural for the multichromatic potential, since clusters are arranged at characteristic positions  and remain relatively fixed in space; so alignment by translation is sufficient to compare different states. This norm can also be useful for other types of interaction potentials, as shown in~\cite{evangelou2024}, but not for every kind of particle dynamics.
On a periodic domain, computing the translation-invariant $L^2$-distance between two concentration profiles reduces to maximizing their circular cross-correlation over all shifts. Using the Fast Fourier Transform (FFT)~\cite{bracewell2000}, this can be evaluated in $\mathcal{O}(K \log K)$ operations for $K$ spatial grid points used to represent a concentration.

\paragraph{Translation-invariant Wasserstein-1 distance.}
In particular, for the Morse potential used in~\Cref{ex2} the $L^2$-metric is not suitable, since cluster centers may drift and merge, and simple point-wise comparison does not adequately capture their relative positions.
Instead, we use the translation-invariant Wasserstein distance, which is sensitive to spatial displacements of mass. Following~\cite{rabin2009}, the Wasserstein-1 distance on the torus is  
\begin{equation} 
    \label{def:Wassertein_alpha}
    W_1^{\mathbb{T}}(c_i,c_j) := \inf_{\alpha \in \mathbb{R}} \big\lVert F_{c_i}-F_{c_j}-\alpha \big \rVert_1 = \inf_{\alpha \in \mathbb{R}} \int_{\mathbb{T}} |F_{c_i}(x)-F_{c_j}(x)-\alpha| \, dx,
\end{equation}
where $F_c$ is the cumulative distribution function of $c$. This definition corresponds to choosing a common cut point on the torus and ensures periodicity. 
To eliminate dependence on absolute positions, we define the translation-invariant version
\begin{equation} 
    \label{def:Wasserstein_trans}
    \delta_W(c_i,c_j) :=  \min_{0\leq \ell <L} W_1^{\mathbb{T}}(c_i(g(\cdot+\ell)),c_j),
\end{equation}
which measures the minimal transport cost after optimally aligning the particle concentrations by a relative shift. The one-dimensional Wasserstein-1 distance can be computed in linear time in $K$, but the additional minimization over all discrete shifts results in an overall complexity of $\mathcal{O}(K^2)$ for $\delta_W$.

\begin{figure}
    \centering
      \begin{subfigure}{0.49\textwidth}
    \includegraphics[width=\textwidth]{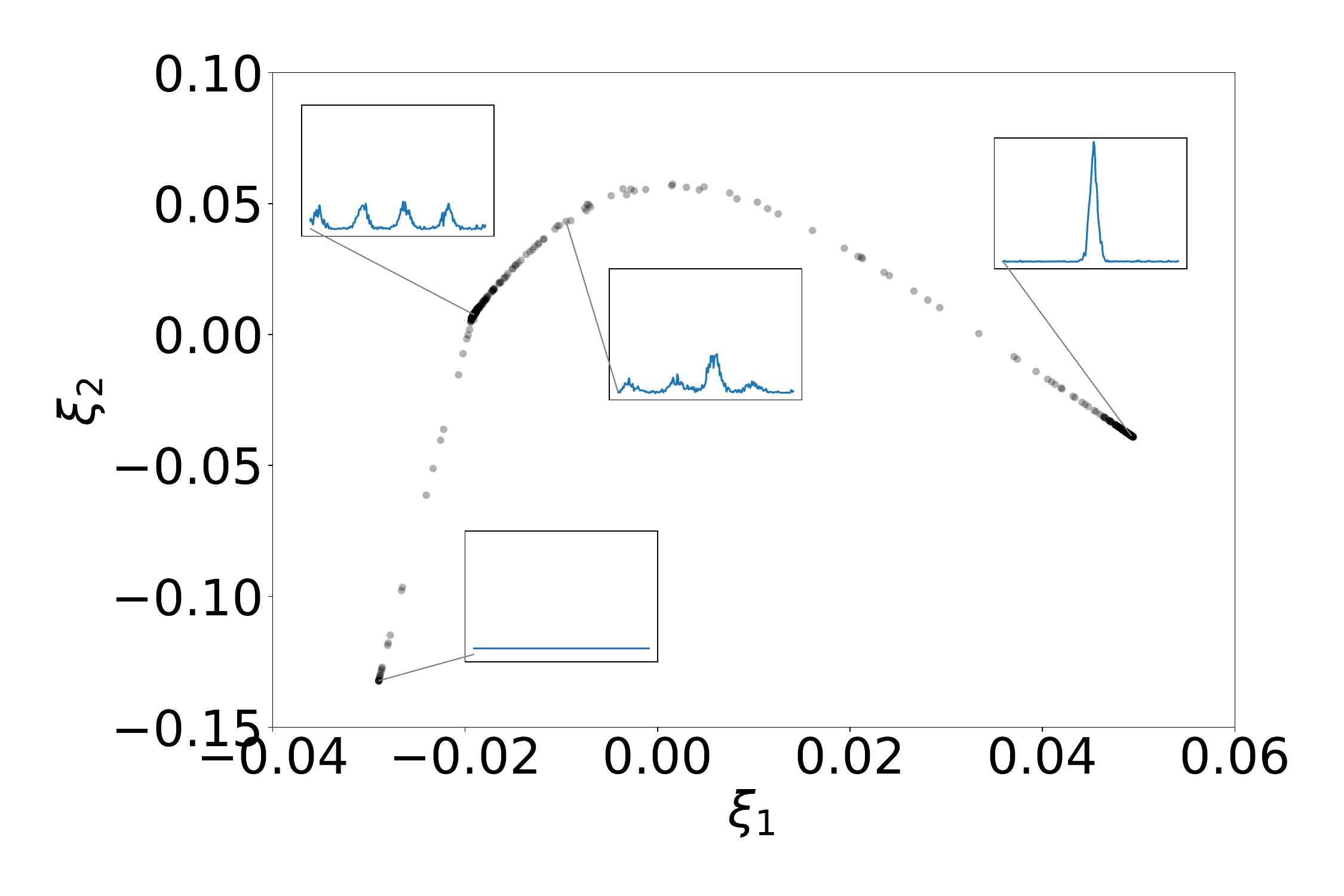} 
    \caption{Multichromatic potential (\Cref{ex1})}
    \label{fig:diffusion_maps_SPDE_multichrom_b}
    \end{subfigure}
       \begin{subfigure}{0.49\textwidth}
    \includegraphics[width=\textwidth]{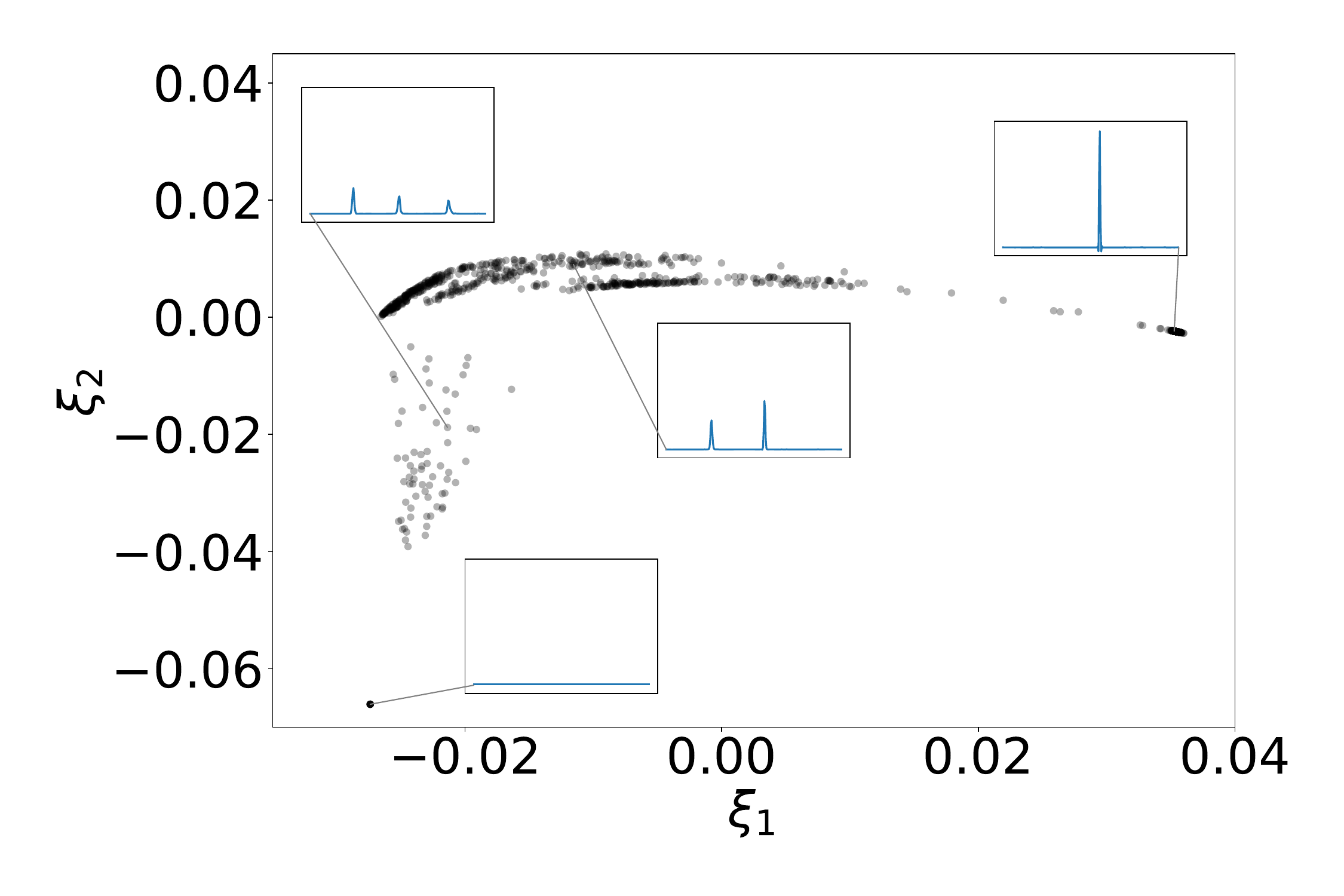}
    \caption{Morse potential (\Cref{ex2})}
    \label{fig:diffusion_maps_SPDE_Morse_b}
    \end{subfigure}
    \caption{\textbf{Diffusion Map embeddings of several trajectories from Examples~\ref{ex1} and \ref{ex2}.} Projections of particle concentrations from (a) five independent SPDE simulations using the multichromatic potential~\eqref{eq:F} (sampling interval length $dt_{\text{diff}}=1$; simulation time $T=120$) and (b) ten independent SPDE simulations using the Morse potential~\eqref{eq:F_Morse} (sampling interval length $dt_{\text{diff}}=20$; simulation time $T=2000$), in both cases projected onto the first two Diffusion Map coordinates.
    In both panels, four exemplary concentration states are shown; these illustrate representative cluster configurations and highlight geometric features of the embedding.}
    \label{fig:diffusion_maps_SPDE_embedding}
\end{figure}

\paragraph{\Cref{ex1} continued.} 
Figure~\ref{fig:diffusion_maps_SPDE_multichrom_b} shows the Diffusion Map embedding for the  multichromatic potential~\eqref{eq:F}. The data are obtained from five independent SPDE simulations starting from a uniform distribution, with particle concentrations recorded at intervals of length $dt_{\text{diff}}=1$. 

According to the criterion in~\cite{coifman2008graph}, $\varepsilon=1$ is a suitable choice for the proximity parameter for the data considered here. This choice of $\varepsilon$ results in all data points lying on a one-dimensional manifold, see Appendix~\ref{app:opt-choice-eps} for further details. For illustration and to enable comparison with the Morse potential, we plot the projection of this one-dimensional manifold onto the first two Diffusion Map coordinates, $\xi_1$ and $\xi_2$. The embedding confirms the one-dimensional structure: all points lie on a smooth curve. 

The exemplary particle concentrations in~\Cref{fig:diffusion_maps_SPDE_multichrom_b} illustrate that the first diffusion coordinate $\xi_1$ encodes both the number of clusters and their uniformity. Small values of $\xi_1$ correspond to the uniform distribution (no clusters). As $\xi_1$ increases, the system passes through a regular four-cluster state, which gradually loses uniformity. Large values of $\xi_1$ represent the one-cluster state. This separation between configurations with four peaks and those with a single peak can also be recovered by applying a standard grouping method (K-means) to the Diffusion Map embedding; see Appendix~\ref{app:cluster detection}. The interpretation of $\xi_2$ is not evident in this example. 

\paragraph{\Cref{ex2} continued.}  

For the Morse potential~\eqref{eq:F_Morse}, \Cref{fig:diffusion_maps_SPDE_Morse_b} shows the Diffusion Map embedding of ten SPDE trajectories starting in the uniform distribution with a distance of $dt_{\text{diff}}=20$ between the time snapshots. This large distance between snapshots is chosen because the computational cost of constructing the Diffusion Map embedding using the translation-invariant Wasserstein metric is high, as discussed above. For the chosen Morse potential parameters and noise strength, the system initially goes from the uniform distribution into a concentration of four clusters, see~\cite{wehlitz2025}. This initial number of clusters can be determined analytically using linear stability analysis~\cite{garnier2016, gerber2025}. However, this four-cluster state only lasts on a short time scale and thus plays no role for the $dt_{\text{diff}}$ chosen here. In the following, only three-, two- and one-cluster states are relevant for the analysis.
    
For $\varepsilon=0.2$ (chosen according to the criterion from \cite{coifman2008graph}) we obtain a 2-dimensional manifold (see Appendix~\ref{app:opt-choice-eps}), so that the clustering dynamics can be represented sufficiently well using the first two projection coordinates $\xi_1$ and $\xi_2$.

The projected manifold in \Cref{fig:diffusion_maps_SPDE_Morse_b} appears to be divided mainly into three groups of points corresponding to three-, two-, and one-peak concentrations. This separation is confirmed by applying K-means to the Diffusion Map embedding; see Appendix~\ref{app:cluster detection}.
Only the point corresponding to the embedded uniform distribution stands out from these three sets. Note that $\xi_1$ rather distinguishes the one-cluster states from the remaining states, i.e., a high $\xi_1$ value corresponds to a concentration state closer to the stationary one-cluster state. $\xi_2$ separates three-cluster states from two- and one-cluster concentration profiles. Within the two-cluster states, different configurations of the two-cluster concentration profiles are distinguished by $\xi_2$.

\begin{rem}[Other initial configurations]
We initialize the dynamics in both examples from a uniform concentration, which provides a neutral starting point without privileging specific configurations. While other choices are possible—for instance, starting from a concentration containing more clusters than those present immediately after formation—such initializations implicitly assume that these configurations are relevant or accessible. 
Based on the intrinsic properties of the two interaction potentials (which are very different in their cluster formation mechanisms), the dynamics, however, quickly relax in both cases into the four-cluster state, so the long-term behavior is essentially the same as under uniform initialization.
\end{rem}

A detailed analysis of the computational effort for the Diffusion Map construction is provided in Appendix~\ref{app:comp_effort_1}.

\subsubsection{From embedding to coarse partition}
\label{sec:partition}

The Diffusion Map embedding from the previous subsections provides a low-dimensional representation $\mathcal{S} \subset \mathbb{R}^d$ of the sampled concentration profiles.  
To obtain the coarse states required for the second Galerkin projection, we discretize this embedding space into $n_S$ disjoint regions $S_1,\dots,S_{n_S}$ such that
\[
   \mathcal{S} = \bigcup_{k=1}^{n_S} S_k,
   \qquad S_k \cap S_\ell = \emptyset \;\; (k\neq \ell).
\]

The embedding and the partition together induce an assignment map
\[
   \kappa : \mathbb{F} \to \{1,\dots,n_S\},
   \qquad
   \kappa(c) = k \;\; \text{if } \xi(c)\in S_k ,
\]
as a realization of the abstract assignment $\kappa$ defined in~\eqref{eq:assign}.
Thus each concentration profile is mapped to the index of the region in
the embedded space that contains its image under $\xi$.  The resulting
coarse sets are
\[
   \mathbb{F}_k = \kappa^{-1}(k)
   = \{\, c \in \mathbb{F} : \xi(c) \in S_k \,\},
\]
providing a data-driven realization of the abstract partition
$\{\mathbb{F}_k\}_{k=1}^{n_S}$ introduced in~\Cref{sec:TO_coarse}.

This discretization identifies the finite set of coarse states on which 
the coarse-grained transfer operator will act.  
The transition probabilities between these coarse states will be 
estimated from dynamical data in~\Cref{sec:galerkin_msm}.

\paragraph{Partitioning strategies.}
We consider two practical ways of partitioning the embedding space:
\begin{enumerate}
\item \textbf{Uniform grid:} A regular discretization based on a fixed grid size, producing non-overlapping, axis-aligned boxes in the embedding space.
\item \textbf{Voronoi cells:} A data-driven partition obtained via the \textit{K-means} algorithm~\cite{lloyd1982,scikit-learn}, in which each data point is assigned to the nearest cluster center.
\end{enumerate}
While other discretization approaches are possible, these two provide a transparent and robust choice for the present examples.

\subsection{Estimation of transition probabilities from dynamical data}
\label{sec:galerkin_msm}

Once the partition is fixed, the transition probabilities between sets are estimated from Monte Carlo simulations of the original dynamics.
We fix a lag time~$\tau$ and generate pairs of consecutive states~$(c_i,c'_i)\in\mathbb{F}\times\mathbb{F}$, $i=1,\ldots,M'$, separated by the time interval~$\tau$, obtained from several simulated trajectories of particle concentrations.
These states are embedded into the reduced space using the out-of-sample extension (see Appendix~\ref{app:out-of-sample}), yielding pairs of embedded coordinates~$(\xi(c_i),\xi(c'_i))\in\mathbb{R}^d\times\mathbb{R}^d$.

A standard procedure to estimate the transition matrix~$P^\tau$ is constructing the maximum-likelihood estimator (MLE) based on transition counts, also known as \textit{Ulam's method}~\cite{norris1998markov, ulam1960, klus2016}:
\begin{equation} 
\label{MLE_P}
    P^\tau_{kl}= \frac{C_{kl}}{\sum_{l'=1}^{n_S} C_{kl'}},
    \qquad
    C_{kl} = \sum_{i=1}^{M'}
    \mathbf{1}_{S_k}\big(\xi(c_i)\big)\,
    \mathbf{1}_{S_l}\big(\xi(c'_i)\big),
\end{equation}
where $C_{kl}$ denotes the number of observed transitions from set~$S_k$ to~$S_l$.
This estimator yields a stochastic matrix satisfying $P^\tau_{kl}\ge0$ and $\sum_l P^\tau_{kl}=1$ for all~$k$.

The choice of lag time $\tau$ and the size of the spatial regions $S_1,\dots,S_{n_s}$ must be balanced in order to obtain a reliable estimate of the transition matrix. 
The lag time should be sufficiently large relative to the box size so that transitions between regions occur with non-negligible probability, but not so large that transition probabilities are insufficiently sampled. 
Conversely, the spatial regions should not be chosen too small relative to the available data density, in order to avoid large statistical errors in the Monte Carlo estimates; see, e.g., \cite[Section~4.2]{koltai2020} and \cite{prinz2011}.

\paragraph{Generation of data pairs.}
The pairs of consecutive states $(c_i,c'_i)$ used in the estimation are obtained from a large ensemble of independent, long SPDE trajectories. 
This procedure ensures that only transitions between regions actually visited by the dynamics are recorded, and that the number of samples associated with each region reflects its empirical visitation frequency.
With $10^3$ trajectories, the resulting transition statistics provide adequate coverage of both frequently and rarely visited regions. Using short trajectories initialized in local equilibrium within each region would, in principle, improve sampling efficiency, but this is not feasible here since the corresponding equilibrium distributions are unknown. 

\paragraph{Enforcing reversibility.}
The underlying interacting particle system~\eqref{eq:PBD_alt} is a gradient diffusion and therefore reversible with respect to its stationary Gibbs measure; under mild conditions it is moreover ergodic. See, e.g.,~\cite{gerber2025} for a discussion of microscopic reversibility in interacting particle systems exhibiting clustering dynamics. 
In practice, however, the empirical matrix~\eqref{MLE_P} may not correspond to a reversible Markov chain, for instance when certain rare transitions are not sampled. In our case, this concerns transitions out of the one-cluster state in both examples. Although such transitions occur only on exponentially long time scales~\cite{gerber2025}, reversibility at the microscopic level implies that clusters can in principle dissolve again, and we therefore enforce reversibility in the reduced model.
To impose detailed balance and obtain a statistically consistent estimator, we employ the \textit{reversibility-constrained maximum-likelihood estimator}~\cite{prinz2011, trendelkamp2015}.
This estimator maximizes the likelihood of the observed transition counts~$C_{kl}$ under the constraints that $P^\tau$ is stochastic and satisfies detailed balance with respect to some stationary distribution $\pi$, i.e., 
\begin{equation}
    \pi_k P^\tau_{kl} = \pi_l P^\tau_{lk} \quad\quad  k,l=1,\ldots,n_S.
\end{equation}
Introducing the symmetric flux variables
\begin{equation}
    m_{kl}:=\pi_k P^\tau_{kl} = \pi_l P^\tau_{lk},
\end{equation}
the optimization problem reduces to finding a symmetric, nonnegative matrix $M=(m_{kl})_{k,l=1,\ldots,n_S}$ that maximizes the log-lilkelihood
\begin{equation}
    \log L(M) = \sum_{k,l} C_{kl} \log \left( \frac{m_{kl}}{\pi_k} \right), \quad  \quad \pi_k := \sum_l m_{kl},
\end{equation}
subject to 
\begin{equation}
    m_{kl}=m_{lk}\geq 0, \quad  \quad \sum_{k,l} m_{kl}= 1.
\end{equation}
In general, this optimization problem has no closed-form solution and must be solved numerically, for instance via fixed-point iteration~\cite[Section~III]{trendelkamp2015}. Once the optimal $M$ is obtained, the reversible transition matrix is reconstructed as
\begin{equation} \label{MLE_P_reversible}
    P^\tau_{kl} = \frac{m_{kl}}{\sum_{l'}m_{kl'}}.
\end{equation}

\quad

The computational cost for estimating the transition probabilities is analyzed in detail in Appendix~\ref{app:comp_effort_2}. The results of the Markov chain construction for our two representative examples and the two types of partitioning---uniform and Voronoi---are presented in the next chapter.

\section{Metastability and implied timescales}
\label{sec:analyzeMSM}

We now analyze the coarse-grained dynamics represented by the Markov chain with transition matrix
\[  P^\tau=\big(P^\tau_{kl}\big)_{k,l=1,\ldots,n_S} \]
which encodes the evolution between the regions~$S_1,\dots,S_{n_S}$ obtained in the previous section.
As before, we identify each region~$S_k$ with its index~$k$ and thus use
\begin{equation}
    \mathbb{S}=\{1,\ldots,n_S\}
\end{equation}
as the state space of the Markov chain. 
Our goal is to characterize the long-term behavior of this reduced process and to identify \textit{metastable structures} that correspond to persistent clustering patterns in the original dynamics.
In particular, we seek to answer questions such as:
Which cluster configurations are comparatively stable?
What are the characteristic timescales for cluster formation and merging? And how long does it take, on average, for the system to reach the one-cluster state starting from a multi-cluster configuration?

The remainder of this section is organized as follows.
We first recall the relevant background on transition rates and timescales in Markov models in~\Cref{sec:background},
and then apply these concepts to our two exemplary systems in~\Cref{sec:results}.

\subsection{Theoretical background} \label{sec:background}

The concepts outlined in this subsection are standard in the analysis of finite-state Markov models and metastable dynamics; see, for example, \cite{norris1998markov, levin2017, metzner2009}. 
We briefly summarize them here to make the presentation self-contained and to fix notation for the subsequent analysis.

\subsubsection{Eigenvalue structure of Markov processes}

The spectral properties of a Markov chain provide essential information about the dynamical behavior of the underlying stochastic process. If the process is reversible, the leading eigenvalues and eigenvectors are real-valued and encode the dominant dynamical features of the process~\cite{levin2017,schuette2023}. Given the transition matrix $P^\tau$ at lag time $\tau$, its dominant left eigenvector gives the stationary distribution, while the rest of the spectrum reflects relaxation processes. For non-reversible processes, one has to analyze its singular values and the related singular vectors, respectively \cite{Fritzsche2007,schuette2023}, or the leading complex-valued eigenvalues and respective elements of the Schur decomposition \cite{DjurdjevacConradWeberSchuette2016}.
Since the particle-based process is reversible and we re-enforce this reversibility by the approach outlined in \Cref{sec:galerkin_msm}, we proceed with the analysis based on dominant eigenvalues and eigenvectors.

If the first $p$ nontrivial eigenvalues are close to one and separated by a spectral gap, the system exhibits $p$ metastable sets: groups of states that mix rapidly internally but exchange probability mass only rarely. The corresponding relaxation timescales are
\begin{equation} 
    \label{T_eigenvalue}
    T_i^\tau = -\frac{\tau}{\log \mu_i}, 
\end{equation}
where $\mu_i$ denotes the $i$-th eigenvalue. These timescales quantify how quickly the corresponding dynamical process decays, and thus how rapidly the system relaxes between metastable regions. The associated eigenvectors provide spatial information, revealing which parts of state space participate in each slow process.

Together, the leading eigenmodes point to a natural partition of the state space into long-lived regions. Next, we briefly recall the clustering method PCCA+, which translates this spectral information into metastable macrostates and will be used in our numerical investigations.

\subsubsection{Detecting metastable regions using PCCA+}

The PCCA+ algorithm (\textit{Robust Perron Cluster Cluster Analysis})~\cite{roeblitz2013,deuflhard2005robust} provides a systematic way to transform dominant right eigenvectors of the transition matrix into membership functions, thereby identifying metastable sets\footnote{An implementation of PCCA+ is included in the Python library \textit{MSMTools}~\cite{msmtools}.}. 

PCCA+ constructs a membership matrix whose rows define fuzzy affiliations of microstates (i.e. states of a Markov chain) to macrostates. A crisp partition is obtained by assigning each microstate to the macrostate with the largest membership value. Unlike generic clustering, PCCA+ exploits the dynamical information in $P^\tau$, ensuring that the resulting partition respects the slow timescales. 

Once the metastable macrostates have been identified, the kinetics between them can be quantified, for example via mean first passage times.

\subsubsection{Mean first passage times}

\textit{Mean first passage times (MFPTs)} offer a simple yet informative measure of transition kinetics: they quantify the average time required for the process to reach a target state (or set) starting from another. 

Formally, the MFPT from $A$ to $B$ is the expected time for the process, initialized in $A$, to reach $B$ for the first time:  
\begin{equation} 
    \label{MFPT}
   T_{A\to B} := \mathbb{E}_A(\tau_B) 
   = \frac{1}{\pi(A)}\sum_{k\in A} \pi_k \, \mathbb{E}_k(\tau_B), 
   \qquad \pi(A):=\sum_{k\in A} \pi_k,
\end{equation}
where  $\tau_B:= \inf\{t\geq 0:Y_t\in B\}$ is the first hitting times of $B$, $\pi$ is the stationary distribution of the chain and $\mathbb{E}_k$ denotes the expectation conditioned on $Y_0=k$~\cite{norris1998markov}.

For a discrete-time Markov chain with lag time $\tau$, the first passage time is measured in multiples of $\tau$, and the MFPT in physical time units is given by
\begin{equation}
   T_{A\to B} = \tau \, \mathbb{E}_A(N_B),
\end{equation}
where $N_B := \inf\{ n \geq 0 : Y_n \in B\}$ denotes the first hitting time in terms of step count.

MFPTs can be computed by solving the linear system
    \begin{equation}
        \mathbb{E}_k(\tau_B) = \tau + \sum_{l\in\mathbb{S}} P^\tau_{kl} \, \mathbb{E}_l(\tau_B), \qquad k\notin B,
    \end{equation}
with boundary condition $\mathbb{E}_k(\tau_B)=0$ for $k\in B$.  
This linear system can be solved numerically \cite{msmtools}.

While MFPTs capture average transition times, they do not provide detailed mechanistic information about how transitions occur. To gain such insight, we turn to transition path theory.

\subsubsection{Transition path theory} 
\label{TPT}

Transition path theory (TPT) \cite{metzner2009} extends the analysis by characterizing the ensemble of reactive trajectories between two disjoint sets $A,B\subset \mathbb{S}$. Beyond average timescales, it identifies transition regions, decomposes probability fluxes, and provides a mechanistic picture of how transitions occur.

The central objects are the \textit{forward} and \textit{backward committor functions}. The forward committor gives the probability that, starting in state $k$, the process reaches $B$ before going to $A$:
\begin{equation}
    q^+(k) = \mathbb{P}\big(\tau^+_B(t) < \tau^+_A(t) \, \big|\, Y_t=k \big), 
\end{equation}
where $\tau^+_S(t):=\inf\{s\geq t:Y_s\in S\}$ is the first hitting time of the set $S\subset \mathbb{S}$ (with $\inf \emptyset := \infty$).
The backward committor encodes the probability the last visited set was $A$ rather than $B$:
\begin{equation}
        q^-(k) = \mathbb{P}\big(\tau^-_A(t) > \tau^-_B(t) \, \big|\, Y_t=k \big), 
\end{equation}
where $\tau^-_S(t):=\sup\{s \leq t: Y_s \in S\}$ is the last exit time from $S$ (with $\sup \emptyset := -\infty$). 
Together, $q^+$ and $q^-$ allow the computation of reactive fluxes, which quantify how probability flows along different pathways from $A$ to $B$.

Finally, TPT also provides a link back to timescales between sets of states~\cite{noe2009}: the \textit{transition time} from $A$ to $B$ can be expressed in terms of the forward committor as
\begin{equation} 
    \label{eq:MFPT_TPT}
    T_{A\to B}^{\text{TPT}} = \frac{1}{k_{AB}}, 
    \qquad   
    k_{AB} := \frac{1}{\tau\,\pi(A)}\sum_{k\in A}\sum_{l\notin A} \pi_k P^\tau_{kl}\,q^+(l), \qquad \pi(A)=\sum_{i\in\mathbb{S}}\pi_iq_i^-,
\end{equation}
where $k_{AB}$ is the transition rate per unit time from $A$ to $B$.\footnote{Numerical implementations of the computation of committors and the resulting timescale are available in \textit{MSMTools}~\cite{msmtools}.}

\begin{rem}[Relation between different timescales]\label{rem:times}
While the MFPT $T_{A\to B}$ quantifies the expected time span that trajectories starting in $A$ need to go to $B$ while it may go back to $A$ during the process, the TPT transition time measures the length of a typical trajectory that starts in $A$ and hits $B$ without ever going back to $A$. The general relation between these to kinetic time spans thus is described by $T_{A\to B}\ge T_{A\to B}^{\text{TPT}}$. While both, $T_{A\to B}$ and $T_{A\to B}^{\text{TPT}}$, are hitting times depending on the choice of the sets $A$ and $B$, the relaxation timescales $T_i^\tau$ are de-correlation times of the entire process and do not depend on any pre-chosen set(s) such that there is no general clear relation between specific relaxation timescales $T_i^\tau$ and the kinetic time spans $T_{A\to B}$ and $T_{A\to B}^{\text{TPT}}$.
\end{rem}

\subsection{Numerical results} 
\label{sec:results}

In this section, we apply the concepts summarized above---eigenvalue analysis, PCCA+, MFPTs, and TPT---to numerical studies of our exemplary settings.

\begin{figure}
    \centering
        \begin{subfigure}{0.3\textwidth}
    \includegraphics[width=\textwidth]{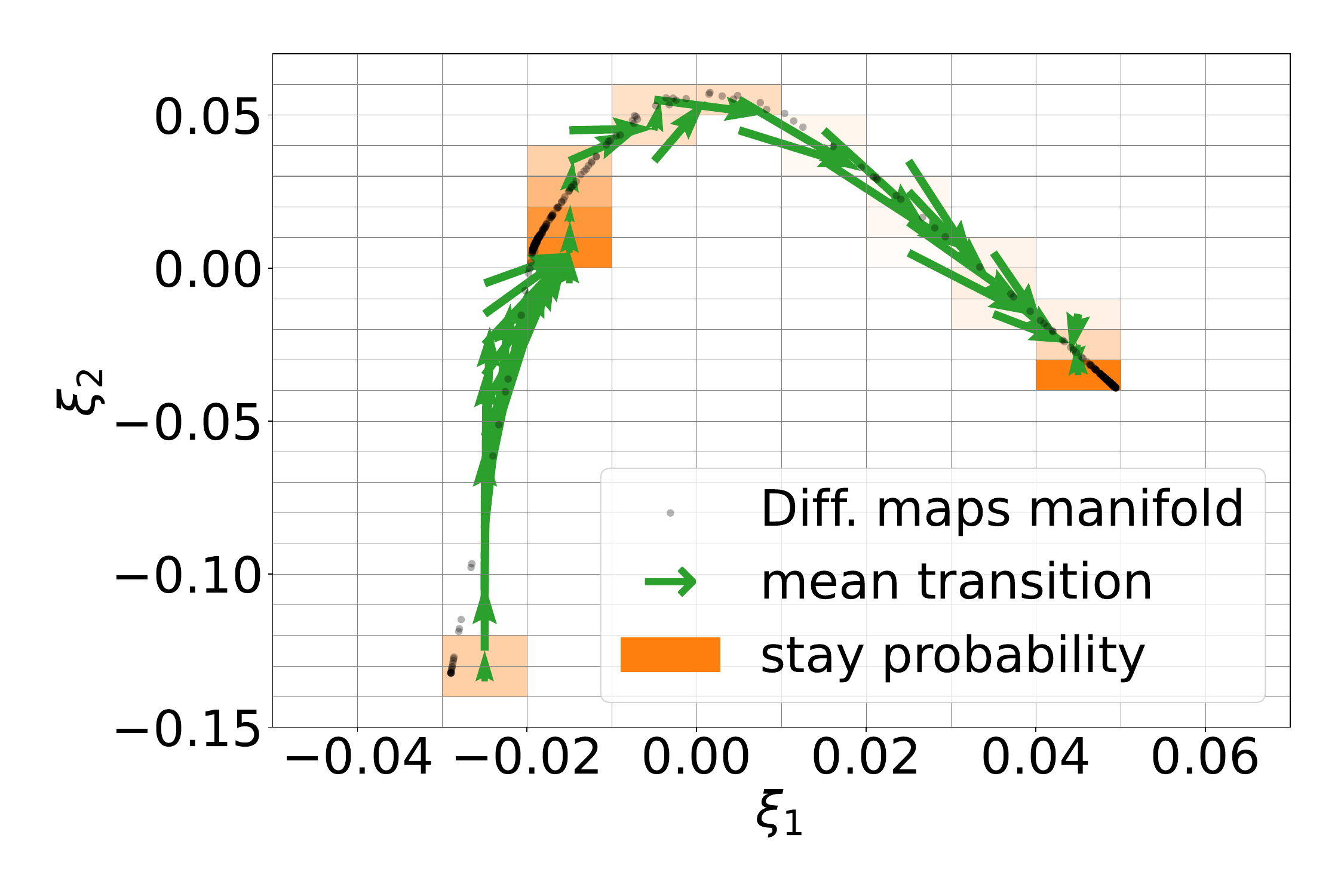}
    \caption{Mean transitions and stay probabilities} 
     \label{fig:diffusion_maps_SPDE_multichrom_MSM_grid_a}
    \end{subfigure}
        \begin{subfigure}{0.34\textwidth}
    \includegraphics[width=\textwidth]{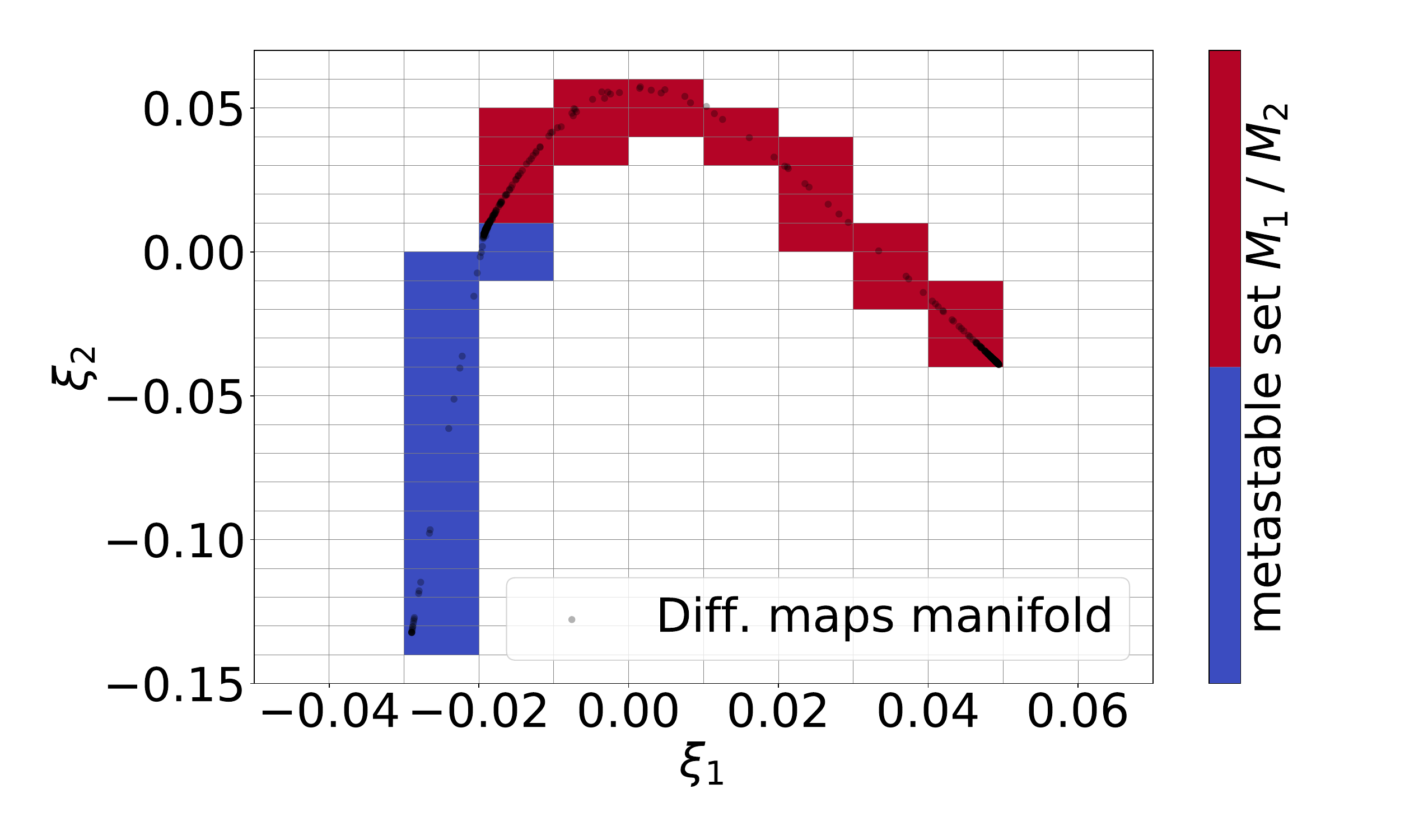}
    \caption{Metastable sets (PCCA+) \newline \,}
    \label{fig:diffusion_maps_SPDE_multichrom_MSM_grid_b}
    \end{subfigure}
       \begin{subfigure}{0.34\textwidth}
    \includegraphics[width=\textwidth]{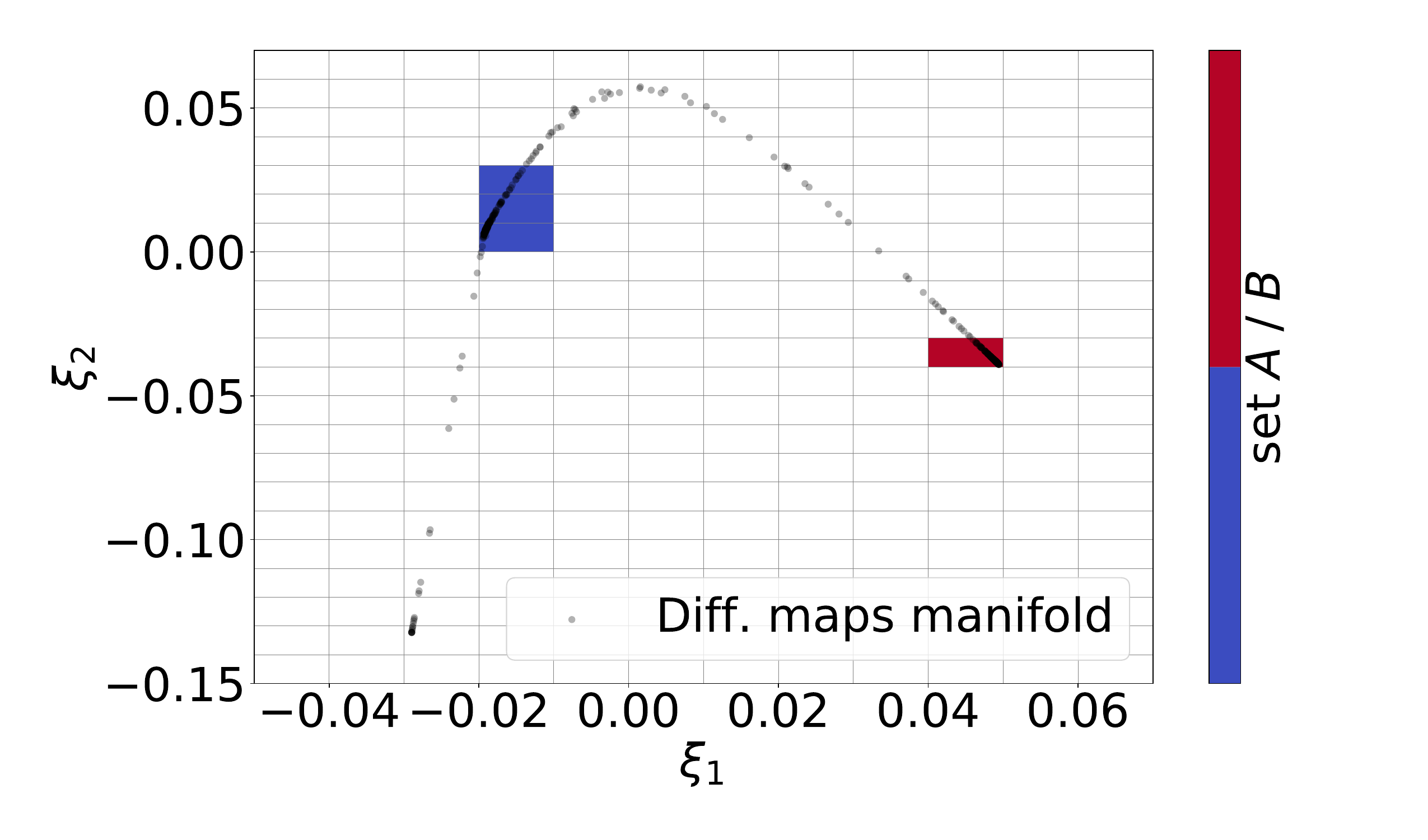}
    \caption{Sets of four-cluster and one-cluster states}
    \label{fig:diffusion_maps_SPDE_multichrom_MSM_grid_c}
    \end{subfigure}
    \caption{\textbf{Multichromatic interaction potential: Dynamics on the reduced space (uniform grid).}
    (a) The relevant boxes are highlighted in orange, with higher transparency indicating lower stay probability. Green arrows illustrate the average transition direction, conditional on the process making a jump.
    (b) Metastable sets of Markov chain states identified using PCCA+ (set $M_1$ in blue, set $M_2$ in red). (c) Sets of Markov chain states representing the four-cluster concentrations ($A$ in blue) and the one-cluster concentrations ($B$ in red).} \label{fig:diffusion_maps_SPDE_multichrom_MSM_grid}
\end{figure}

\subsubsection{Results for \Cref{ex1}} 

To estimate the transition matrix for the example of the multichromatic potential~\eqref{eq:F}, we simulated $10^3$ SPDE trajectories starting from the uniform distribution, running up to $T = 120$ with lag time $\tau = 1$.

\paragraph{The reduced Markov chain.}
For the non-reversible transition matrix $P^\tau$ (see~\Cref{MLE_P}), we obtain that all states are transient except for the one containing the one-cluster state. This behavior occurs for both regular grid and Voronoi discretization, provided that the boxes are not chosen too small. Consequently, the stationary distribution is zero everywhere except for the absorbing state.
For the reversibility-constrained estimator\footnote{\label{fn:iteration}The reversibility-constrained estimator is obtained by the fixed-point iteration from \cite{trendelkamp2015} until ${\|P^{\tau,k+1}-P^{\tau,k}\|_F<10^{-9}}$.}~\eqref{MLE_P_reversible}, the situation is similar: the box corresponding to the one-cluster state is not perfectly but nearly absorbing. In any case, the exact behavior may vary slightly depending on the size and shape of the discretization boxes. In the following, we consider the reversible case.

\begin{figure}
    \centering
       \begin{subfigure}{0.3\textwidth}
    \includegraphics[width=\textwidth]{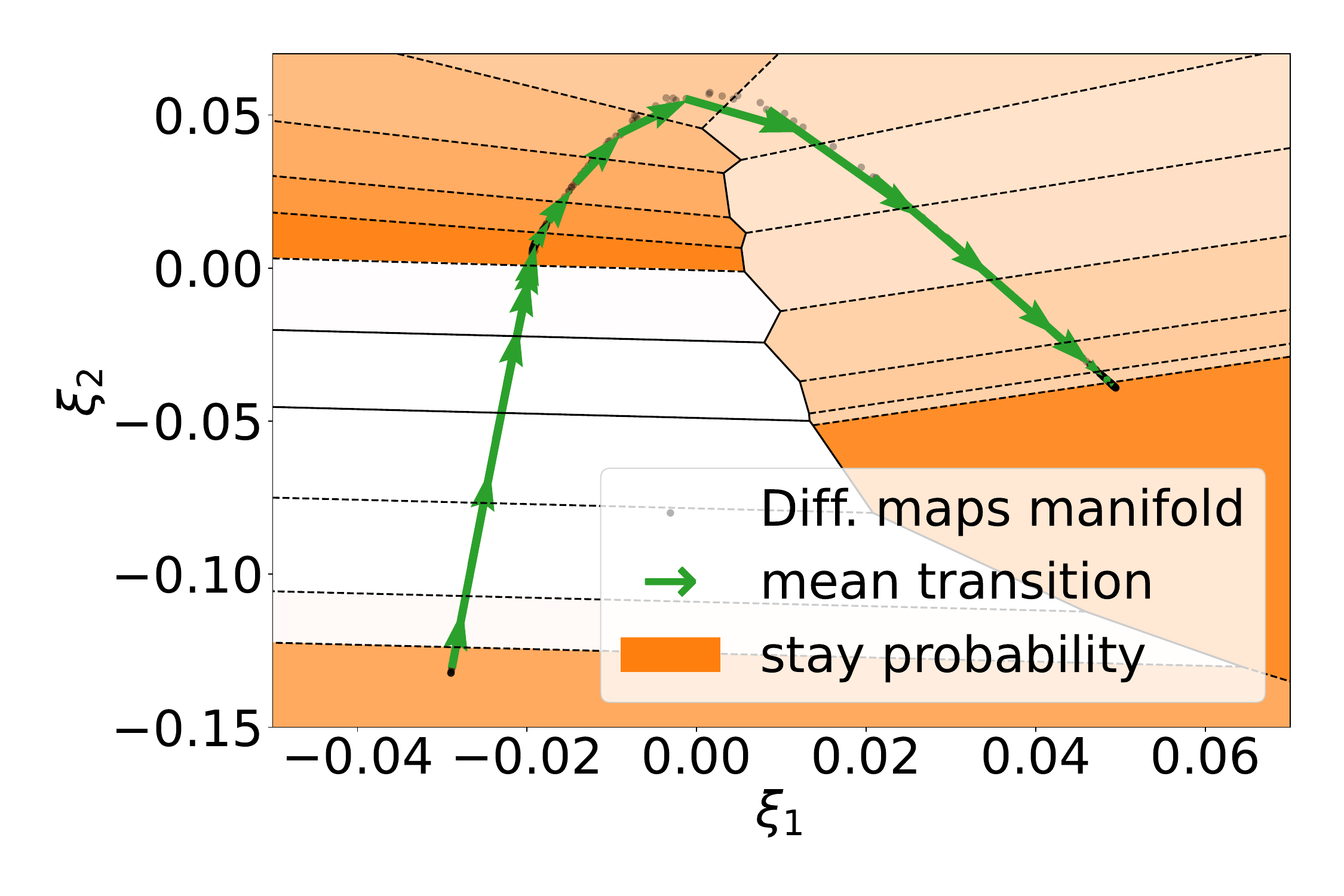}
    \caption{Mean transitions and stay probabilities}
    \label{fig:diffusion_maps_SPDE_multichrom_MSM_Voronoi_a}
    \end{subfigure}
       \begin{subfigure}{0.34\textwidth}
    \includegraphics[width=\textwidth]{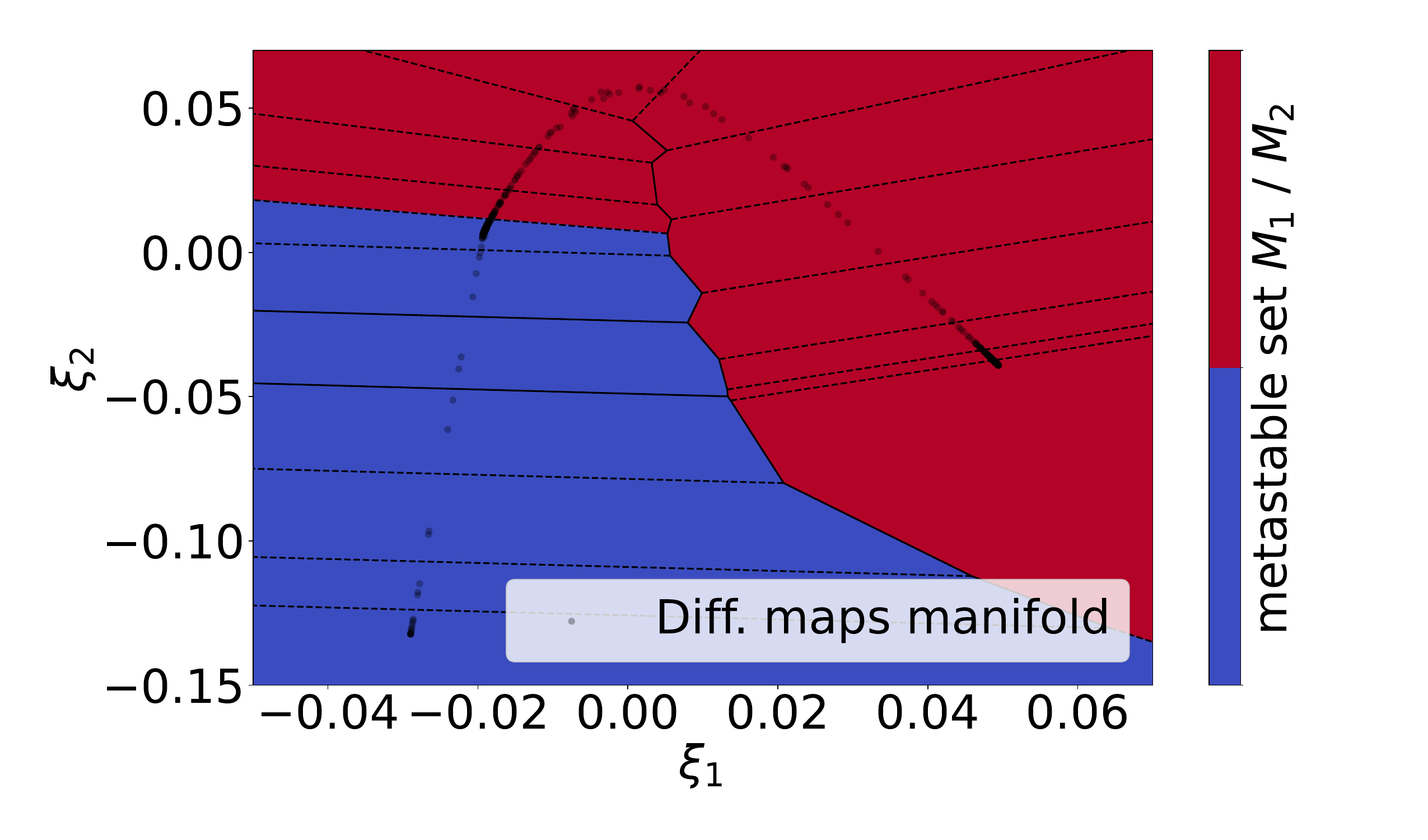}
    \caption{Metastable sets (PCCA+) \newline \,}
    \label{fig:diffusion_maps_SPDE_multichrom_MSM_Voronoi_b}
    \end{subfigure}
       \begin{subfigure}{0.34\textwidth}
    \includegraphics[width=\textwidth]{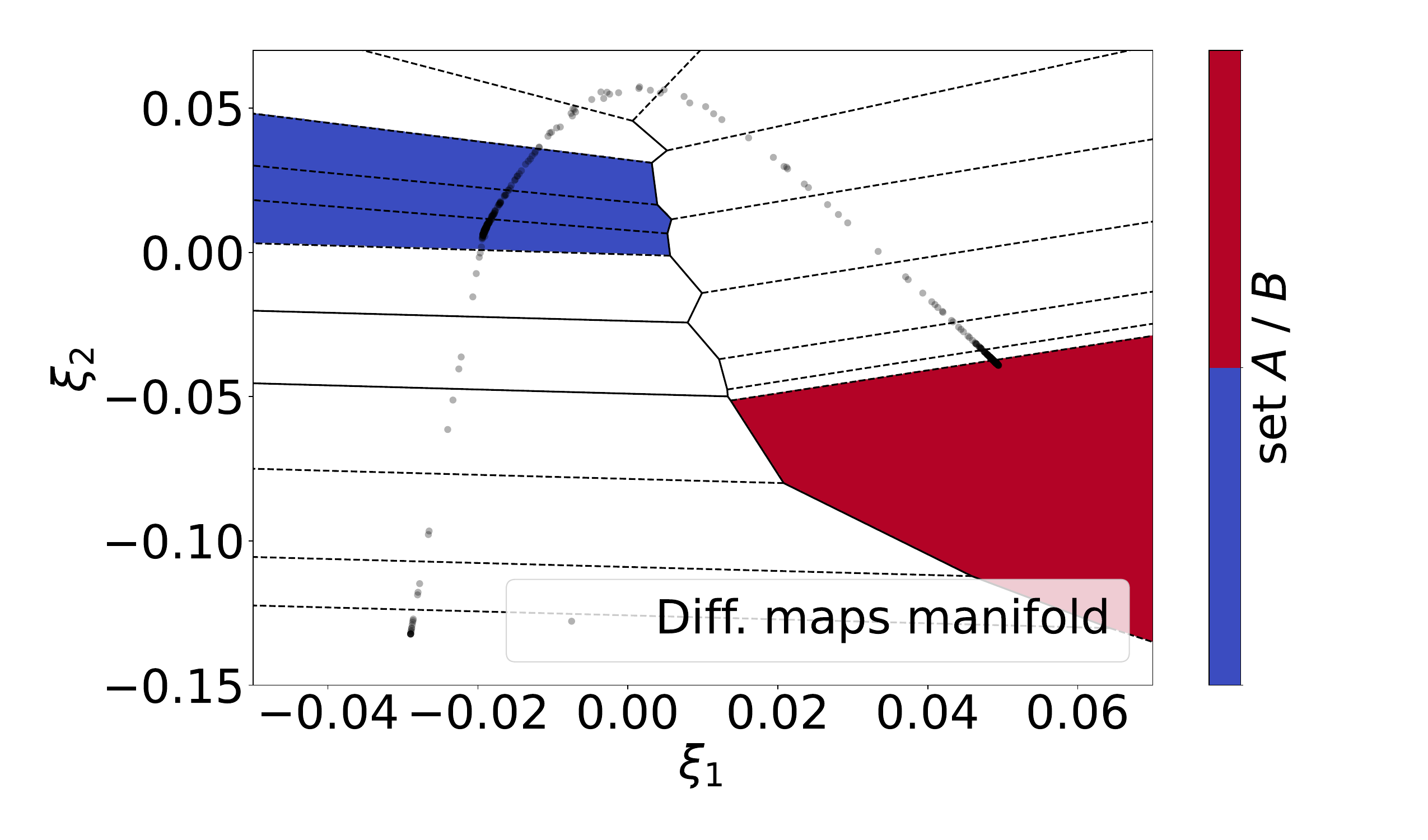}
    \caption{Sets of four-cluster and one-cluster states}
    \label{fig:diffusion_maps_SPDE_multichrom_MSM_Voronoi_c}
    \end{subfigure}
    \caption{\textbf{Multichromatic interaction potential: Dynamics on the reduced space (Voronoi cells).}
    (a) The cells are highlighted in orange, with higher transparency indicating lower stay probability. Green arrows illustrate the average transition direction, conditional on the process making a jump.
    (b) Metastable sets of Markov chain states identified using PCCA+ (set $M_1$ in blue, set $M_2$ in red). (c) Sets of Markov chain states representing the four-cluster concentrations ($A$ in blue) and the one-cluster concentrations ($B$ in red).}   \label{fig:diffusion_maps_SPDE_multichrom_MSM_Voronoi}
\end{figure}

Considering the spatial discretization by a regular grid, the transition probabilities of the resulting Markov chain in \Cref{fig:diffusion_maps_SPDE_multichrom_MSM_grid_a} demonstrate that the dynamics are strongly unidirectional. It can also be observed that the highest probabilities of staying are found in the boxes representing the four-cluster states and the one-cluster states, which are also marked as sets $A$ and $B$ in \Cref{fig:diffusion_maps_SPDE_multichrom_MSM_grid_c}. A similar behavior is observed for partitioning into Voronoi cells in \Cref{fig:diffusion_maps_SPDE_multichrom_MSM_Voronoi_a}.

\paragraph{Implied timescales.}
For both discretizations, the spectrum exhibits a clear gap after the first non-trivial eigenvalue, indicating the presence of two metastable regions.
The corresponding relaxation timescales $T_1^\tau$ computed from the leading non-trivial eigenvalue (see~\eqref{T_eigenvalue}) are listed in \Cref{tab:multichrom_timescales} (first column).
Note that the timescales obtained from the uniform grid and the Voronoi discretization agree closely.

The macrostates $M_1,M_2$ identified by PCCA+\footnote{\label{fn:pcca}The number of metastable sets is fixed a priori to two based on the spectral gap.} for the regular grid are shown in \Cref{fig:diffusion_maps_SPDE_multichrom_MSM_grid_b}. Interestingly, the four-cluster configurations are not all assigned to the same macrostate. Instead, the method separates configurations with four clusters of nearly equal size from those in which the cluster masses are strongly unbalanced. This suggests that the latter lie dynamically close to the one-cluster state. In other words, within the chosen Diffusion Map projection, the transition from four evenly divided clusters to four unevenly divided clusters is already a rare event. 
A similar picture arises for the discretization into Voronoi cells in \Cref{fig:diffusion_maps_SPDE_multichrom_MSM_Voronoi_b}. An example trajectory is shown in \Cref{fig:PCCA_partition_traj_a}, illustrating the classification with respect to the metastable regions identified from the global spectral analysis. This separation between balanced and unbalanced four-cluster configurations suggests an interpretation as a natural \textit{early-warning signal}: unbalanced four-cluster configurations appear as intermediate states that
precede the collapse into a single cluster.  
Their placement near the boundary between the metastable regions suggests that
they act as precursors of the imminent transition. 

We define state sets $A$ and $B$---for example, boxes associated mainly with four-cluster and one-cluster states, respectively---with an intermediate transition region between them, see
Figures~\ref{fig:diffusion_maps_SPDE_multichrom_MSM_grid_c} and
\ref{fig:diffusion_maps_SPDE_multichrom_MSM_Voronoi_c}. In this case, both the MFPT $T_{A \to B}$ and the transition time $T^{\mathrm{TPT}}_{A \to B}$ are well-defined and yield values of comparable order of magnitude (second and third columns of Table~\ref{tab:multichrom_timescales}). 

\begin{table}[htbp]
    \centering
    \begin{tabularx}{\textwidth}{l|c|c|c}
        \toprule
        \textbf{Timescales} & \textbf{relaxation time \boldmath $T_1^\tau$ \eqref{T_eigenvalue}} & \textbf{MFPT \boldmath $T_{A\to B}$ \eqref{MFPT}} & \textbf{transition time \boldmath $T^{\text{TPT}}_{A\to B}$ \eqref{eq:MFPT_TPT}} \\
        \midrule
        uniform grid & 20.70  & 13.48 & 10.46 \\
        Voronoi cells & 21.38 & 33.15 & 26.28 \\
        \bottomrule
    \end{tabularx}
        \caption{\textbf{Multichromatic interaction potential: Overview of different timescales.} Values of relaxation timescale $T_1^\tau$, mean first passage time $T_{A \to B}$ and transition time $T^{\text{TPT}}_{A \to B}$ between sets $A$ and $B$ (see Figures~\ref{fig:diffusion_maps_SPDE_multichrom_MSM_grid_c} and \ref{fig:diffusion_maps_SPDE_multichrom_MSM_Voronoi_c}). These timescales should have the same order of magnitude, see Remark~\ref{rem:times} for the relation between them.}
    \label{tab:multichrom_timescales}
\end{table}

\subsubsection{Results for \Cref{ex2}} 

\begin{figure}
    \centering
        \begin{subfigure}{0.3\textwidth}
    \includegraphics[width=\textwidth]{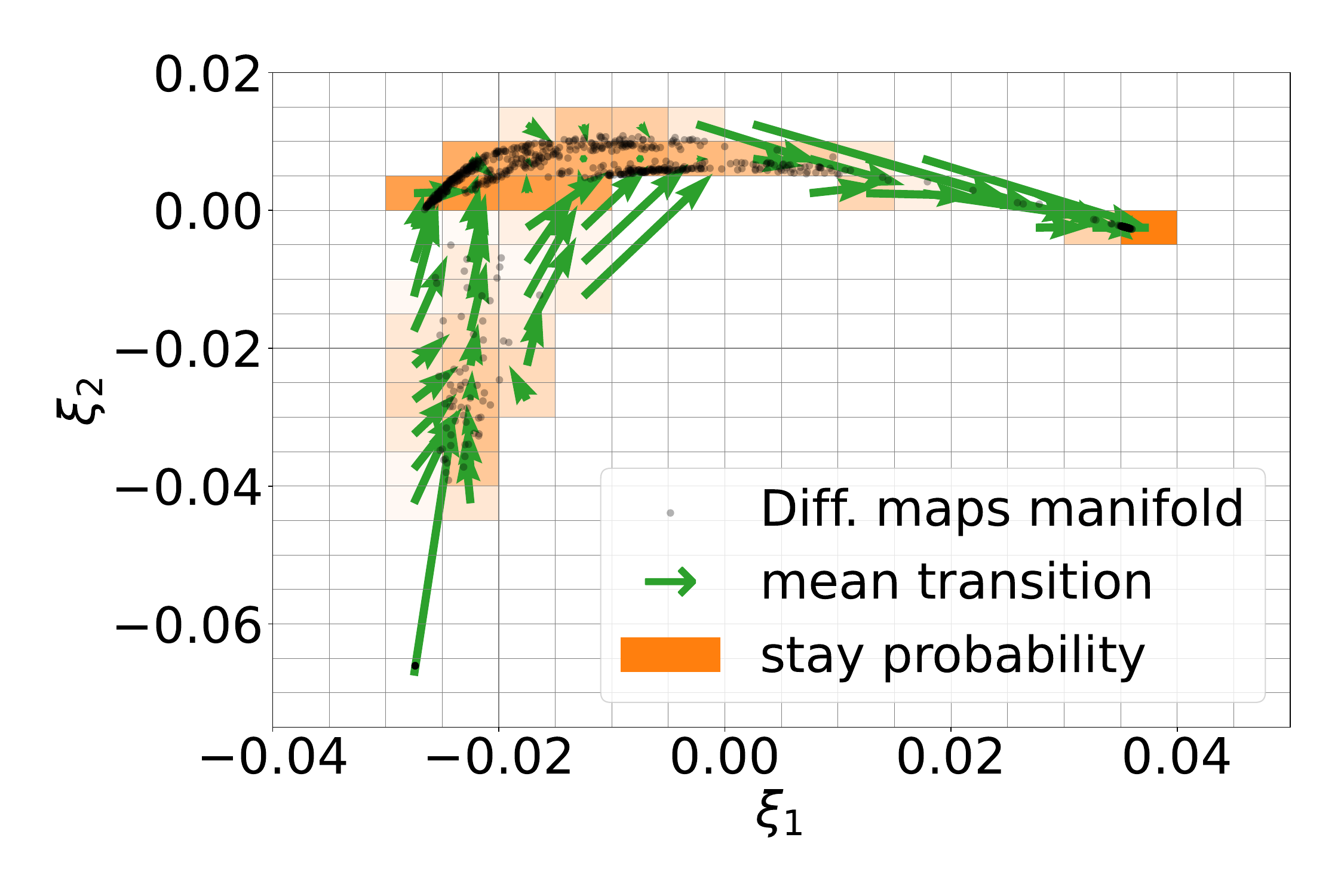}
    \caption{Mean transitions and stay probabilities}
    \label{fig:diffusion_maps_SPDE_Morse_MSM_grid_a}
    \end{subfigure}
        \begin{subfigure}{0.34\textwidth}
    \includegraphics[width=\textwidth]{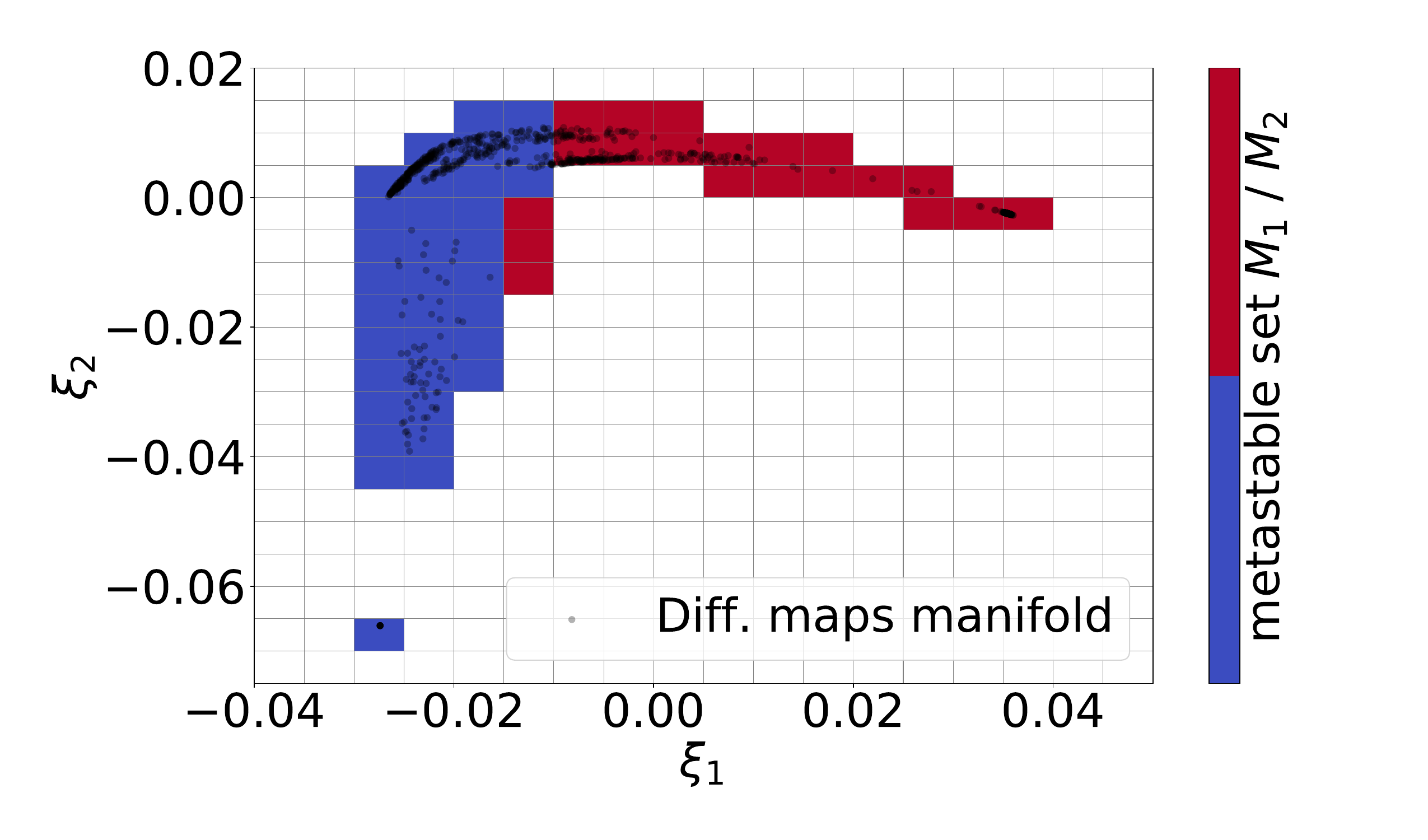}
    \caption{Metastable sets (PCCA+) \newline \,}
    \label{fig:diffusion_maps_SPDE_Morse_MSM_grid_b}
    \end{subfigure}
           \begin{subfigure}{0.34\textwidth}
    \includegraphics[width=\textwidth]{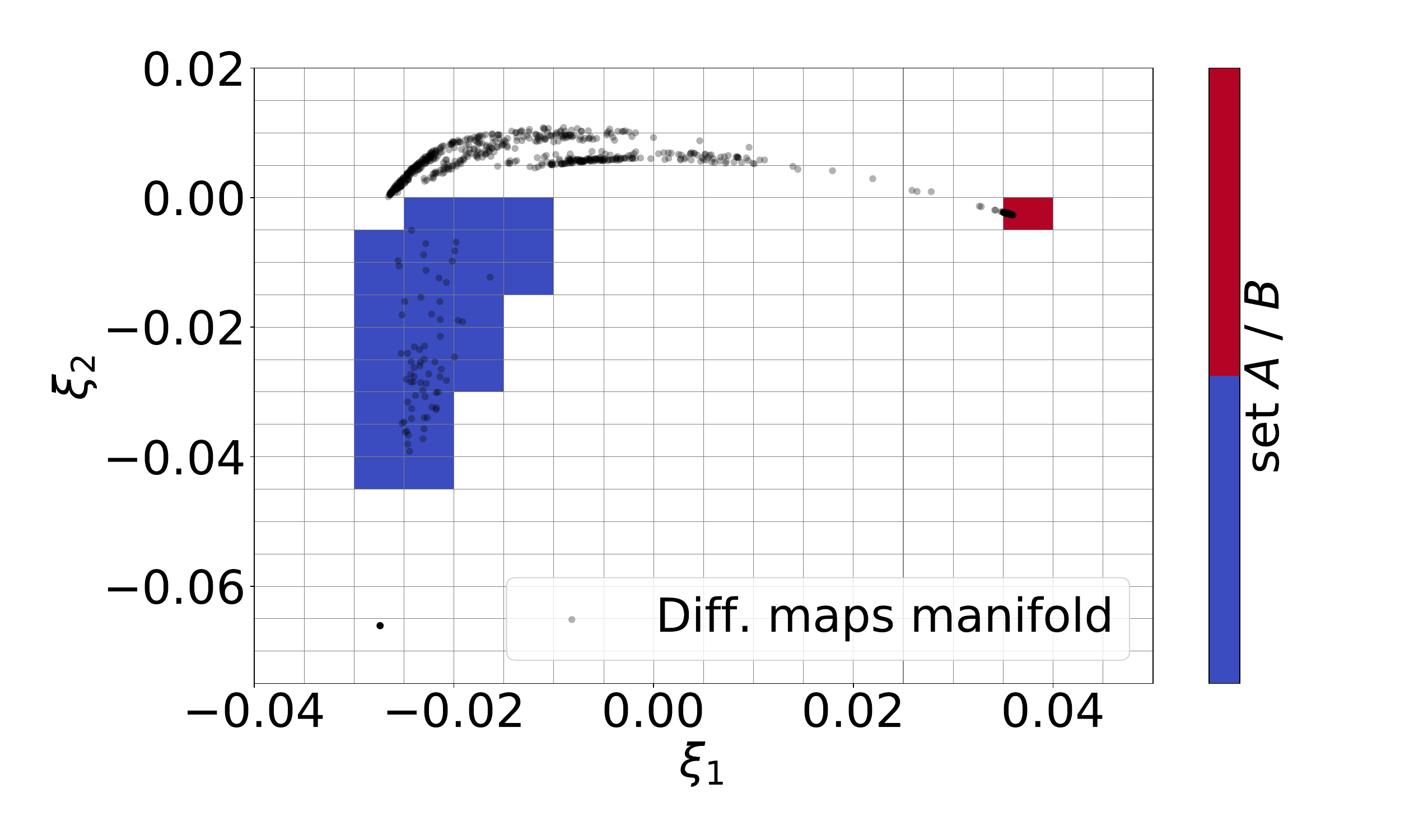}
    \caption{Sets of three-cluster and one-cluster states}
    \label{fig:diffusion_maps_SPDE_Morse_MSM_grid_c}
    \end{subfigure}
    \caption{\textbf{Morse potential: Dynamics on the reduced space (uniform grid).} (a) The relevant boxes are highlighted in orange, with higher transparency indicating lower stay probability. Green arrows illustrate the average transition direction, conditional on the process making a jump. (b) Metastable sets of Markov chain states identified using PCCA+ (set $M_1$ in blue, set $M_2$ in red). (c) Sets of Markov chain states representing the three-cluster concentrations ($A$ in blue) and the one-cluster concentrations ($B$ in red). 
    } \label{fig:diffusion_maps_SPDE_Morse_MSM_grid}
\end{figure}

For the example of the Morse potential~\eqref{eq:F_Morse}, a total of $10^3$ SPDE simulations were run up to $T = 2000$, with concentrations at intervals of $\tau=20$ being considered for estimating the transition matrix. The simulations were started at uniform concentration.

\begin{figure}
    \centering
       \begin{subfigure}{0.3\textwidth}
    \includegraphics[width=\textwidth]{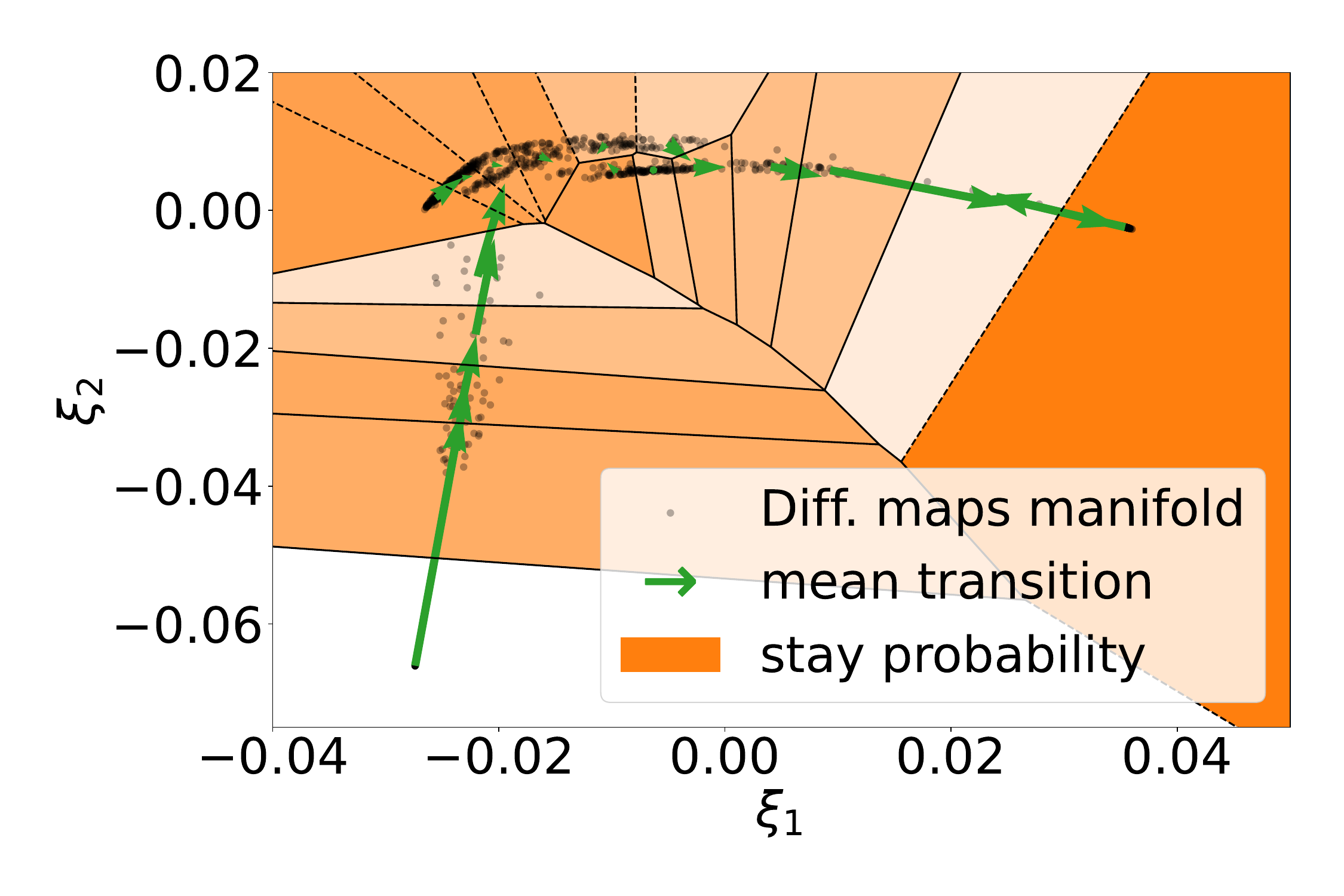}
    \caption{Mean transitions and stay probabilities}
    \label{fig:diffusion_maps_SPDE_Morse_MSM_Voronoi_a}
    \end{subfigure}
     \begin{subfigure}{0.34\textwidth}
    \includegraphics[width=\textwidth]{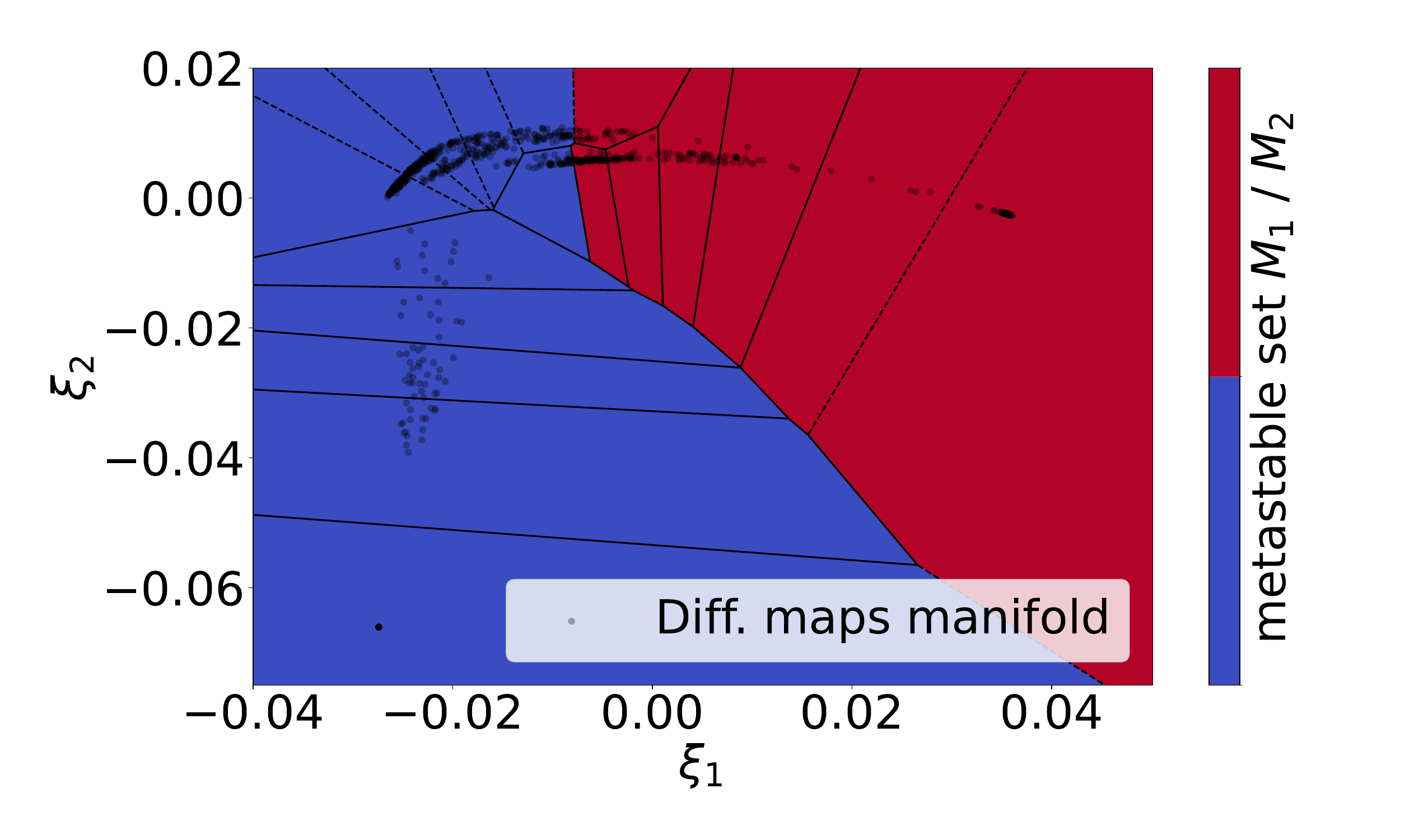}
    \caption{Metastable sets (PCCA+) \newline \,}
    \label{fig:diffusion_maps_SPDE_Morse_MSM_Voronoi_b}
    \end{subfigure}
       \begin{subfigure}{0.34\textwidth}
    \includegraphics[width=\textwidth]{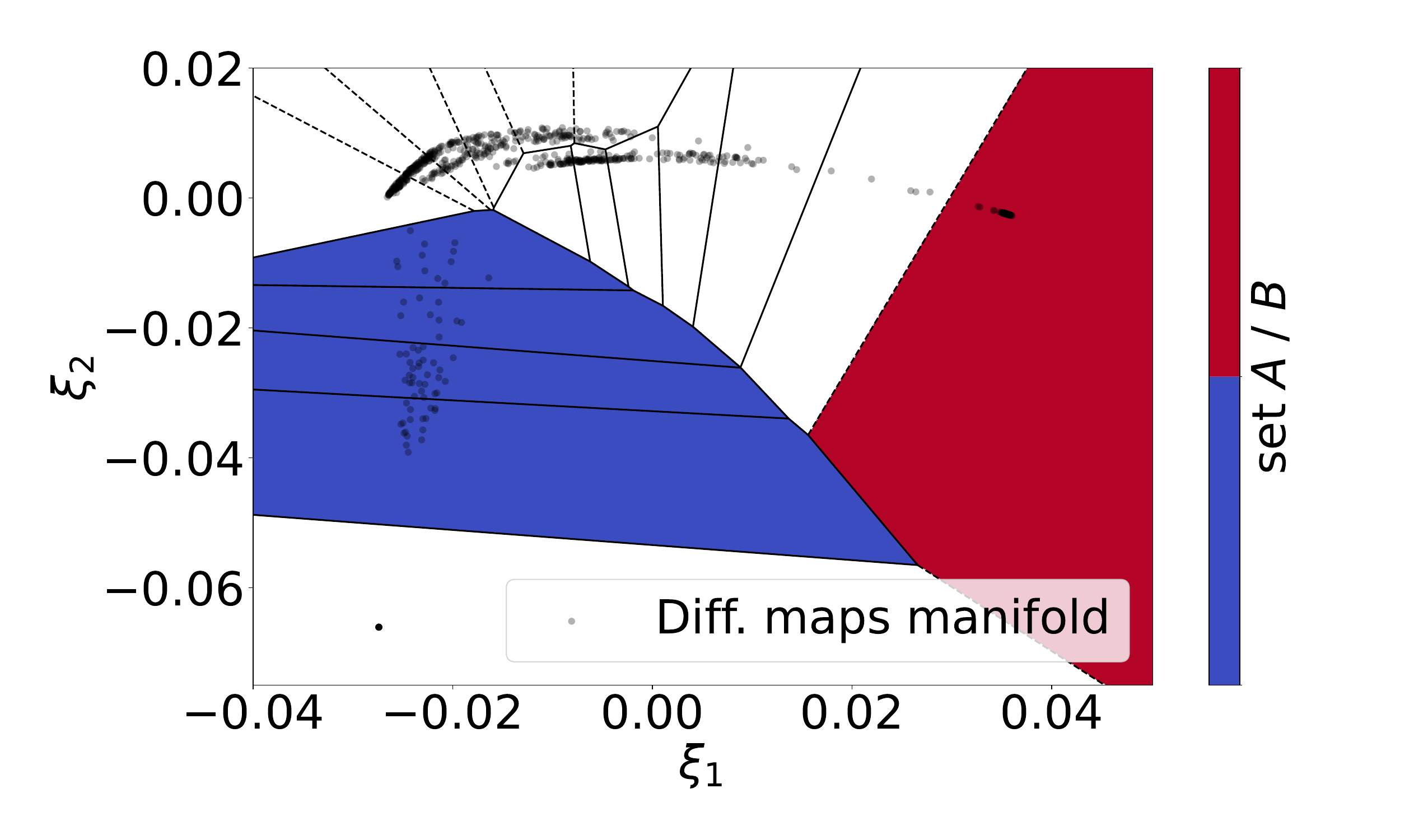}
    \caption{Sets of three-cluster and one-cluster states}
    \label{fig:diffusion_maps_SPDE_Morse_MSM_Voronoi_c}
    \end{subfigure}
    \caption{\textbf{Morse potential: Dynamics on the reduced space (Voronoi cells).}
    (a) The cells are highlighted in orange, with higher transparency indicating lower stay probability. Green arrows illustrate the average transition direction, conditional on the process making a jump.  (b) Metastable sets of Markov chain states identified using PCCA+ (set $M_1$ in blue, set $M_2$ in red). (c) Sets of Markov chain states representing the three-cluster concentrations ($A$ in blue) and the one-cluster concentrations ($B$ in red).   
    }   \label{fig:diffusion_maps_SPDE_Morse_MSM_Voronoi}
\end{figure}

\paragraph{The reduced Markov chain.}
Similar to \Cref{ex1}, the non-reversible Markov chain for the Morse potential exhibits an absorbing state, namely that of the one-cluster concentration state. The reversibility-constrained estimator\footref{fn:iteration}, which is used in the following, deviates only minimally from this absorbing behavior. 
However, the dynamics on the way to the almost absorbing one-cluster state differs from \Cref{ex1}. This can be observed in Figures~\ref{fig:diffusion_maps_SPDE_Morse_MSM_grid_a} and \ref{fig:diffusion_maps_SPDE_Morse_MSM_Voronoi_a}, where transitions to previously visited states and circular transitions are found within the two-cluster region. This is not directly visible in the mean-transition representation, but boxes or cells where the mean transition does not point in a clear direction actually have similarly probable transitions in a wide range of directions, indicating more reversible dynamics than in \Cref{ex1}. The two-cluster region is located between the three-cluster and one-cluster regions, which are highlighted in Figures~\ref{fig:diffusion_maps_SPDE_Morse_MSM_grid_c} and \ref{fig:diffusion_maps_SPDE_Morse_MSM_Voronoi_c}. The division into these three regions is based on the first non-trivial left eigenvector of the transition matrix, which suggests almost-invariant sets of states. 

\paragraph{Implied timescales.} For the two discretizations, the eigenvalue spectra exhibit a gap following the first non-trivial eigenvalue, indicating the presence of two metastable regions (and not three, which could be expected given the three regions described before). The corresponding relaxation timescales $T_1^{\tau}$ are reported in Table~\ref{tab:Morse_timescales} (first column). These timescales show good agreement across the discretizations.

\begin{figure}
    \centering
     \begin{subfigure}{0.49\textwidth}
    \includegraphics[width=\textwidth]{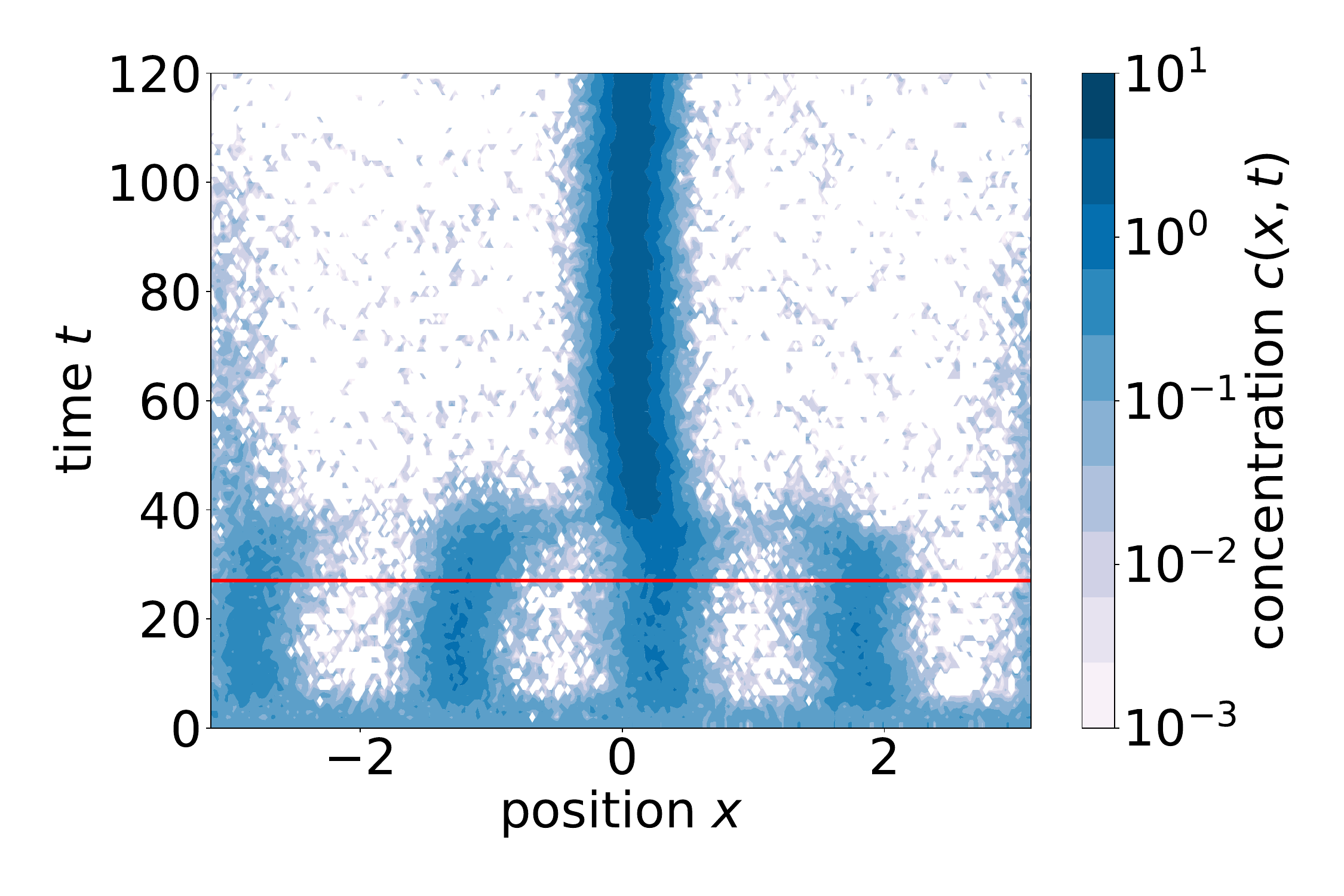}
    \caption{Multichromatic potential (\Cref{ex1})}
     \label{fig:PCCA_partition_traj_a}
    \end{subfigure}
    \begin{subfigure}{0.49\textwidth}
    \includegraphics[width=\textwidth]{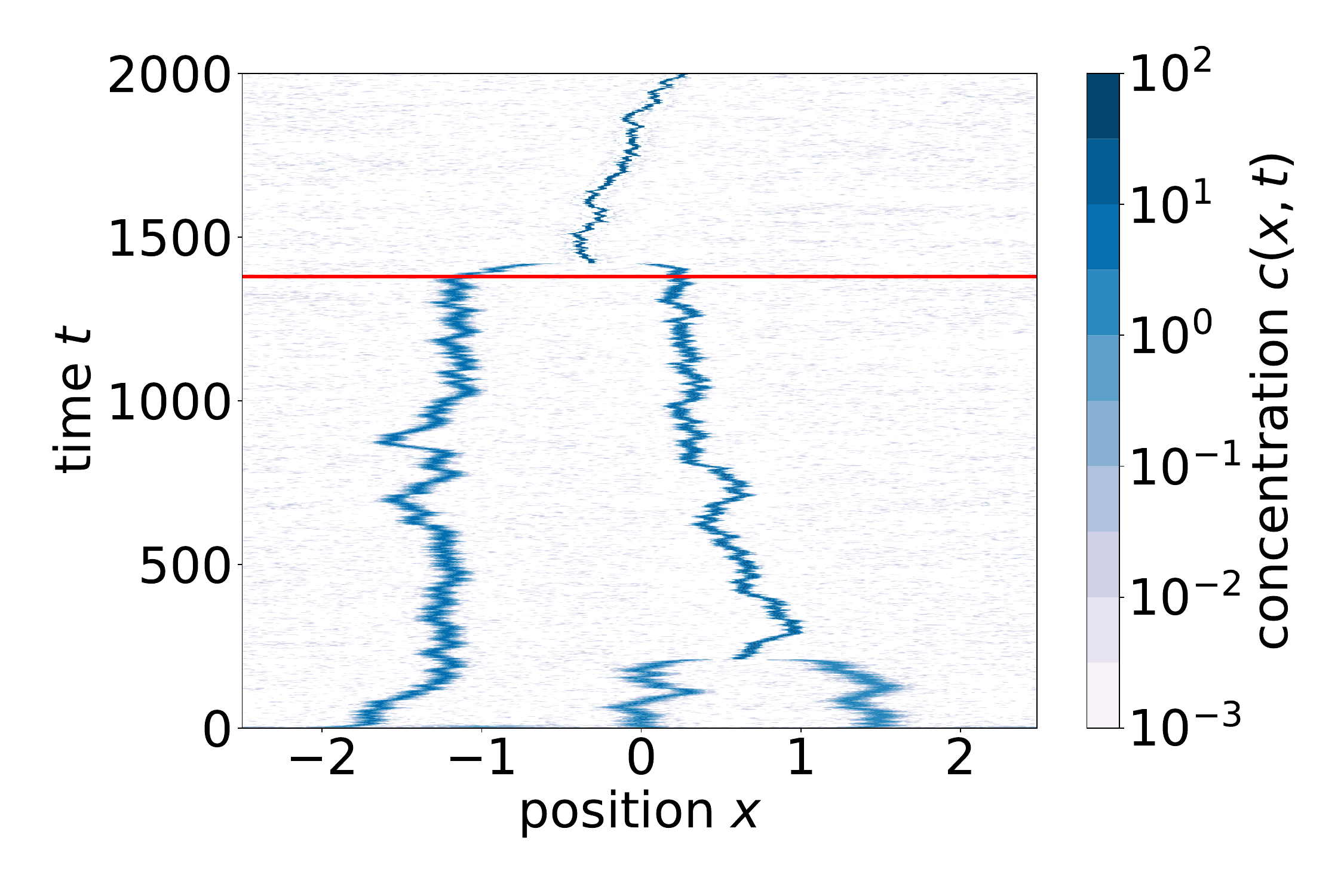}
    \caption{Morse potential (\Cref{ex2})}
     \label{fig:PCCA_partition_traj_b}
    \end{subfigure}
    \caption{\textbf{Partitioning of trajectories into metastable sets.} Division of exemplary trajectories (from Figures~\ref{fig:diffusion_maps_SPDE_multichrom_a} and \ref{fig:diffusion_maps_SPDE_Morse_a}) into metastable sets $M_1$ and $M_2$ (PCCA+) in Voronoi discretization ($M_1$ before red line, $M_2$ after red line) for (a) multichromatic potential and (b) Morse potential. The metastable sets are obtained from the spectral analysis of the full transition matrix (see Figures~\ref{fig:diffusion_maps_SPDE_multichrom_MSM_Voronoi_b} and \ref{fig:diffusion_maps_SPDE_Morse_MSM_Voronoi_b}) and are here illustrated along two representative trajectories. 
    Note that there can be multiple transitions across the boundary between $M_1$ and $M_2$ due to the reversibility of the dynamics; only the last transition is plotted here.} \label{fig:PCCA_partition_traj}
\end{figure}

\Cref{fig:diffusion_maps_SPDE_Morse_MSM_grid_b} shows the metastable sets $M_1$ and $M_2$ identified by PCCA+ for the uniform grid discretization. As suggested by the eigenvalue spectrum, the analysis reveals only two macrostates instead of the three nominal cluster configurations.  
Moreover, the two-cluster states are split across these two macrostates: PCCA+ distinguishes between configurations in which the two clusters are far apart and those where the clusters are closer together, indicating that the latter are dynamically closer to the one-cluster states. A similar pattern emerges for the discretization into Voronoi cells shown in \Cref{fig:diffusion_maps_SPDE_Morse_MSM_Voronoi_b}. The trajectory in \Cref{fig:PCCA_partition_traj_b} illustrates a transition between the two macrostates determined from the ensemble-based analysis.
As in \Cref{ex1}, this separation of seemingly similar configurations reflects the system's proximity to a critical transition and provides a natural early-warning indicator of the collapse event.

Choosing $A$ and $B$ such that $A$ corresponds to the set of all three-cluster states and $B$ to the set of one-cluster states, the transition region consists of the two-cluster states, see Figures~\ref{fig:diffusion_maps_SPDE_Morse_MSM_grid_c} and \ref{fig:diffusion_maps_SPDE_Morse_MSM_Voronoi_c}. For this choice, the corresponding MFPT $T_{A \to B}$ can be reliably calculated using~\eqref{MFPT}, while the transition time $T^{\text{TPT}}_{A \to B}$ can be obtained from~\eqref{eq:MFPT_TPT}, with both approaches yielding similar values, see Table~\ref{tab:Morse_timescales}~(second and third column). 

\begin{table}[htbp]
    \centering
    \begin{tabularx}{\textwidth}{l|c|c|c}
        \toprule
        \textbf{Timescales} & \textbf{relaxation time \boldmath $T_1^\tau$ \eqref{T_eigenvalue}} &  \textbf{MFPT \boldmath $T_{A\to B}$ \eqref{MFPT}} & \textbf{transition time \boldmath $T^{\text{TPT}}_{A\to B}$ \eqref{eq:MFPT_TPT}} 
        \\
        \midrule
        uniform grid & 1893.12 & 1397.27 & 1383.71   \\
        Voronoi cells & 1864.99 & 2133.36 & 2128.34 \\
        \bottomrule
    \end{tabularx}
        \caption{\textbf{Morse potential: Overview of different timescales.} Values of relaxation timescale $T_1^\tau$, mean first passage time $T_{A \to B}$ and transition time $T^{\text{TPT}}_{A \to B}$ between sets $A$ and $B$ (see Figures~\ref{fig:diffusion_maps_SPDE_Morse_MSM_grid_c} and \ref{fig:diffusion_maps_SPDE_Morse_MSM_Voronoi_c}). These timescales should have the same order of magnitude, see Remark~\ref{rem:times} for the relation between them.
        }
    \label{tab:Morse_timescales}
\end{table}

\begin{rem}[Limitation of TPT]
\label{rem:TPT}
TPT characterizes reactive trajectories in the \textit{stationary} regime of an ergodic Markov process. In our system, the stationary distribution places overwhelming weight on the one-cluster set $B$, while the multi–cluster set $A$ is visited only very rarely (in the enforced reversible formulation, returns from $B$ to $A$ are theoretically possible but extremely unlikely), cf. Remark~\ref{rem:times}.  
Consequently, the ensemble of reactive trajectories form $A$ to $B$ is not rich enough for reliable statistics.

In contrast, the mean first-passage time we seek is a \textit{transient} quantity: it refers to trajectories that are \textit{initialized in $A$} and terminated upon reaching $B$.  Such first-passage times are given by the MFPT formula~\eqref{MFPT}, not directly by standard TPT, which assumes a stationary ensemble rather than a prescribed initial condition.  
Extensions of TPT to time-dependent or finite-time settings~\cite{helfmann2020extending} do not provide an alternative closed expression for the MFPT, since the MFPT is inherently a time-independent expectation defined with respect to an initial distribution supported~in~$A$.
\end{rem}

\section{Discussion and outlook}

In this work, we applied a known coarse-graining strategy---combining manifold learning with the construction of a Markov model---to interacting particle dynamics in which clustering emerges from pairwise forces. While the underlying motivation comes from particle-based systems, in practice we approximate these dynamics by discretized particle concentrations and generate the corresponding data using simulations of the Dean--Kawasaki SPDE.
Interpreted through the transfer-operator viewpoint, the approach follows a multi-stage reduction of the Perron--Frobenius operator: first projecting the particle-level operator onto concentration space and then onto a coarse partition of that space. 
Our data-driven framework embeds the concentration data into a low-dimensional manifold using Diffusion Maps and then builds a Markov chain on disjoint regions of this manifold, yielding a reduced model that approximates the Perron--Frobenius operator of the underlying dynamics. The resulting coarse-grained transfer operator preserves the key features of the clustering process---including the number, size, and spatial arrangement of clusters---demonstrating that the approach provides a systematic and effective reduction of complex particle-based dynamics. 
Using standard tools for analyzing Markov processes, we further examined the emergent dynamical structure in terms of time scales and metastability.

For two basic but representative exemplary settings, we uncovered the following main insights:
\begin{itemize}
    \item The effective dynamics evolve on a low-dimensional manifold and can be approximated by a Markov process with only a small number of discrete states. 
    \item The approximate Markov process is nearly irreversible because escapes from the one-cluster state are extremely rare. This makes the use of transition path theory delicate.
    \item A metastable decomposition obtained via PCCA+ identifies a partition before the one-cluster state is reached, which can be interpreted as an early-warning signal indicating that the system has crossed a point of no return.
\end{itemize}

Our study serves as a proof of principle and opens several directions for further research. These include exploring different interaction potentials, extending the analysis to two- and three-dimensional settings, and considering alternative physical domains, boundary conditions, and parameter regimes---particularly those enforcing reversibility. The methodology could further be adapted to non-stationary or externally forced systems with slowly varying parameters or time-dependent interaction strengths. Beyond particle-based models, the approach may also prove valuable for network dynamics in which synchronization plays a role analogous to clustering, such as neuronal networks where synchronized firing is associated with epileptic seizure onset~\cite{gerster2020}, or opinion-dynamics models where consensus formation resembles aggregation into coherent groups~\cite{djurdjevacconrad2024}.

\section*{Code availability}
The Python code used to produce simulations and plots in this paper is available on \url{https://doi.org/10.5281/zenodo.17710015}. 

\section*{Acknowledgment}
This research has been funded by Deutsche Forschungsgemeinschaft
(DFG) through grant \textit{CRC 1114 Scaling Cascades in Complex Systems} (Project
No. 235221301), Project C03: Multiscale modelling and simulation for spatiotemporal master equations, and under Germany's Excellence Strategy MATH+: Berlin Mathematics Research Center (EXC 2046/1, project 390685689). G.A.P. is partially supported by an ERC-EPSRC Frontier Research Guarantee through Grant No. EP/X038645, ERC Advanced Grant No. 247031 and a Leverhulme Trust Senior Research Fellowship, SRF$\backslash$R1$\backslash$241055. 

\appendix

\section{Appendix}

\addcontentsline{toc}{section}{Appendix}

\renewcommand{\thesubsection}{A.\arabic{subsection}}

\subsection{Dimension of Diffusion Map embedding} \label{app:opt-choice-eps}

\begin{figure}
    \centering
    \begin{subfigure}{0.32\textwidth}
    \includegraphics[width=\textwidth]{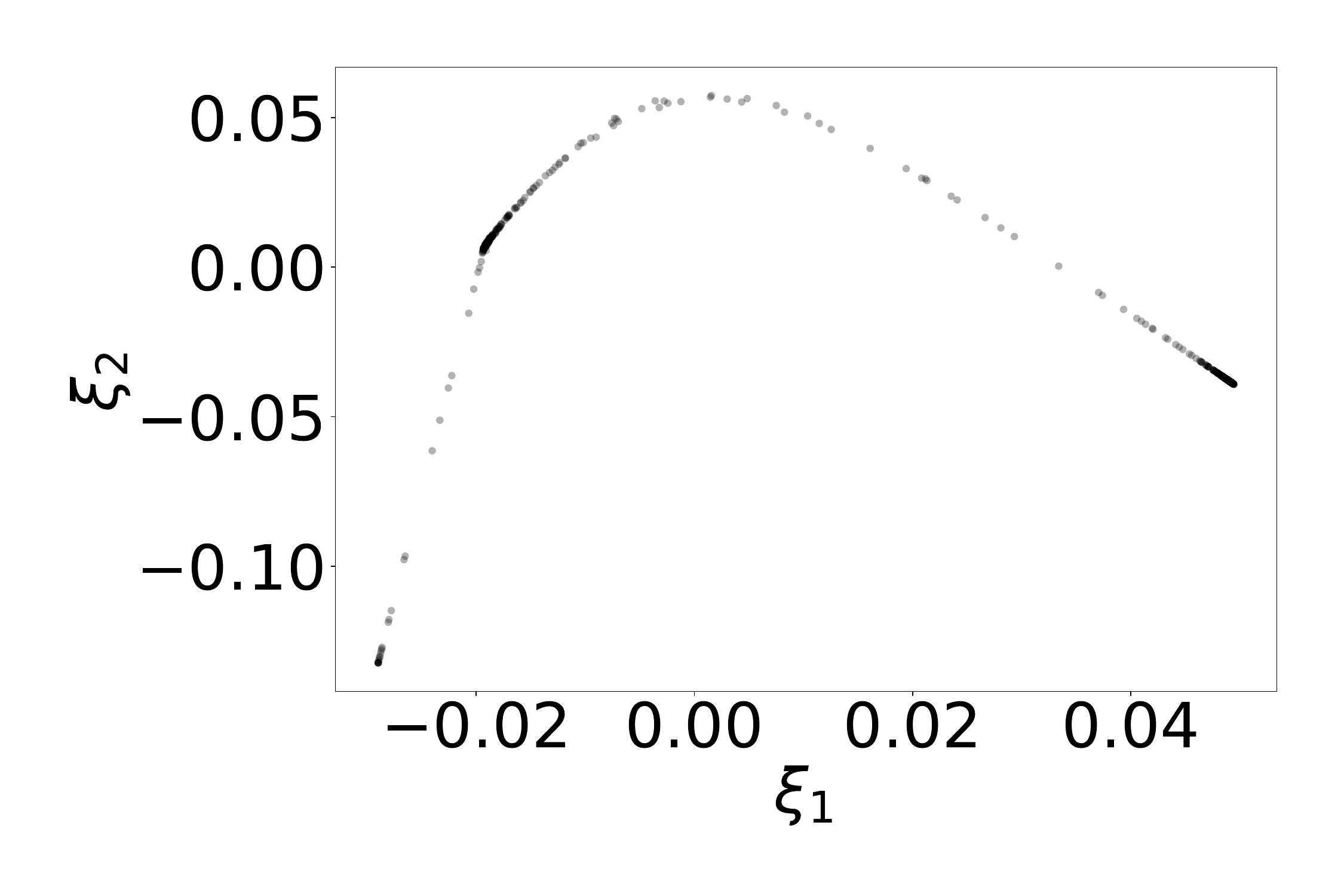}
    \caption{}
    \label{fig:diffusion_maps_SPDE_multichrom_eps_b}
    \end{subfigure}
    \begin{subfigure}{0.32\textwidth}
    \includegraphics[width=\textwidth]{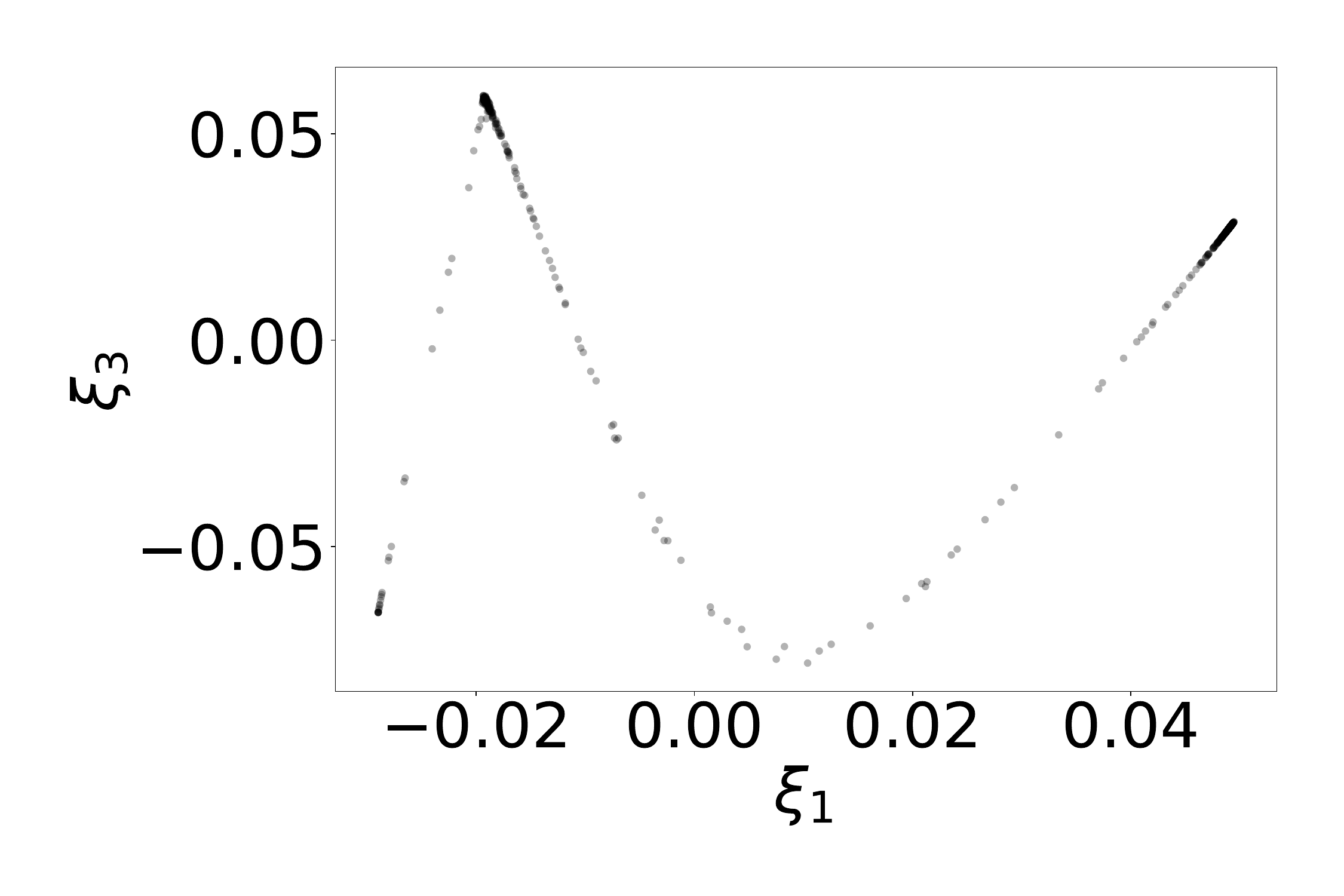}
    \caption{}
    \label{fig:diffusion_maps_SPDE_multichrom_eps_c}
    \end{subfigure}
     \begin{subfigure}{0.32\textwidth}
    \includegraphics[width=\textwidth]{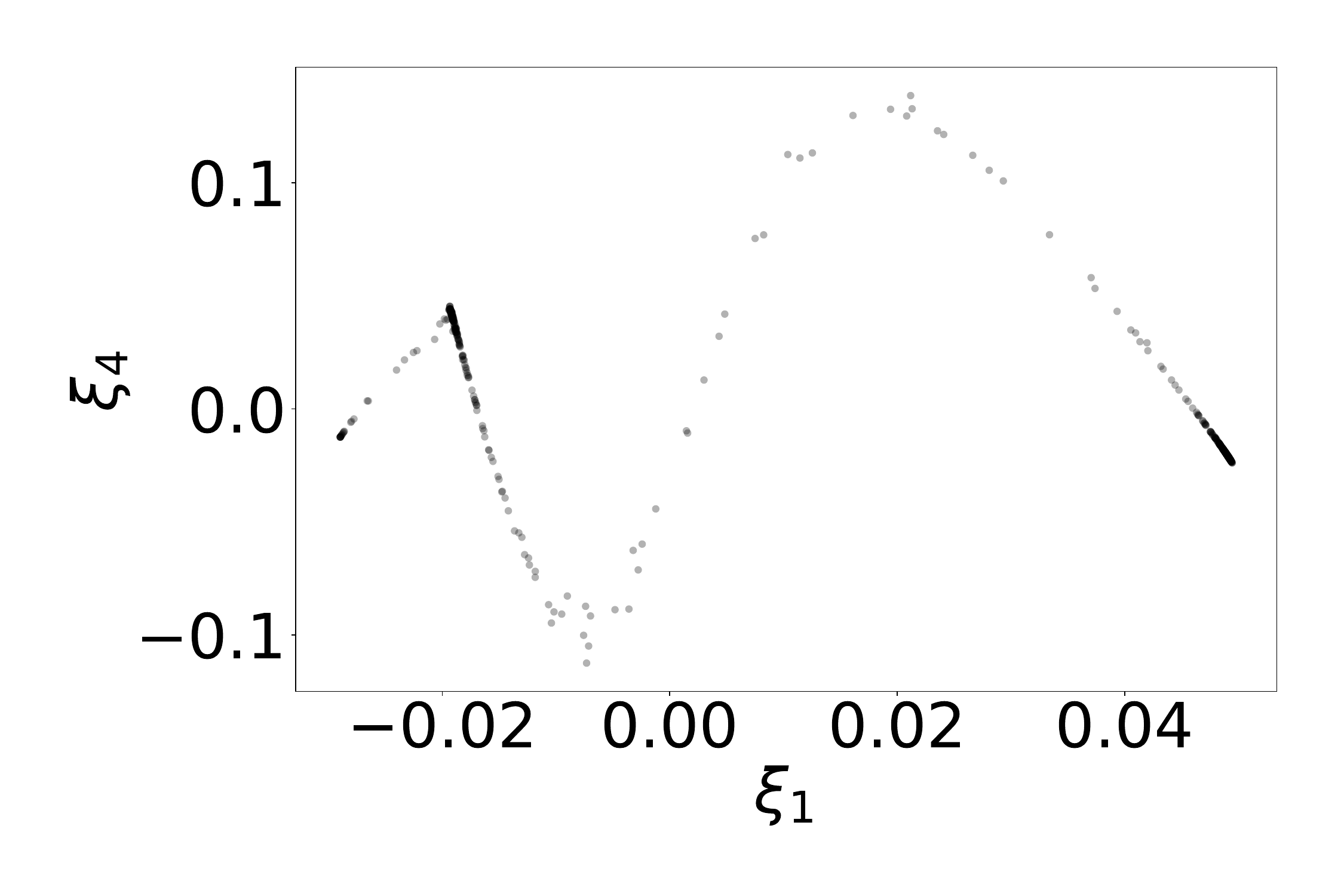}
    \caption{}
    \label{fig:diffusion_maps_SPDE_multichrom_eps_d}
    \end{subfigure}
    \caption{\textbf{Multichromatic interaction potential: Dimension of Diffusion Map embedding.} Different combinations of Diffusion Map coordinates $(\xi_1, \xi_i), i=2,3, 4$.}
    \label{fig:diffusion_maps_SPDE_multichrom_eps}
\end{figure}

For the multichromatic potential of \Cref{ex1}, setting the proximity parameter to $\varepsilon =1$ and testing different combinations of Diffusion Map coordinates in two dimensions (Figures~\ref{fig:diffusion_maps_SPDE_multichrom_eps_b}-\ref{fig:diffusion_maps_SPDE_multichrom_eps_d}) leads to the observation that the lower-ranked eigenvectors are one-dimensional curves (harmonics) of the first non-trivial eigenvector. This indicates that the embedded particle concentrations form a one-dimensional manifold.

\begin{figure}
    \centering
     \begin{subfigure}{0.32\textwidth}
    \includegraphics[width=\textwidth]{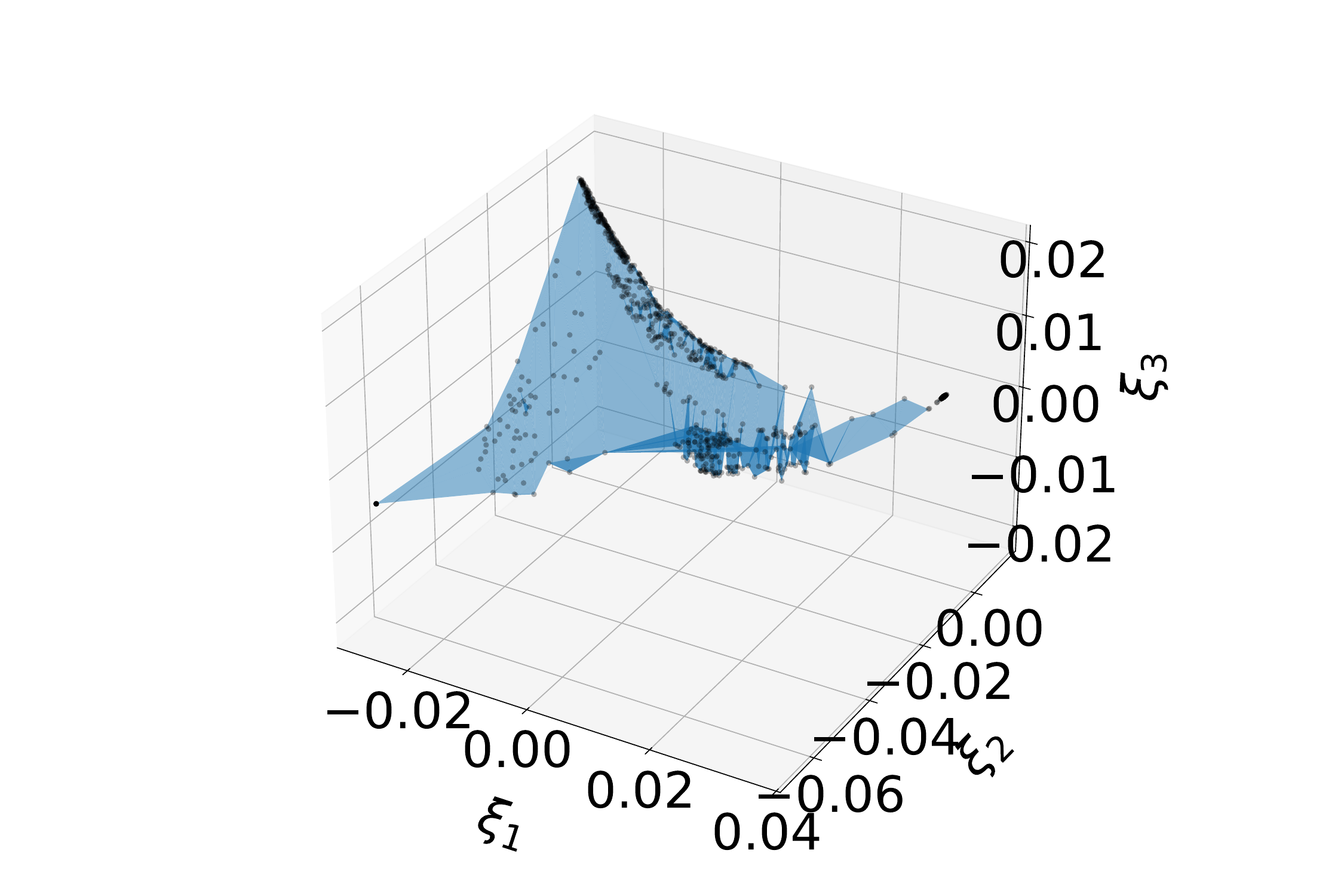}
    
    \caption{}
    \label{fig:diffusion_maps_SPDE_Morse_eps_b}
    \end{subfigure}
    \begin{subfigure}{0.32\textwidth}
    \includegraphics[width=\textwidth]{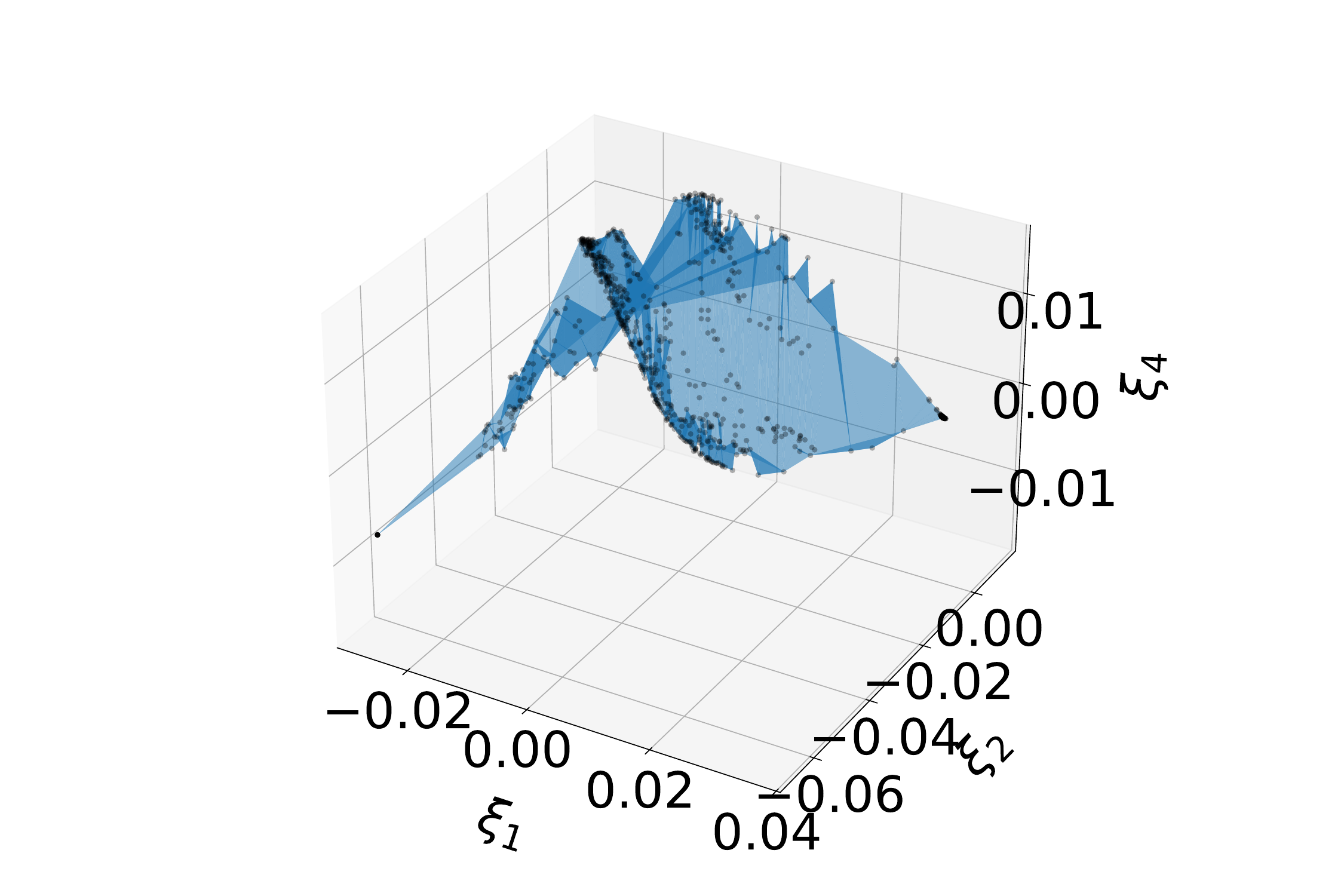}
    \caption{}
    \label{fig:diffusion_maps_SPDE_Morse_eps_c}
    \end{subfigure}
     \begin{subfigure}{0.32\textwidth}
    \includegraphics[width=\textwidth]{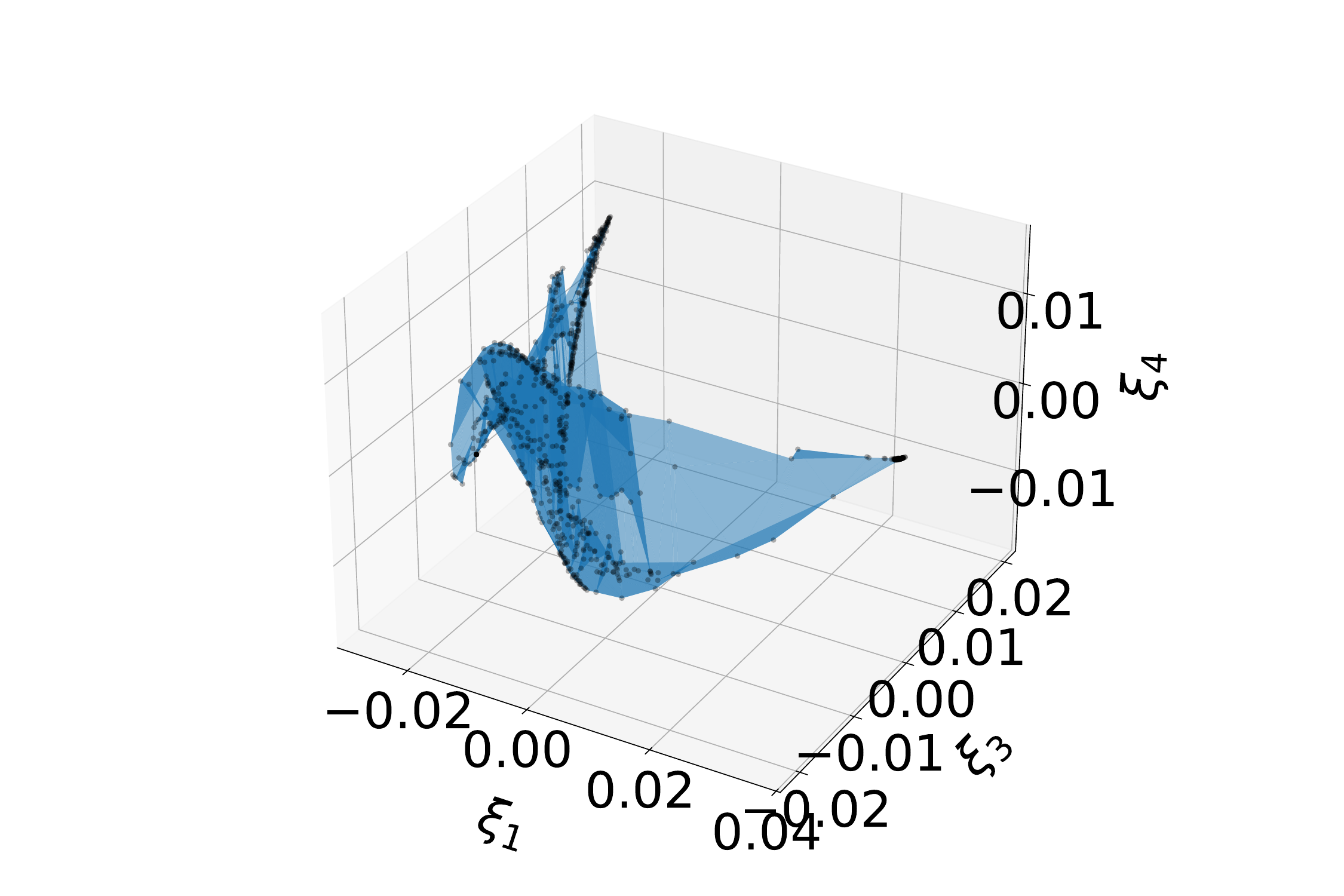}
    \caption{}
    \label{fig:diffusion_maps_SPDE_Morse_eps_d}
    \end{subfigure}
    \begin{subfigure}{0.24\textwidth}
    \end{subfigure}
    \caption{\textbf{Morse potential: Dimension of Diffusion Map embedding.} Different combinations of Diffusion Map coordinates $(\xi_1, \xi_j, \xi_l), j, l\in\{2,3, 4\}$.}
    \label{fig:diffusion_maps_SPDE_Morse_eps}
\end{figure}

Considering the combinations of Diffusion Map coordinates for the Morse potential of \Cref{ex2} ($\varepsilon=0.2$) shown in Figures~\ref{fig:diffusion_maps_SPDE_Morse_eps_b}-\ref{fig:diffusion_maps_SPDE_Morse_eps_d} suggest that the manifold is two-dimensional. Therefore, the first two projection coordinates provide a sufficient representation of its intrinsic geometry.

\subsection{Grouping of concentration states in the Diffusion Map embedding}\label{app:cluster detection}

\begin{figure}
    \centering
      \begin{subfigure}{0.49\textwidth}
    \includegraphics[width=\textwidth]{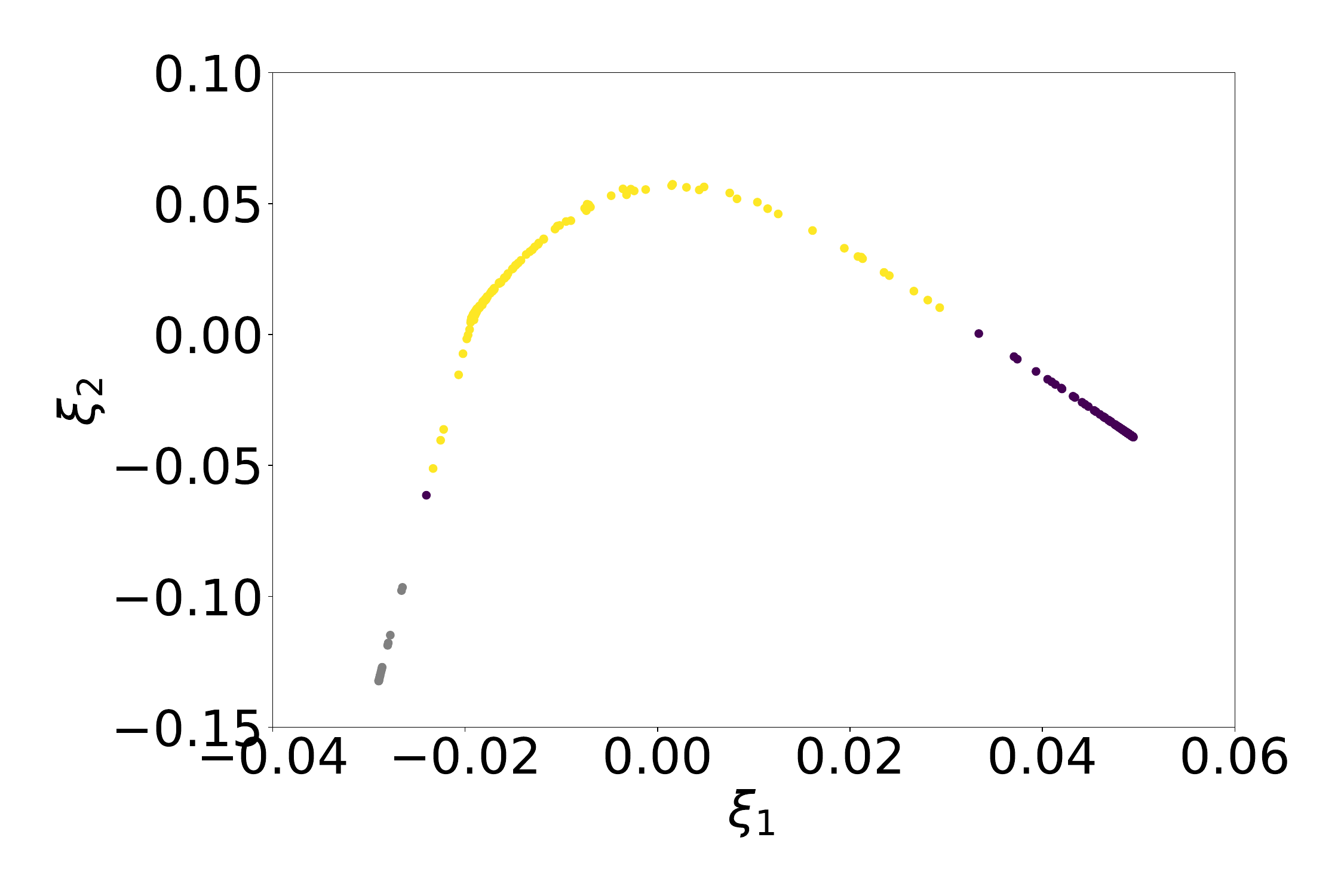} 
    \caption{Multichromatic potential (\Cref{ex1})}
    \label{fig:diffusion_maps_SPDE_multichrom_cluster_b}
    \end{subfigure}
       \begin{subfigure}{0.49\textwidth}
    \includegraphics[width=\textwidth]{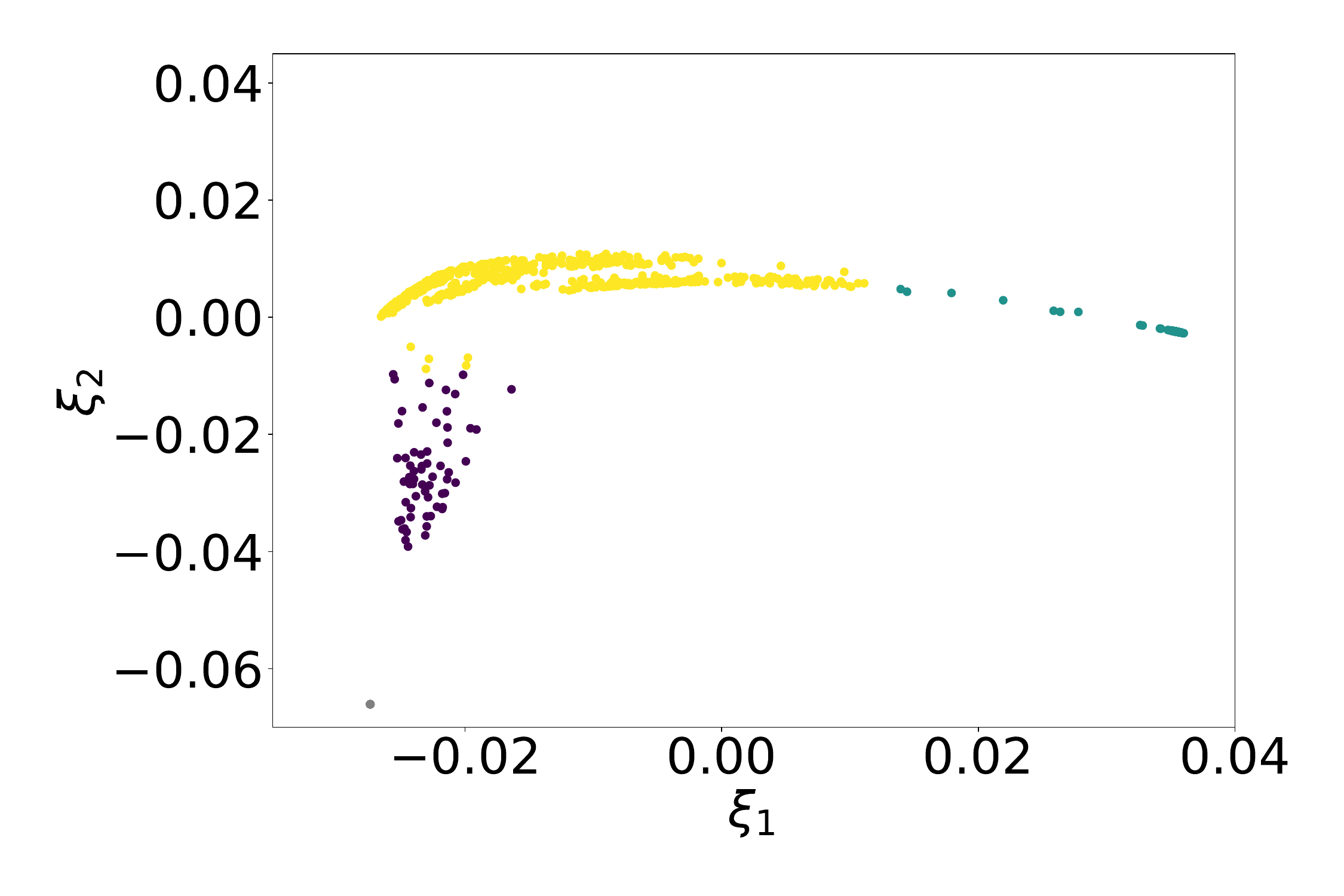}
    \caption{Morse potential (\Cref{ex2})}
    \label{fig:diffusion_maps_SPDE_Morse_cluster_b}
    \end{subfigure}
    \caption{\textbf{Grouping of Diffusion Map embeddings.} K-means is applied to the Diffusion Map embeddings shown in \Cref{fig:diffusion_maps_SPDE_embedding}, with a predefined number of groups $n_{\text{clu}}=2$ in (a) and $n_{\text{clu}}=3$ in (b). The grey dots mark the projections of the initial concentrations (uniform concentration and very similar ones), which are omitted from the K-means grouping.} \label{fig:diffusion_maps_SPDE_embedding_cluster}
\end{figure}

Projecting trajectories into a low-dimensional space via Diffusion Maps, as shown in \Cref{fig:diffusion_maps_SPDE_embedding}, allows one to visually identify groups of similar concentration configurations. As a simple geometric analysis, we apply the standard \textit{K-means} algorithm~\cite{lloyd1982,scikit-learn} to the Diffusion Map embeddings. Since K-means requires the number of groups to be specified in advance, we choose this number according to the dominant concentration patterns observed in the simulations. The initial uniform concentrations are excluded from this analysis, since they correspond to prescribed initial conditions and occur only at the very beginning of the trajectories. Including these transient states would bias the grouping, as they are far from the concentration patterns that dominate the dynamics and would therefore influence the partition determined by K-means. The resulting groupings for Examples~\ref{ex1} and~\ref{ex2} are shown in \Cref{fig:diffusion_maps_SPDE_embedding_cluster}.

In Example~\ref{ex1}, the algorithm separates configurations with four peaks from those with a single peak. In Example~\ref{ex2}, configurations with three peaks are distinguished from those with two peaks and those with a single peak. These results indicate that the distances defined in \eqref{eq:L2-dist} and \eqref{def:Wasserstein_trans} between concentrations are mapped to informative Euclidean distances in the Diffusion Map embedding.

\subsection{Out-of-sample extension for Diffusion Maps} \label{app:out-of-sample}

The goal is to embed a new data point  $c_{\text{new}}$ into the low-dimensional space obtained by applying Diffusion Maps to the data points $c_1, \dots, c_M$. 

From the procedure described in \Cref{subsec:emb_proc}, we require, in addition to the data points, the quantities 
$q_\varepsilon(c_j) = \sum_{m=1}^M k_\varepsilon(c_j, c_m)$, $ j = 1, \dots, M $, 
as well as the eigenvalues $ \lambda_1, \dots, \lambda_d $ and eigenvectors 
$\psi_1, \dots, \psi_d$ of the matrix $Q_\varepsilon$ given in~\eqref{eq:diff_map_matrix}, in order to perform 
the out-of-sample extension (\textit{Nyström formula}) \cite{coifman2006geometric},
    \begin{equation}
    \psi_l(c_{\text{new}})=\frac{1}{\lambda_l}\sum_{j=1}^{M}Q_\varepsilon(c_{\text{new}}, c_j)(\psi_l)_j.
\end{equation}

The following vectors are computed analogously to the Diffusion Maps procedure, using $c_{\text{new}}$ as an input:
\begin{equation} 
\label{eq:kernel_OOSE}
        k_\varepsilon(c_{\text{new}},c_j) = \exp\left(-\frac{\delta(c_{\text{new}},c_j)^2}{\varepsilon}\right),
    \end{equation}
    \begin{equation}
        \tilde{k}_\varepsilon(c_{\text{new}},c_j) = \frac{k_\varepsilon(c_{\text{new}},c_j)}{q_\varepsilon(c_{\text{new}})\,q_\varepsilon(c_j)},
    \end{equation}
    for $q_\varepsilon(c_{\text{new}}) = \sum_{m=1}^M k_\varepsilon(c_{\text{new}}, c_m)$ and 
    \begin{equation}
        Q_\varepsilon(c_{\text{new}},c_j)= \frac{\tilde{k}_\varepsilon(c_{\text{new}},c_j)}{s_\varepsilon(c_{\text{new}})}
    \end{equation}
    for $s_\varepsilon(c_{\text{new}})=\sum_{m=1}^M \tilde{k}_\varepsilon(c_{\text{new}},c_m)$.     
    Finally, the embedding of $c_{\text{new}}$ is given by $(\lambda_1 \psi_1(c_{\text{new}}), \dots, \lambda_d\psi_d(c_{\text{new}}))$.

\subsection{Computational effort for data-driven approximation of the transfer operator}
\label{app:comp_effort}

In this section, we briefly discuss the numerical cost of the procedure used to approximate the transfer operator from data. Generating the concentration data itself constitutes a substantial part of the computational effort, especially when long simulations are required to capture the relevant dynamics. In many cases, SPDE simulations provide a more efficient way to generate such data than particle-based simulations~\cite{cornalba2023dean}. In the following, we treat the data used for the analysis as given (e.g., they may also originate from experiments) and focus on the cost of the two subsequent steps: constructing the Diffusion Map embedding and estimating the Markov chain.
For both steps, the main computational costs scale with the dimension of the state space $\mathbb{T}$, which is reflected in the number of spatial grid points $K$.

\subsubsection{Computational costs for Diffusion Maps}\label{app:comp_effort_1}
The computational cost of the Diffusion Maps construction in \Cref{sec:galerkin_diffmap} is dominated by the evaluation of the pairwise kernel entries given in~\eqref{kernel}. For $N_{\text{train}}$ concentration snapshots represented on $K$ spatial grid points, this requires $\mathcal{O}(N_{\text{train}}^{2} K \log K)$ operations when using the translation-invariant $L^2$-distance~\eqref{eq:L2-dist} (computed via FFT) or $\mathcal{O}(N_{\text{train}}^{2} K^{2})$ when using the translation-invariant Wasserstein-1 distance~\eqref{def:Wasserstein_trans}. The subsequent normalization of the kernel matrix requires $\mathcal{O}(N_{\text{train}}^{2})$ operations. Finally, computing the eigenvalues and eigenvectors of the diffusion map matrix scales as $\mathcal{O}(N_{\text{train}}^{3})$, which becomes negligible compared to the kernel construction if $N_{\text{train}} \ll K\log K$ or $N_{\text{train}} \ll K^{2}$, respectively. 

\subsubsection{Computational costs for estimating the transition probabilities}\label{app:comp_effort_2}
Let $N_{\text{s}} \gg N_{\text{train}}$ denote the number of concentration snapshots used to estimate the Markov chain in \Cref{sec:galerkin_msm}. In the out-of-sample extension, the dominant computational cost arises from evaluating the kernel entries between the $N_{\text{train}}$ training snapshots and the $N_{\text{s}}$ additional samples, see Equation~\eqref{eq:kernel_OOSE} in Appendix~\ref{app:out-of-sample}. This requires $\mathcal{O}(N_{\text{train}} N_{\text{s}} K\log K)$ operations when using the translation-invariant $L^2$-distance and $\mathcal{O}(N_{\text{train}} N_{\text{s}} K^2)$ operations when using the translation-invariant Wasserstein-1 distance. The purpose of the out-of-sample extension is to reduce the computational effort to this cost, instead of the $\mathcal{O}(N_{\text{s}}^2 K\log K)$ or $\mathcal{O}(N_{\text{s}}^2 K^2)$ operations that would arise if Diffusion Maps were recomputed from scratch for all $N_{\text{s}}$ samples. The subsequent normalization of the kernel and the computation of the new projection coordinates both require $\mathcal{O}(N_{\text{train}} N_{\text{s}})$ operations and are therefore negligible compared to the kernel evaluation. Estimating the transition probabilities in Ulam's method~\eqref{MLE_P} from pairs of consecutive embedded data points has computational complexity $\mathcal{O}(N_{\text{s}}n_{S})$, where $n_S$ is the number of disjoint regions in the spatially discretized embedding space (see \Cref{sec:partition}).

\bibliography{references}

@article{bertoli2025,
    author = {Bertoli, Benedetta and Goddard, Benjamin D and Pavliotis, Grigorios A},
    title = {Stability of stationary states for mean field models with multichromatic interaction potentials},
    journal = {IMA Journal of Applied Mathematics},
    volume = {89},
    number = {5},
    pages = {833--859},
    year = {2025},
    doi = {10.1093/imamat/hxaf001}
}

@article{Fritzsche2007,
    author    = {David Fritzsche and Volker Mehrmann and Daniel B. Szyld and Elena Virnik},
    title     = {An {SVD} approach to identifying metastable states of {Markov} chains},
    journal   = {Electronic Transactions on Numerical Analysis},
    volume    = {29},
    pages     = {46--69},
    year      = {2007}
}

@article{DjurdjevacConradWeberSchuette2016,
    author    = {Natasa Djurdjevac-Conrad and Marcus Weber and Christof Sch\"utte},
    title     = {Finding dominant structures of nonreversible {Markov} processes},
    journal   = {SIAM Interdisciplinary Journal on Multiscale Modeling and Simulation},
    year      = {2016},
    volume    = {14},
    number    = {4},
    pages     = {1319--1340},
    doi       = {10.1137/15M1032272}
}

@incollection{DeuflhardDellnitzJungeSchuette1999,
  author    = {Peter Deuflhard and Michael Dellnitz and Oliver Junge and Christof Sch\"utte},
  title     = {Computation of Essential Molecular Dynamics by Subdivision Techniques},
  booktitle = {Computational Molecular Dynamics: Challenges, Methods, Ideas},
  editor    = {Peter Deuflhard and J. Hermans and B. Leimkuhler and A. Mark and S. Reich and B. Skeel},
  publisher = {Lecture Notes in Computational Science and Engineering},
  volume    = {4},
  pages     = {98--115},
  year      = {1999}
}

@article{SchuetteFischerHuisingaDeuflhard1999,
  author    = {Christof Sch{\"u}tte and Alexander Fischer and Wilhelm Huisinga and Peter Deuflhard},
  title     = {A Direct Approach to Conformational Dynamics Based on Hybrid {Monte Carlo}},
  journal   = {Journal of Computational Physics},
  volume    = {151},
  number    = {1},
  pages     = {146--168},
  year      = {1999},
  doi       = {10.1006/jcph.1999.6231}
}

@article{carrillo2020,
  title={Long-time behaviour and phase transitions for the {McKean}--{Vlasov} equation on the torus},
  author={Carrillo, Jos{\'e} A and Gvalani, RS and Pavliotis, GA and Schlichting, A},
  journal={Archive for Rational Mechanics and Analysis},
  volume={235},
  number={1},
  pages={635--690},
  year={2020},
  publisher={Springer},
  doi={10.1007/s00205-019-01430-4}
}

@article{wehlitz2025,
    author = {Wehlitz, Nathalie and Sadeghi, Mohsen and Montefusco, Alberto and Sch\"{u}tte, Christof and Pavliotis, Grigorios A. and Winkelmann, Stefanie},
    title = {Approximating Particle-Based Clustering Dynamics by Stochastic {PDE}s},
    journal = {SIAM Journal on Applied Dynamical Systems},
    volume = {24},
    number = {2},
    pages = {1231--1250},
    year = {2025},
    doi = {10.1137/24M1676661}
}

@article{helfmann2021,
    author = {Helfmann, Luzie and Heitzig, Jobst and Koltai, Péter and Kurths, Jürgen and Schütte, Christof},
    year = {2021},
    pages = {3249--3271},
    title = {Statistical analysis of tipping pathways in agent-based models},
    volume = {230},
    journal = {The European Physical Journal Special Topics},
    doi = {10.1140/epjs/s11734-021-00191-0}
}

@inproceedings{carrillo2019,
  title={Aggregation-diffusion equations: {D}ynamics, asymptotics, and singular limits},
  author={Carrillo, Jos{\'e} A and Craig, Katy and Yao, Yao},
  booktitle={Active Particles, Volume 2: Advances in Theory, Models, and Applications},
  year = {2019},
  pages = {65--108},
  publisher={Springer International Publishing},
  doi = {10.1007/978-3-030-20297-2_3}
}

@article{dorsogna2006,
  author = {D'Orsogna, Maria and Chuang, Yao-Li and Bertozzi, Andrea and Chayes, L},
  title = {Self-Propelled Particles with Soft-Core Interactions: {P}atterns, Stability, and Collapse},
  journal = {Physical Review Letters},
  volume = {96},
  issue = {10},
  pages = {104302},
  year = {2006},
  publisher = {American Physical Society},
  doi = {10.1103/PhysRevLett.96.104302}
}

@article{leverentz2009,
  author  = {Leverentz, Andrew J. and Topaz, Chad M. and Bernoff, Andrew J.},
  title   = {Asymptotic Dynamics of Attractive-Repulsive Swarms},
  journal = {SIAM Journal on Applied Dynamical Systems},
  volume  = {8},
  number  = {3},
  pages   = {880--908},
  year    = {2009},
  doi     = {10.1137/090749037}
}

@article{topaz2008,
  author  = {Topaz, Chad M. and Bernoff, Andrew J. and Logan, Sheldon and Toolson, Wyatt},
  title   = {A model for rolling swarms of locusts},
  journal = {The European Physical Journal Special Topics},
  volume  = {157},
  year    = {2008},
  pages   = {93--109},
  doi     = {10.1140/epjst/e2008-00633-y}
}

@article{dawson1983,
  title={Critical dynamics and fluctuations for a mean-field model of cooperative behavior},
  author={Donald A. Dawson},
  journal={Journal of Statistical Physics},
  year={1983},
  volume={31},
  pages={29--85},
  doi={10.1007/BF01010922}
}

@article{gaertner1988,
  title={On the {McKean‐Vlasov} Limit for Interacting Diffusions},
  author={J{\"u}rgen G{\"a}rtner},
  journal={Mathematische Nachrichten},
  year={1988},
  volume={137},
  pages={197--248},
  number = {1},
  doi ={10.1002/mana.19881370116}
}

@article{dean1996,
  title={Langevin equation for the density of a system of interacting {Langevin} processes},
  author={Dean, David S},
  journal={Journal of Physics A: Mathematical and General},
  volume={29},
  number={24},
  pages={L613},
  year={1996},
  publisher={IOP Publishing},
  doi={10.1088/0305-4470/29/24/001}
}

@article{kawasaki1998,
  title={Microscopic Analyses of the Dynamical Density Functional Equation of Dense Fluids},
  author={Kawasaki, Kyozi},
  journal={Journal of Statistical Physics},
  volume={93},
  pages={527--546},
  year={1998},
  doi = {10.1023/B:JOSS.0000033240.66359.6c}
}

@article{koltai2020,
    doi = {10.1088/1361-6544/ab6a76},
    year = {2020},
    publisher = {IOP Publishing},
    volume = {33},
    number = {4},
    pages = {1723},
    author = {Koltai, Péter and Weiss, Stephan},
    title = {Diffusion maps embedding and transition matrix analysis of the large-scale flow structure in turbulent {Rayleigh–Bénard} convection},
    journal = {Nonlinearity}
}

@article{evangelou2024,
  title = {Machine learning for the identification of phase transitions in interacting agent-based systems: {A Desai-Zwanzig} example},
  author = {Evangelou, Nikolaos and Giovanis, Dimitris G. and Kevrekidis, George A. and Pavliotis, Grigorios A. and Kevrekidis, Ioannis G.},
  journal = {Physical Review E},
  volume = {110},
  issue = {1},
  pages = {014121},
  year = {2024},
  publisher = {American Physical Society},
  doi = {10.1103/PhysRevE.110.014121}
}

@unpublished{adams2025,
    title={Separation of time scales in weakly interacting diffusions}, 
    author={Zachary P. Adams and Maximilian Engel and Rishabh S. Gvalani},
    year={2025},
    note = {Working paper or preprint},  
    url={https://arxiv.org/abs/2502.12881}
}

@article{coifman2008graph,
  author={Coifman, Ronald R. and Shkolnisky, Yoel and Sigworth, Fred J. and Singer, Amit},
  journal={IEEE Transactions on Image Processing}, 
  title={Graph {L}aplacian Tomography From Unknown Random Projections}, 
  year={2008},
  volume={17},
  number={10},
  pages={1891--1899},
  doi={10.1109/TIP.2008.2002305}
}

@article{garnier2016,
    author = {Garnier, Josselin and Papanicolaou, George and Yang, Tzu-Wei},
    year = {2017},
    pages = {51--75},
    title = {Consensus Convergence with Stochastic Effects},
    volume = {45},
    journal = {Vietnam Journal of Mathematics},
    doi = {10.1007/s10013-016-0190-2}
}

@article{prinz2011,
    author = {Prinz, Jan-Hendrik and Wu, Hao and Sarich, Marco and Keller, Bettina and Senne, Martin and Held, Martin and Chodera, John and Schütte, Christof and Noé, Frank},
    year = {2011},
    pages = {174105},
    title = {Markov models of molecular kinetics: {G}eneration and validation},
    volume = {134},
    journal = {The Journal of Chemical Physics},
    doi = {10.1063/1.3565032}
}

@article{metzner2009,
    author = {Metzner, Philipp and Sch\"{u}tte, Christof and Vanden-Eijnden, Eric},
    title = {Transition Path Theory for {M}arkov Jump Processes},
    journal = {Multiscale Modeling \& Simulation},
    volume = {7},
    number = {3},
    pages = {1192--1219},
    year = {2009},
    doi = {10.1137/070699500}
}

@article{trendelkamp2015,
    author = {Trendelkamp-Schroer, Benjamin and Wu, Hao and Paul, Fabian and Noé, Frank},
    title = {Estimation and uncertainty of reversible {M}arkov models},
    journal = {The Journal of Chemical Physics},
    volume = {143},
    number = {17},
    pages = {174101},
    year = {2015},
    doi = {10.1063/1.4934536}
}

@article{roeblitz2013,
    author = {R\"{o}blitz, Susanna and Weber, Marcus},
    title = {Fuzzy spectral clustering by {PCCA+}: {A}pplication to {M}arkov state models and data classification},
    year = {2013},
    publisher = {Springer},
    address = {Berlin, Heidelberg},
    volume = {7},
    number = {2},
    doi = {10.1007/s11634-013-0134-6},
    journal = {Advances in Data Analysis and Classification},
    pages = {147--179}
}

@article{noe2009,
    author = {Frank Noé  and Christof Schütte  and Eric Vanden-Eijnden  and Lothar Reich  and Thomas R. Weikl },
    title = {Constructing the equilibrium ensemble of folding pathways from short off-equilibrium simulations},
    journal = {Proceedings of the National Academy of Sciences},
    volume = {106},
    number = {45},
    pages = {19011--19016},
    year = {2009},
    doi = {10.1073/pnas.0905466106}
}

@article{helfmann2021interacting,
    title={From interacting agents to density-based modeling with stochastic {PDEs}},
    author={Helfmann, Luzie and Djurdjevac Conrad, Nata{\v{s}}a and Djurdjevac, Ana and Winkelmann, Stefanie and Sch{\"u}tte, Christof},
    journal={Communications in Applied Mathematics and Computational Science},
    volume={16},
    number={1},
    pages={1--32},
    year={2021},
    publisher={Mathematical Sciences Publishers},
    doi = {10.2140/camcos.2021.16.1}
}

@article{cornalba2023dean,
    title={The {D}ean--{K}awasaki equation and the structure of density fluctuations in systems of diffusing particles},
    author={Cornalba, Federico and Fischer, Julian},
    journal={Archive for Rational Mechanics and Analysis},
    volume={247},
    number={5},
    pages={76},
    year={2023},
    publisher={Springer},
    doi = {10.1007/s00205-023-01903-7}
}

@article{coifman2006diffusion,
    title = {Diffusion maps},
    journal = {Applied and Computational Harmonic Analysis},
    volume = {21},
    number = {1},
    pages = {5--30},
    year = {2006},
    note = {Special Issue: Diffusion Maps and Wavelets},
    doi = {10.1016/j.acha.2006.04.006},
    author = {Ronald R. Coifman and Stéphane Lafon}
}

@article{coifman2005geometric,
    author = {R. R. Coifman  and S. Lafon  and A. B. Lee  and M. Maggioni  and B. Nadler  and F. Warner  and S. W. Zucker},
    title = {Geometric diffusions as a tool for harmonic analysis and structure definition of data: {D}iffusion maps},
    journal = {Proceedings of the National Academy of Sciences},
    volume = {102},
    number = {21},
    pages = {7426--7431},
    year = {2005},
    doi = {10.1073/pnas.0500334102}
}

@article{coifman2006geometric,
    title = {Geometric harmonics: {A} novel tool for multiscale out-of-sample extension of empirical functions},
    journal = {Applied and Computational Harmonic Analysis},
    volume = {21},
    number = {1},
    pages = {31--52},
    year = {2006},
    note = {Special Issue: Diffusion Maps and Wavelets},
    doi = {10.1016/j.acha.2005.07.005},
    author = {Ronald R. Coifman and Stéphane Lafon}
}

@article{rabin2009,
    author = {Rabin, Julien and Delon, Julie and Gousseau, Yann},
    year = {2009},
    title = {Transportation Distances on the Circle},
    volume = {41},
    journal = {Journal of Mathematical Imaging and Vision},
    doi = {10.1007/s10851-011-0284-0}
}

@article{scikit-learn,
  title={Scikit-learn: Machine Learning in {P}ython},
  author={Pedregosa, F. and Varoquaux, G. and Gramfort, A. and Michel, V.
          and Thirion, B. and Grisel, O. and Blondel, M. and Prettenhofer, P.
          and Weiss, R. and Dubourg, V. and Vanderplas, J. and Passos, A. and
          Cournapeau, D. and Brucher, M. and Perrot, M. and Duchesnay, E.},
  journal={Journal of Machine Learning Research},
  volume={12},
  pages={2825--2830},
  year={2011}
}

@article{lloyd1982,
  author={Lloyd, S.},
  journal={IEEE Transactions on Information Theory}, 
  title={Least squares quantization in {PCM}}, 
  year={1982},
  volume={28},
  number={2},
  pages={129--137},
  doi={10.1109/TIT.1982.1056489}
}

@article{deuflhard2005robust,
    title={Robust {P}erron cluster analysis in conformation dynamics},
    author={Deuflhard, Peter and Weber, Marcus},
    journal = {Linear Algebra and its Applications},
    volume = {398},
    pages = {161--184},
    year = {2005},
    note = {Special Issue on Matrices and Mathematical Biology},
    doi = {10.1016/j.laa.2004.10.026}
}

@book{norris1998markov,
  title={Markov Chains},
  author={Norris, James R},
  year={1998},
  publisher={Cambridge University Press}
}

@unpublished{ioannou2025data,
  title={Data-driven approximation of transfer operators for mean-field stochastic differential equations},
  author={Ioannou, Eirini and Klus, Stefan and Reis, Gon{\c{c}}alo dos},
  journal={arXiv preprint arXiv:2509.09891},
  year={2025},
  note = {Working paper or preprint},
  url={https://arxiv.org/abs/2509.09891} 
}

@article{Sadeghi2023,
    author = {Sadeghi, Mohsen},
    title = {Formation of membrane invaginations by curvature-inducing peripheral proteins: {F}ree energy profiles, kinetics, and membrane-mediated effects},
    year = {2023},
    doi = {10.1101/2022.11.09.515891},
    publisher = {Cold Spring Harbor Laboratory},
    journal = {bioRxiv}
}

@book{winkelmann2020stochastic,
  title={Stochastic Dynamics in Computational Biology},
  author={Winkelmann, Stefanie and Sch{\"u}tte, Christof},
  volume={645},
  year={2020},
  publisher={Springer}
}

@article{sadeghi2021,
    author = {Sadeghi, Mohsen and Noé, Frank},
    year = {2021},
    pages = {10497--10504},
    title = {Thermodynamics and Kinetics of Aggregation of Flexible Peripheral Membrane Proteins},
    volume = {12},
    number = {43},
    journal = {The Journal of Physical Chemistry Letters},
    doi = {10.1021/acs.jpclett.1c02954}
}

@Misc{msmtools,
  title        = {{MSMTools: Tools} for Estimating and Analyzing {M}arkov State Models},
  author       = {Martin K. Scherer and contributors},
  howpublished = {\url{https://github.com/markovmodel/msmtools}},
  year         = {2021},
  version      = {1.2.4},
  note         = {Open-source Python package, LGPLv3+},
}

@article{helfmann2020extending,
    title={Extending transition path theory: {P}eriodically driven and finite-time dynamics},
    author={Helfmann, Luzie and Ribera Borrell, Enric and Sch{\"u}tte, Christof and Koltai, P{\'e}ter},
    journal={Journal of Nonlinear Science},
    volume={30},
    number={6},
    pages={3321--3366},
    year={2020},
    publisher={Springer},
    doi = {10.1007/s00332-020-09652-7}
}

@unpublished{gerber2025,
    title={Formation of clusters and coarsening in weakly interacting diffusions}, 
    author={Nicolai Gerber and Rishabh Gvalani and Martin Hairer and Greg Pavliotis and André Schlichting},
    year={2025},
    note = {Working paper or preprint},
    url={https://arxiv.org/abs/2510.17629}
}

@article{klus2016,
    title = {On the numerical approximation of the {Perron-Frobenius and Koopman} operator},
    journal = {Journal of Computational Dynamics},
    volume = {3},
    number = {1},
    pages = {51--79},
    year = {2016},
    doi = {10.3934/jcd.2016003},
    author = {Stefan  Klus and Péter Koltai and Christof  Schütte}
}

@book{gardiner2009stochastic,
  title={Handbook of Stochastic Methods},
  author={Gardiner, Crispin},
  edition={3rd},
  year={2004},
  publisher={Springer Berlin Heidelberg},
  chapter={4}
}

@unpublished{leimkuhler2025cluster,
  title={Cluster Formation in Diffusive Systems},
  author={Leimkuhler, Benedict and Lohmann, Ren{\'e} and Pavliotis, Grigorios A and Whalley, Peter A},
  year={2025},
  note = {Working paper or preprint},
  url={https://arxiv.org/abs/2510.25034},
}

@book{ulam1960,
    author = {Ulam, Stanislaw M.},
    year = {1960},
    title = {A Collection of Mathematical Problems},
    publisher = {Interscience Publisher NY}
}

@book{lasota1994,
    title={Chaos, Fractals, and Noise: {S}tochastic Aspects of Dynamics},
    author={Andrzej Lasota and Michael C. Mackey},
    series={Applied
    Mathematical Sciences},
    year={1994},
    edition ={2nd},
    publisher={Springer New York}
}

@book{levin2017,
    title={Markov Chains and Mixing Times: Second Edition},
    author={David A. Levin and Yuval Peres},
    year={2017},
    publisher={American Mathematical Society}
}

@Inbook{carrillo2010swarm,
    author="Carrillo, Jos{\'e} A.
    and Fornasier, Massimo
    and Toscani, Giuseppe
    and Vecil, Francesco",
    title="Particle, kinetic, and hydrodynamic models of swarming",
    bookTitle="Mathematical Modeling of Collective Behavior in Socio-Economic and Life Sciences",
    year="2010",
    publisher="Birkh{\"a}user Boston",
    address="Boston",
    pages="297--336",
    doi="10.1007/978-0-8176-4946-3_12"
}

@article{schneide2019,
  title = {Lagrangian coherent sets in turbulent {R}ayleigh-{B}\'enard convection},
  author = {Schneide, Christiane and Stahn, Martin and Pandey, Ambrish and Junge, Oliver and Koltai, P\'eter and Padberg-Gehle, Kathrin and Schumacher, J\"org},
  journal = {Physical Review E},
  volume = {100},
  issue = {5},
  pages = {053103},
  year = {2019},
  publisher = {American Physical Society},
  doi = {10.1103/PhysRevE.100.053103}
}

@unpublished{kluenker2025,
    title={Lagrangian description and quantification of scalar mixing in fluid flows from particle tracks}, 
    author={Anna Klünker and Alexandra von Kameke and Kathrin Padberg-Gehle},
    year={2025},
    note = {Working paper or preprint},
    url={https://arxiv.org/abs/2509.25030} 
}

@article{schuette2023, 
    title={Overcoming the timescale barrier in molecular dynamics: {T}ransfer operators, variational principles and machine learning}, 
    volume={32}, 
    DOI={10.1017/S0962492923000016}, 
    journal={Acta Numerica}, 
    author={Schütte, Christof and Klus, Stefan and Hartmann, Carsten}, 
    year={2023}, 
    pages={517--673}
}

@article{blaskovic2025,
    author = {Blašković, Filip and Conrad, Tim O. F. and Klus, Stefan and Djurdjevac Conrad, Nataša},
    title = {Random walk based snapshot clustering for detecting community dynamics in temporal networks},
    journal = {Scientific Reports},
    pages = {24414},
    volume = {15},
    issue = {1},
    year = {2025},
    doi = {10.1038/s41598-025-09340-0}
}

@article{konarovskyi2019,
    author = {Konarovskyi, Vitalii and Lehmann, Tobias and Renesse, Max},
    year = {2019},
    pages = {1--9},
    title = {{Dean--Kawasaki} dynamics: ill-posedness vs. triviality},
    volume = {24},
    journal = {Electronic Communications in Probability},
    doi = {10.1214/19-ECP208}
}

@book{bracewell2000,
  title={The {F}ourier Transform and Its Applications},
  author={Bracewell, R.N.},
  series={Circuits and systems},
  year={2000},
  publisher={McGraw--Hill},
  edition={3rd}
}

@article{gerster2020,
    author = {Gerster, Moritz and Berner, Rico and Sawicki, Jakub and Zakharova, Anna and Škoch, Antonín and Hlinka, Jaroslav and Lehnertz, Klaus and Schöll, Eckehard},
    title = {{FitzHugh--Nagumo} oscillators on complex networks mimic epileptic-seizure-related synchronization phenomena},
    journal = {Chaos: An Interdisciplinary Journal of Nonlinear Science},
    volume = {30},
    number = {12},
    pages = {123130},
    year = {2020},
    doi = {10.1063/5.0021420}
}

@article{djurdjevacconrad2024,
    author = {Djurdjevac Conrad, Nataša and Quang Vu, Nhu and Nagel, Sören},
    title = {Co-evolving networks for opinion and social dynamics in agent-based models},
    journal = {Chaos: An Interdisciplinary Journal of Nonlinear Science},
    volume = {34},
    number = {9},
    pages = {093116},
    year = {2024},
    doi = {10.1063/5.0226054}
}
\bibliographystyle{abbrvnat}

\end{document}